\documentclass[aps,pre,epsfigure,twocolumn,reprint,showpacs,superscriptaddress,nofootinbib,floatfix]{revtex4-1}
\usepackage[colorlinks=true,linkcolor=blue,urlcolor=blue,citecolor=blue,pdfusetitle]{hyperref}
\usepackage[utf8]{inputenc}
\usepackage[english]{babel}
\usepackage{amsmath}
\usepackage[caption = false]{subfig}
\usepackage{graphicx,epstopdf}
\usepackage{blindtext}
\usepackage{lipsum}
\usepackage{amsfonts}
\usepackage{bbm}
\usepackage{amssymb}
\usepackage{enumerate}
\usepackage{color}
\usepackage{latexsym}
\usepackage{physics}
\usepackage{times,txfonts}

\newcommand{\1}{\mathbbm{1}}

\begin{document}

\title{Quantum battery charging by non-equilibrium steady-state currents} 

\author{F. H. Kamin}
\email{F.Hatami@uok.ac.ir}
\affiliation{Department of Physics, University of Kurdistan, P.O.Box 66177-15175 , Sanandaj, Iran}

\author{Z. Abuali}
\email{Z.Abuali@uok.ac.ir}
\affiliation{Department of Physics, University of Kurdistan, P.O.Box 66177-15175 , Sanandaj, Iran}

\author{H. Ness}
\email{herve.ness@kcl.ac.uk}
\affiliation{Department of Physics, King's College London, Strand, London WC2R 2LS, UK}

\author{S. Salimi}
\email{ShSalimi@uok.ac.ir}
\affiliation{Department of Physics, University of Kurdistan, P.O.Box 66177-15175 , Sanandaj, Iran}

\begin{abstract}
We present an analysis of the availability and maximum extractable work of quantum batteries in the presence of charge and/or heat steady-state currents. Quantum batteries are modeled as non-interacting open quantum systems (mesoscopic systems) strongly coupled to two thermal and particle reservoirs within the framework of non-equilibrium Green's function theory in a steady-state regime. We found that the battery can be charged manifestly by a steady-state charge current compared to heat one, especially, in an off-resonant transport regime. It allows us to reliably access the performance of the quantum batteries in the high bias-charging regime.
\end{abstract}

\maketitle

\section{Introduction}
Quantum thermodynamics is concerned with the interchange of energy and matter between microscopic systems and their environments, as well as their description in terms of thermodynamic quantities such as heat, work, entropy, etc. \cite{Alicki2018}. 
	In recent decades, quantum transport has received a lot of attention, e.g., heat and charge transport through molecular junctions \cite{Nitzan:1,pop:1,Bergfield:1}. At the atomic level, a temperature (chemical potential) gradient causes charge carriers in materials to disperse from hot (high potential) to cold (low potential) and this effect can be utilized to measure temperature, generate electricity, and so on. It is no secret that transport phenomena are of great importance to various types of scientific research, including physics. Also, quantum transport has been extensively studied in order to continue progress in nanofabrication. Moreover, recent advances in nanoscale fabrication techniques have led to the theoretical and experimental developments of non-equilibrium (NE) quantum impurity systems \cite{goldhaber:1,Cronenwett:1,SCHMID1998182,lerner:1,Glazman:1}.  Quantum impurities are commonly known as quantum dots. In this type of system with an initial NE state, energy and particles are exchanged between the system and the environment to restore equilibrium. This equilibrium is well understood for classical systems, where it usually leads to thermal stability. Therefore, a NE steady-state current occurs across a quantum dot (central region) when it is connected to several leads at different temperatures and chemical potentials. Studies of NE steady-states have shown that they continuously dissipate energy to their surroundings, in contrast with equilibrium states. Consequently, this leads to continuous entropy production and a time-reversal symmetry breakdown.

Today, quantum batteries (QBs) represent a vital field of research that concerns designing optimal energy storage protocols for the transfer to quantum devices. As now, a variety of theoretical efforts have been made, including examining how quantum resources affect 
QB performance \cite{Campaioli:1,Andolina:1,Kamin:3,Cruz:1,Rossini:2020,Gyhm:2022,Bozkurt:2018},  
presenting models for achieving optimal mechanisms for batteries such as high charging and capacity \cite{Ferraro:1,Rossini:1,Crescente:1,Le:1,Chen:1}, slow erosion \cite{Bai:1,pirmoradian:1}, to discussing the environmental effects on charging and discharging of QBs \cite{Barra:1,Farina:1,Liu:1,Quach:1,Kamin:1,Tabesh:1,Carrega:1}. Furthermore, several experimental platforms have been studied to realize operational quantum batteries \cite{James:1,wenniger:1,Hu:1,Zheng:1,Gemme:1,Joshi:1}. In this regard, we can address the use of an organic semiconductor that is composed of two-levels systems connected to a microcavity \cite{James:1}. Alternatively, QBs can be represented by semiconductor quantum dots embedded within optical microcavities, where energy is exchanged between the solid-state qubit and light fields during charging and discharging \cite{wenniger:1}. Superconducting circuits are also another field of experimental research for quantum batteries \cite{Hu:1,Zheng:1}. An example is the transmon qutrit QB, which is composed of a three-level transmon coupled to an external field. In this model, to avoid unwanted spontaneous discharge or attenuation, a stimulated Raman adiabatic passage is incorporated into the charging to ensure a stable charging process \cite{Hu:1}. Moreover, IBM quantum chips have been introduced as stable and optimal quantum batteries regarding charging time and stored energy \cite{Gemme:1}, and nuclear magnetic resonance (NMR) architecture has been employed to investigate quantum benefits in collectively charging spin systems \cite{Joshi:1}.

In a variety of different scenarios, a cyclic unitary process is usually employed as the best method of maximum work extraction. Nonetheless, stable charging and optimal energy transfer processes are critical for QBs. Typically, the quantum system, which is a battery or charger, interacts with the external environment, leading to decoherence and quantum resource destruction. Due to this interaction, the entropy level of the battery increases, and so unitary evolution applied to the system tend not to be sufficient to rectify any entropy production and ultimately stabilize the system. In fact, the presence of decoherence effects of the environment during the charging process plays a negative role in the performance of operational QBs \cite{pirmoradian:1,Barra:1,Farina:1,Liu:1,Quach:1,Kamin:1,Tabesh:1,Carrega:1}. Moreover, the self-discharge phenomenon is the result of such interactions \cite{Kamin:1,Santos:1}. So far, attempts have been made to avoid the inevitable interactions of the QB with the environment, which may lead to its deactivation over time \cite{pirmoradian:1}. 
	However, some approaches can change the destructive role of the environment from negative to positive. Where compared with a cyclic unitary process, non-unitary discharging provides more charge through availability or exergy \cite{Kamin:2}. In such situations, the steady-state of an open quantum system provides the desired quantum resource. It is therefore possible to design QBs that do not wear out in the presence of environmental effects. These considerations open up a new path for using quantum impurity models to study quantum battery charging by converting energy into extractable work. Such transformations are executed in many steady-state mesoscopic or nanoscale systems through a steady-state current of microscopic particles such as electrons and photons \cite{Datta:1,Ventra:1}. Thus, new insights into steady-state electron currents can be used to discover how various QBs are charged. In this sense, for example, batteries can be characterized as energy-converting devices. \cite{Goldsmid:1,Rowe:1,komoto:1,Dresselhaus:1}. Also, it is worthwhile to emphasize that the protocol described here can be applied to practically all existing impurity systems, from single quantum dots, to double quantum dots, etc. As a matter of fact, finding ways to fully charge quantum batteries from a quantum thermodynamic perspective will be highly crucial, which is our main purpose.
%

Typically, we consider the open batteries with a few degree of freedom, but presume that they strongly interact with macroscopic heat reservoirs. We present a model for charge and heat transfer based on the non-equilibrium Green’s function (NEGF) formalism \cite{Maciejko:1,Stefanucci:1}. Moreover, we fix the boundary values $(T_{L},\mu_{L})$ and $(T_{R},\mu_{R})$ for the temperature and chemical potential of left and right reservoirs, respectively. In light of these statements, we investigate the charging process of a QB via a transport set consisting of a central quantum system (quantum battery) in contact with a pair of electron reservoirs in the local equilibrium. We suggest an optimal bias-charging process of the QBs by applying the appropriate bias (chemical potential difference), and consequently the charge current, in a specific transport regime.

The paper is organized as follows. In Sec. \ref{II} we present the general procedure for construction of the work extraction for the QB. The special model of battery and reservoirs is discussed in Sec. \ref{III}.
In Sec. \ref{IV} we discuss some illustrative results of model. Conclusions are presented in Sec. \ref{V}. In Appendix \ref{appA}, we compare our method with an alternative approach based on reduced density matrices. Finally, a detailed discussion of charge and energy currents is included in Appendix \ref{appB}.

\section{Figure of merit}\label{II}

A QB with non-equilibrium steady-state characteristics is proposed in this work. Where, by interacting with the environment, it makes use of resources such as heat and/or charge currents to accomplish useful and accessible work. 

\begin{figure}[t] 
	\includegraphics[scale=0.55]{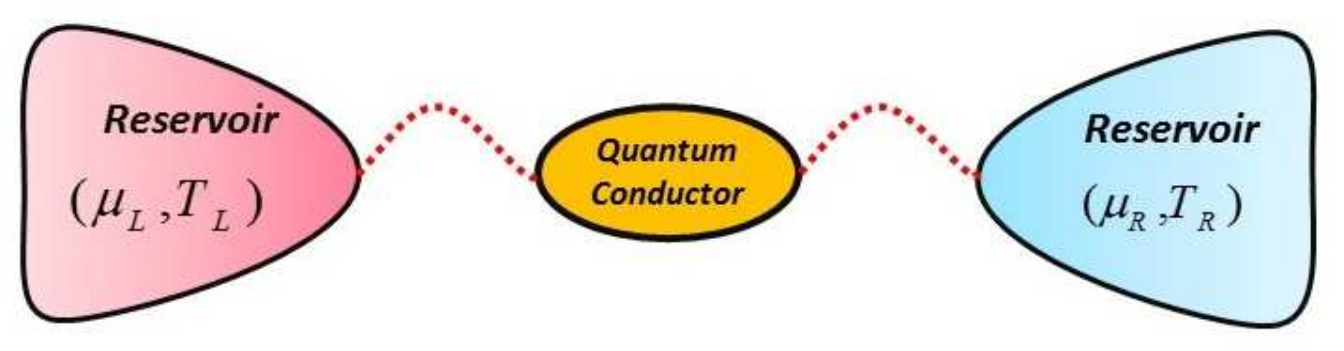}
	\caption{A mesoscopic system, composed of a quantum conductor (QB) connected to two particle and heat reservoirs, left $L$ and
		righ $R$, at their own local equilibrium.}
	\label{Meso}
\end{figure}

To start, we consider a simplified model of the QB as a central conductor connecting two electron reservoirs $L$ and $R$ in their own local thermal equilibrium state.
The reservoirs are characterized by a density matrix $\hat\rho_{i}= Z_{i}^{-1}e^{-\beta_{i}(\hat{H}_{i}-\mu_{i}\hat{N}_{i})}~(i=L,R)$ at two different inverse temperatures $\beta_{L}=\frac{1}{kT_{L}}$ and $\beta_{R}=\frac{1}{kT_{R}}$ and with two chemical potentials $\mu_{L}$ and $\mu_{R}$ (see Fig.~\ref{Meso}). 
Here, $Z_{i}=tr_{i}[e^{-\beta_{i}(\hat{H}_{i}-\mu_{i}\hat{N}_{i})}]$ is the partition function, and $\hat{H}_{i}$ and $\hat{N}_{i}$ are the Hamiltonian and particle number 
operators for each reservoir $i$, respectively. 
The initial density matrix $\hat{\rho}_{0}$ of the decoupled L-QB-R (left reservoir–battery–right reservoir) system is given by the product state $\hat{\rho}_{0}=\hat{\rho}_{L}\otimes\hat{\rho}_{QB}\otimes\hat{\rho}_{R}$, where $\hat{\rho}_{QB},~\hat{\rho}_{L}$, and $\hat{\rho}_{R}$ denote the density matrix of battery, left and right reservoir, respectively. Since quantum battery QB is not in a thermodynamic limit, $\hat{\rho}_{QB}$ is assumed to be arbitrary. 
Once the battery is connected to the reservoirs, the time-evolution of the entire system is ruled by the full Hamiltonian $\hat{H}=\hat{H}_{0}+\hat{H}_{int}$, where $\hat{H}_{0}=\hat{H}_{L}+\hat{H}_{QB}+\hat{H}_{R}$ displays the sum of independent free Hamiltonians in each part and $\hat{H}_{int}$ characterizes the interaction between QB and $i$th reservoir. In the long time limit, the reservoirs drive the system into a global non-equilibrium (NE) steady-state.

The NE steady-state regime is described as the Gibbs-like ensembles that can be obtained either by using the McLennan-Zubarev approaches \cite{McLennan:1,McLennan:2,Zubarev:1,Zubarev:2} or the NE density matrix approach developed by Hershfield in Ref.~\cite{Hershfield:1}, which provides a thorough description of the NE steady-state behavior. Recently, the full equivalence between the McLennan-Zubarev NE statistical operator and Hershfield's approach for the NE steady-state has been shown in Ref.~\cite{Ness:1,Ness:1a}. By definition, the NE density matrix is given by ${\hat{\rho}}^{NE}={\hat{\Omega}}^{(+)}\hat{\rho}_{0}\hat{\Omega}^{(+)^{-1}}$, where $\hat{\Omega}^{(+)}=\lim_{\tau\rightarrow\infty}e^{i \hat{H}\tau}e^{-i \hat{H}_{0}\tau}$ is the Moeller operator \cite{Gell-Mann:1,Bohm:1,Baute:1} and characterizes the asymptotic steady state.

Moreover, NE steady-state can be distinguished by its non-zero entropy production rate as well as its ability to maintain non-zero mean currents in the system \cite{meixner:1,Bruch:2}. Where the NE entropy production rate $\sigma$ is directly related to the asymptotic NE steady-state current of particles $I_{Q}$ and energy $J_{E}$ as $\sigma=\Delta_{\mu}I_{Q}-(\beta_{L}-\beta_{R}) J_{E}$ with $\Delta_{\mu}=\beta_{L}\mu_{L}-\beta_{R}\mu_{R}$ \cite{Ness:2,Guarnieri:1}. 

Recent studies have shown that the battery performance suffers from some energy loss due to the non-unitary effects on the QB when we consider the coupling with external heat reservoirs \cite{Kamin:2}. As a result, cyclic unitary transformations cannot retrieve part of the total energy in the system since part of it is not stored as ergotropy. This new perspective assesses the amount of residual energy that cannot be extracted as useful work from open quantum batteries by unitary processes. 
	A non-unitary extraction process can provide maximum  work through availability or exergy \cite{Kamin:2,Friedrich:1}. In thermodynamic terms, a arbitrary system (associated with a density matrix $\hat{\rho}$) that is out of equilibrium with its environment can exchange work and heat with its surroundings. Furthermore, it can transfer pure work energy to a third external system. In this manner, the maximum available work is the so-called availability or exergy \cite{Friedrich:1}, which is defined here as follows
\begin{align}\label{1}
	W_{ext}=\Lambda(\hat{\rho})-\Lambda(\hat{\rho}^{eq})~.
\end{align}
where $\Lambda(\hat{\rho})=E-\mu N-T S(\hat{\rho})$ is the non-equilibrium grand potential with $E=tr(\hat{\rho}~\hat{H})$, $S(\hat{\rho})=-tr(\hat{\rho}\ln{\hat{\rho}})$, and $N=tr(\hat{\rho}~\hat{N})$ being the energy, von Neumann entropy, and particle number of the system, respectively. The superscript $"eq"$ designates the values belonging to the equilibrium state $\hat{\rho}^{eq}=Z^{-1}e^{-\beta~(\hat{H}-\mu\hat{N})}$ of the system at the inverse temperature $\beta=1/kT$. As a result, it can be concluded that $\Lambda(\hat{\rho}^{eq})=-\beta^{-1}\ln{Z}$ for the equilibrium state.

Assuming that $\hat{\rho}$ represents a momentary state of the system, we obtain following Ref.~[\onlinecite{Friedrich:1}]
(with $k=1$)
\begin{align}\label{2}
	S(\hat{\rho}~\|~\hat{\rho}^{eq})=\frac{1}{T}(E-E^{eq})+\frac{\mu}{T}(N-N^{eq})-(S-S^{eq})~, 
\end{align} 
where the relative entropy $S(\hat{\rho}~\|~\hat{\rho}^{eq})=tr[\hat{\rho}(\ln\hat{\rho}-\ln\hat{\rho}^{eq})]$ is the information gain of system. 
Therefore one can find that \cite{Friedrich:1}
\begin{align}\label{3}
	\beta W_{ext}=S(\hat{\rho}~\|~\hat{\rho}^{eq})~. 
\end{align}
Evidently, availability (exergy) is the information gain, up to a factor $\beta$ and it is considered a fundamental quantity both in statistics and physics. 

The next step is to achieve a NE steady-state reduced density matrix for the central system by partial trace of global NE steady-state $\hat{\rho}^{NE}$. However, it may be challenging to determine the exact form of $\hat{\rho}_{QB}$ in a long time limit. In this paper, we use a NE thermodynamic description of 
the system based on the non-equilibrium Green functions (NEGF) theory. This method provides standard definitions for energy, charge and heat currents, system entropy, and exergy (see Sec.~\ref{III}).

\section{Model}\label{III}
A simple QB can be modeled as a single-level quantum dot connected to two one-dimensional electron reservoirs (tight-binding approach to non-interacting electrons). The free Hamiltonian of the L-QB-R system is given by
\begin{eqnarray}
	H_{0}&=& \varepsilon_{QB} \hat{d}^{\dagger}\hat{d}
	+ \sum_{\alpha=L,R}\sum _{k=0}^{\infty}\varepsilon_{\alpha}\hat{c}^{\dagger}_{\alpha,k}\hat{c}_{\alpha,k}	
	- h_{\alpha}(\hat{c}^{\dagger}_{\alpha,k-1}\hat{c}_{\alpha,k}+c.c.)~,
\end{eqnarray}
with the energy electron level $\varepsilon_{QB}$ and the creation and annihilation operators $\hat d^{\dagger},\hat d$ of an electron on the QB. 
$\varepsilon_{\alpha}$ is the energy and $\hat c_{\alpha,k}$, $\hat c^{\dagger}_{\alpha,k}$ are the fermionic annihilation and creation operators of $k$th mode of the $\alpha=L,R$ reservoir, respectively. The central system (QB) interacts with the reservoirs by electron tunneling term
\begin{align} 
	H_{int}= -\sum_{\alpha=L,R}\nu_{\alpha}(\hat{c}^{\dagger}_{\alpha 0}\hat{d}+ \hat{d}^{\dagger}\hat{c}_{\alpha 0})~,
\end{align}
where $\nu_{\alpha}$ is the interaction strength between the QB and the $\alpha=L,R$ reservoir. It is worth noting that our proposed scenario can be used to model the charge transmission phenomenon. 
The electrons in the reservoirs are described by Fermi-Dirac (FD) equilibrium distribution functions $f_{\alpha}^{eq}(\omega;\mu_{\alpha},T_{\alpha})=[e^{\beta_{\alpha}(\hbar\omega-\mu_{\alpha})}+1]^{-1}$. 
Applying a temperature gradient $\Delta T =T_L -T_R$ or/and bias voltage $\Delta\mu=\mu_L-\mu_R$ leads to heat or/and electron transfer between 
the reservoirs. 

We analyze the typical transport in which a QB is connected to two reservoirs that are first stored at different temperatures and chemical potentials. 
Over a long period of time, the system reaches a NE steady-state with a mean rate of non-interacting quantum charge current $I_{Q}$ and energy current $J_{E}$. 
The Landauer-B\"uttiker formalism describes the NE steady-state regime by a transmission probability   
\begin{align}
	\mathcal{\tau}(\omega)=G^{r}(\omega)\Gamma_{L}(\omega)G^{a}(\omega)\Gamma_{R}(\omega)~,
\end{align}
for charge and energy transport at energy $\omega~(\hbar=1)$.
It can be obtained from the retarded Green function $G^{r}(\omega)=[\omega-\varepsilon_{QB}-\Sigma^{r}_{L}(\omega)-\Sigma^{r}_{R}(\omega)]^{-1}$ and advanced Green function $G^{a}(\omega) = [G^r(\omega)]^*$ of the QB and the so-called leads (reservoirs) self-energies $\Sigma^{r}_\alpha(\omega)$
	where $\Sigma^{r}_{\alpha}(\omega)=\nu^{2}_{\alpha}e^{-ik_{\alpha}(\omega)}/h_{\alpha}$ 
	using the energy dispersion relation $\omega=\varepsilon_{QB}-2h_{\alpha}\cos(k_{\alpha})$ of the $\alpha=L,R$ reservoir \cite{Ventra:1,Ness:2}. Each reservoir is associated with a spectral function $\Gamma_{\alpha}(\omega)=-2\Im(\Sigma^{r}_{\alpha}(\omega))$.

The charge $I_Q$ and energy $J_E$ currents are given by ($e=1,\hbar=1$) \cite{Datta:1,Esposito:1,Ness:2}
\begin{align}
	I_{Q}=\frac{1}{2\pi}\int~d\omega~\mathcal{\tau}(\omega)~(f_{L}(\omega)-f_{R}(\omega))~,
\end{align}
and
\begin{align}
	J_{E}=\frac{1}{2\pi}\int~d\omega~\mathcal{\tau}(\omega)~\omega~(f_{L}(\omega)-f_{R}(\omega))~,
\end{align}
respectively. Thus, we can define the heat current as $J^{\alpha}_{H}=J_{E}-\mu_{\alpha}I_{Q}$ for each reservoir $\alpha$.

We are interested in calculating the NE thermodynamical properties of the QB connected to the reservoirs. The number of particles, the energy, the entropy and the exergy in the QB can be obtained from the reduced density matrix $\hat{\rho}_{QB}={\rm Tr}_{L,R}[\hat{\rho}^{NE}]$, which is derived from the trace of the full density matrix $\hat{\rho}^{NE}$ (defined in Section \ref{II}) over the degrees of freedom of the
	$L$ and $R$ reservoirs. As mentioned above, summing over the infinite number of degrees of freedom in the reservoirs can be a difficult task (this is even more true in the presence of particle interaction in the QB).
	An alternative way is to use the so-called NE distribution function of the QB connected to the reservoirs. Such a distribution function is well defined within the NEGF formalism \cite{Meir:1992,Ness:2014}, and it describes the correct statistics of the electron(s) in the QB under the NE (steady-state) conditions.
	Note that, in Appendix \ref{appA}, we discuss in detail the differences between calculations performed with the correct NE distribution function and with an approximated expression for the reduced density matrix $\hat{\rho}_{QB}$.

For the model system we consider here, i.e. a single electron level connected to two reservoirs, the NE distribution function $f_{QB}^{NE}$ of the QB is given by a linear combination of the FD distributions of reservoirs weighted by the ``strength'' of the coupling of the QB and the reservoirs. It is expressed as follows \cite{Meir:1992,Ness:2014}:
	\begin{align}
		f_{QB}^{NE}(\omega)=\frac{\Gamma_{L}(\omega)f_{L}(\omega)+\Gamma_{R}(\omega)f_{R}(\omega)}{\Gamma_{L}(\omega)+\Gamma_{R}(\omega)}~,
	\end{align}
	where the reservoir spectral functions $\Gamma_{L,R}$ are defined above.

We can now calculate the number of particle $N_{QB}^{\rm NE}$, the energy $E_{QB}^{\rm NE}$, the entropy $S_{QB}^{\rm NE}$ of the $QB$ using the spectral function 
	$A_{QB}(\omega)=-\Im G^{r}(\omega)/\pi$ 
	and $f_{QB}^{NE}$ as follows \cite{Esposito:1}:
	\begin{align}\label{N_B}
		N_{QB}^{\rm NE}&=
		\int {{\rm d}\omega}\ A_{QB}(\omega)\ f_{QB}^{\rm NE}(\omega) \\
		E_{QB}^{\rm NE}&=
		\int {{\rm d}\omega}\ \omega\ A_{QB}(\omega)\ f_{QB}^{\rm NE}(\omega)
	\end{align}
	and 
	\begin{align}\label{10}
		S_{QB}^{\rm NE}&=
		- \int {{\rm d}\omega}\ A_{QB}(\omega) \nonumber\\
		&\left[ f_{QB}^{\rm NE}(\omega) \ln f_{QB}^{\rm NE} + (1-f_{QB}^{\rm NE}(\omega)) \ln (1-f_{QB}^{\rm NE}) 
		\right] \ .
	\end{align}
	Finally, from the general definitions Eqs.~\eqref{1} and \eqref{2} and following Ref.~[\onlinecite{Esposito:1}], one obtains a compact expression for the exergy $W_{QB}^{ext}$, expressed in terms of the spectral function $A_{QB}$, the NE distribution function $f_{QB}^{\rm NE}$ and the equilibrium FD distribution $f_{QB}^{eq}=f^{eq}_\alpha$ of the QB
	\begin{align}\label{9}
		\beta W_{ext}&= 
		\int d\omega ~A_{QB}(\omega)\nonumber\\
		&\left[
		f_{QB}^{NE}\ln(\frac{f_{QB}^{NE}}{f_{QB}^{eq}})+(1-f_{QB}^{NE})\ln(\frac{1-f_{QB}^{NE}}{1-f_{QB}^{eq}})
		\right]~. 
	\end{align}
	Now we have all the expressions needed to perform numerical calculations. 
	We are looking for understanding the NE effects of charge and heat currents on the performance of the QB. More specifically, we want to know the advantages of applying a bias along with (or without) a temperature gradient on the charging protocol of the QBs, and find the best conditions to optimize the exergy (and/or the entropy) in the QB. Ultimately, we would like to provide an optimal charging protocol for the QBs.

\section{Results}\label{IV}

Our model system contains several adjustable parameters. For convenience, we choose that the parameters $\varepsilon_{\alpha},~h_{\alpha},~\nu_{\alpha}$ describing the
	$\alpha=L,R$ reservoir are the same for both $L$ and $R$ reservoirs.
	Additionally, in order to avoid atypical results due to the (electron-hole) symmetry of the spectral function $A_{QB}(\omega)$, namely $A_{QB}(\omega)=A_{QB}(-\omega)$, and/or of the distribution functions, we choose that the equilibrium chemical potential $\mu_{eq}$ is different from the band-center $\varepsilon_{\alpha}$ of the reservoirs.
	\subsection{The equilibrium case}
	First, it is instructive to consider the equilibrium case, for which $\mu_L=\mu_R=\mu_{eq}$ and $T_L=T_R=T_{eq}$. At equilibrium, there is no currents since hence $f_L=f_R=f^{eq}_\alpha$. There is also no exergy in the QB since $f_{QB}^{NE}=f_{QB}^{eq}=f^{eq}_\alpha$.

\begin{figure}
	\centering
	{\includegraphics[scale=0.45]{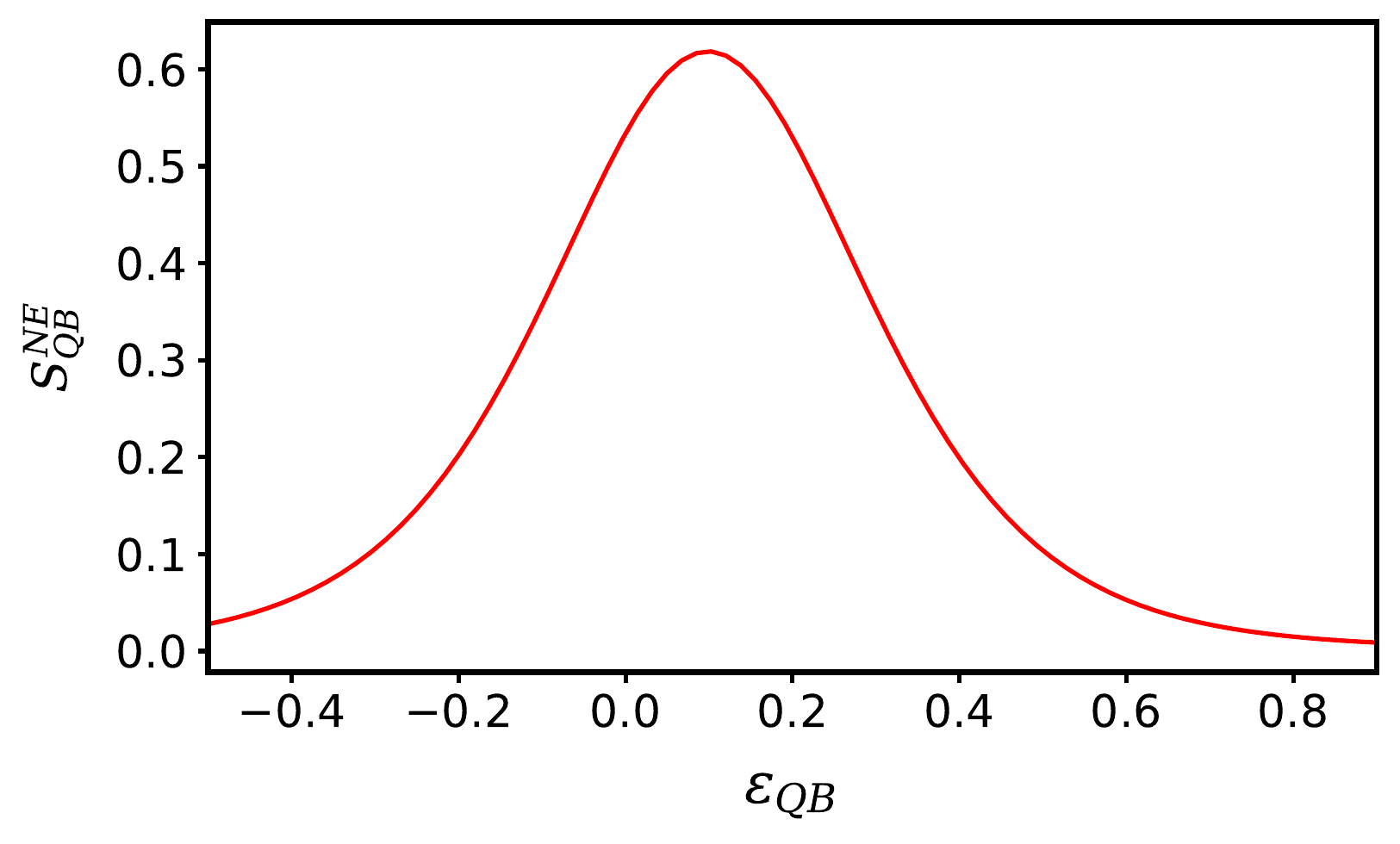}}~
	\caption{Equilibrium entropy in the QB versus the energy level $\varepsilon_{QB}$. The other parameters are $h_{\alpha}=1.0,~\nu_{\alpha}=0.12$,
			$T_L=T_R=T_{eq}=0.1$, and $\mu_L=\mu_R=\mu_{eq}=0.1$.}
	\label{fig:Sequi}
\end{figure}

However, the number of particles, the energy and the entropy of the QB have a finite value, 
which depends on the position of the energy level $\varepsilon_{QB}$ relative to the equilibrium Fermi level $\mu_{eq}$.

We define three regimes:
(1) the resonant transport regime where $\varepsilon_{QB}=\mu_{eq}$, 
i.e. the equilibrium chemical potential $\mu_{eq}$ is located at the peak of the spectral function $A_{QB}(\omega)$, which is also the peak of the transmission probability 
$\mathcal{\tau}(\omega=\varepsilon_{QB})\sim 1$.
One should note that, for the one energy-level model considered here, the spectral function $A_{QB}(\omega)$ as well as 
the transmission $\mathcal{\tau}(\omega)$ are essentially peaked functions, with a maximum at $\omega=\varepsilon_{QB}$ and a width 
given approximatively by $\sim (\Gamma_L(\varepsilon_{QB})+\Gamma_R(\varepsilon_{QB}))/2$.
(2) the off-resonant-empty regime where $\varepsilon_{QB} \gg \mu_{eq}$. The equilibrium chemical potential $\mu_{eq}$ is 
located in the (ascending) tail of the spectral function $A_{QB}(\omega)$, and the energy level $\varepsilon_{QB}$ is mostly empty at equilibrium, i.e. $\int {\rm d}\omega f_{QB}^{NE} A_{QB} \ll 1$.
(3) the off-resonant-full regime where $\varepsilon_{QB} \ll \mu_{eq}$. The equilibrium chemical potential $\mu_{eq}$ is 
located in the (descending) tail of the spectral function $A_{QB}(\omega)$, and the energy level $\varepsilon_{QB}$ is almost full
at equilibrium, i.e. $\int {\rm d}\omega f_{QB}^{NE} A_{QB} \sim 1$.

Fig~\eqref{fig:Sequi} shows the equilibrium entropy versus $\varepsilon_{QB}$. 
	The equilibrium entropy has a maximum value for half filling of the electronic level, i.e. when $\varepsilon_{QB}=\mu_{eq}$, as expected.
	The maximum value is close to the value of $-\ln(1/2)=\ln(2) \sim 0.69$ given by the Landauer's principle.
	The entropy is zero when the electronic level is completely empty ($\varepsilon_{QB}\ll\mu_{eq}$) or completely full ($\varepsilon_{QB}\gg\mu_{eq}$).
	Note that this is simply a quantum electron equivalent of a classical problem in statistical mechanics
\footnote{For the classical problem of $N$ "particles" put on $M$ sites, the
		entropy is given by the logarithm of the number of possible combination to arrange the $N$ particles on
		the $M$ sites. The entropy is maximum when $N$ is at/around half the number of sites.
		When all sites are occupied (or empty), there is only one combination possible, and therefore the entropy is zero.}.

Driving the QB out of equilibrium by applying a bias $\Delta\mu=\mu_L-\mu_R$ or a temperature gradient $\Delta T=T_L-T_R$ will lead to particle/energy transfer between the two reservoirs.

There will be an energy window for which $f_L \ne f_R$. Most of the transfer of particle/energy happens within this energy window. The dependence of the entropy (and of the exergy) versus $\varepsilon_{QB}$ will be modified from the equilibrium case when the energy level $\varepsilon_{QB}$ is located within this energy window as shown in the next section.
\begin{figure}
	\centering
	{\includegraphics[scale=0.45]{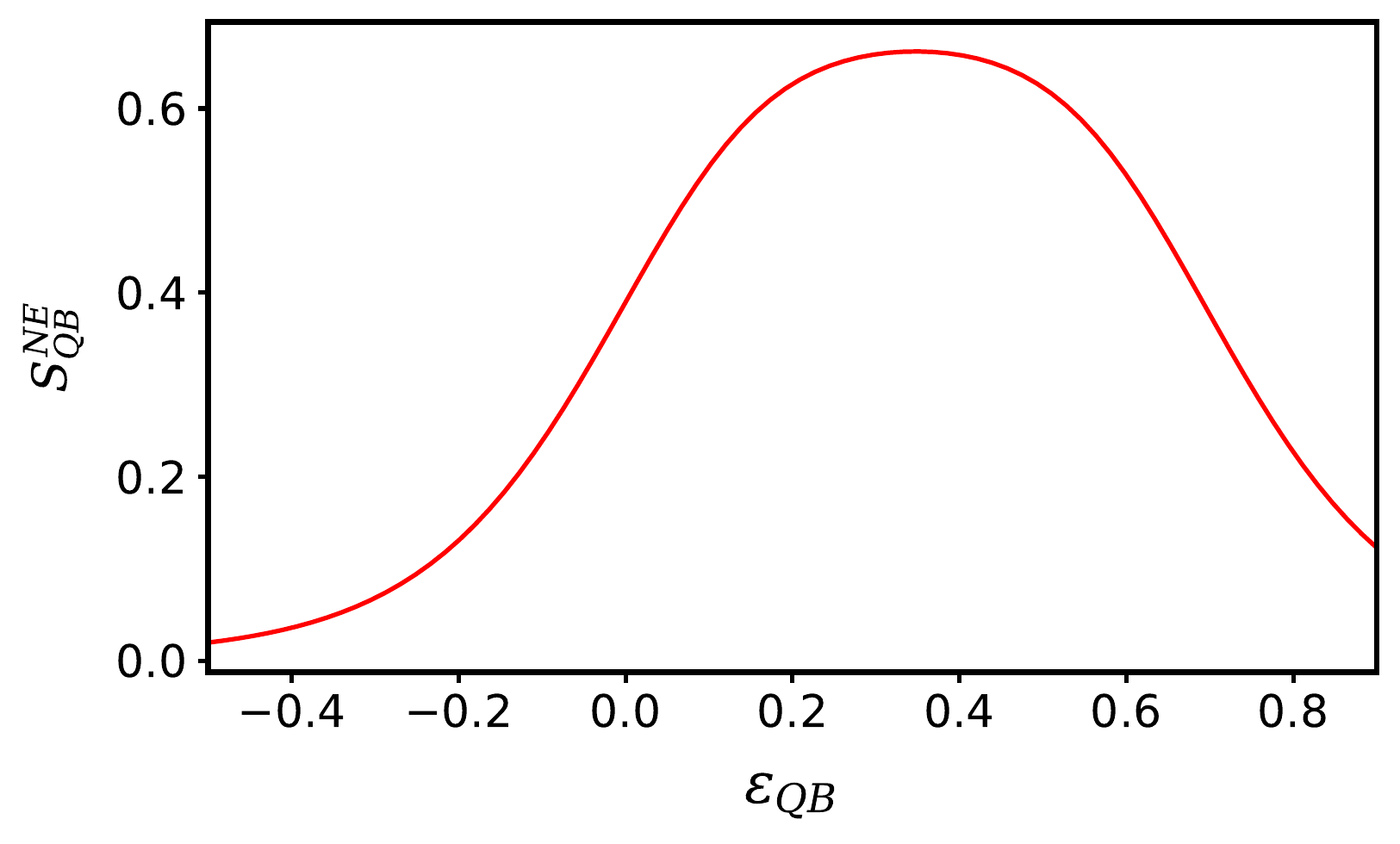}}~
	\caption{Non-equilibrium entropy in the QB versus the energy level $\varepsilon_{QB}$, for an applied bias $\Delta\mu=0.5$
			and $T_L=T_R=T_{eq}$. Other parameters are the same as in Fig.~\ref{fig:Sequi}.}
	\label{fig:SNE_e0}
\end{figure}

\subsection{Non-equilibrium cases}
As a first step towards understanding the QB charging, we will consider that the bias and/or the temperature gradient is applied only on one side of the L-QB-R junctions.
The chemical potential $\mu_R$ and the temperature $T_R$ of the right reservoir is kept constant and equal to the equilibrium values, i.e. $\mu_R=\mu_{eq}$ and $T_R=T_{eq}$.
While a finite bias $\Delta\mu$ and/or temperature gradient $\Delta T$ is applied on the left reservoir, with chemical potential $\mu_L = \mu_{eq} + \Delta\mu$ and temperature $T_L = T_{eq} +\Delta T$.

We have performed calculations for the three regimes (mentioned in the previous section) in the presence of a temperature gradient (with/without applied bias) and in the presence of an applied bias (with/without temperature gradient). All the results for the charge and energy/heat currents are shown in Appendix \ref{appB}. 

As expected, applying a bias $\Delta\mu$ or a temperature gradient $\Delta T$ will generate a charge or an energy current through the QB. 
	This, in turn, increases the entropy and the exergy in the battery. 
	The increase in entropy originates from the increase of possible electron/hole combinations in the QB due to the applied bias or temperature gradient.
	The increase in exergy originates from the charge and energy currents since the latter can be seen 
	as extra work created in the QB by the NE conditions.
As a general trend, the larger $\Delta\mu$ or $\Delta T$, the larger the currents (see Appendix \ref{appB}) and consequently the larger the entropy and the exergy.

Figure \ref{fig:SNE_e0} shows the NE entropy in the QB versus the energy level $\varepsilon_{QB}$ for an applied bias. 
	In comparison to the equilibrium case Fig.~\ref{fig:Sequi}, the NE entropy increases for $\varepsilon_{QB}$ values included in an energy window corresponding to the applied bias. This is to be expected since for these energies $\omega$,
	we have $f_L(\omega)\ne f_R(\omega)$ and correlatively $f^{\rm NE}_{QB}(\omega)\ne f_\alpha(\omega)$ with values 
	allowing for a wider range of energies for which the QB energy level is quasi half-filled.

\begin{figure*}[t!]
	\centering
	\subfloat[ ]{\includegraphics[scale=0.4]{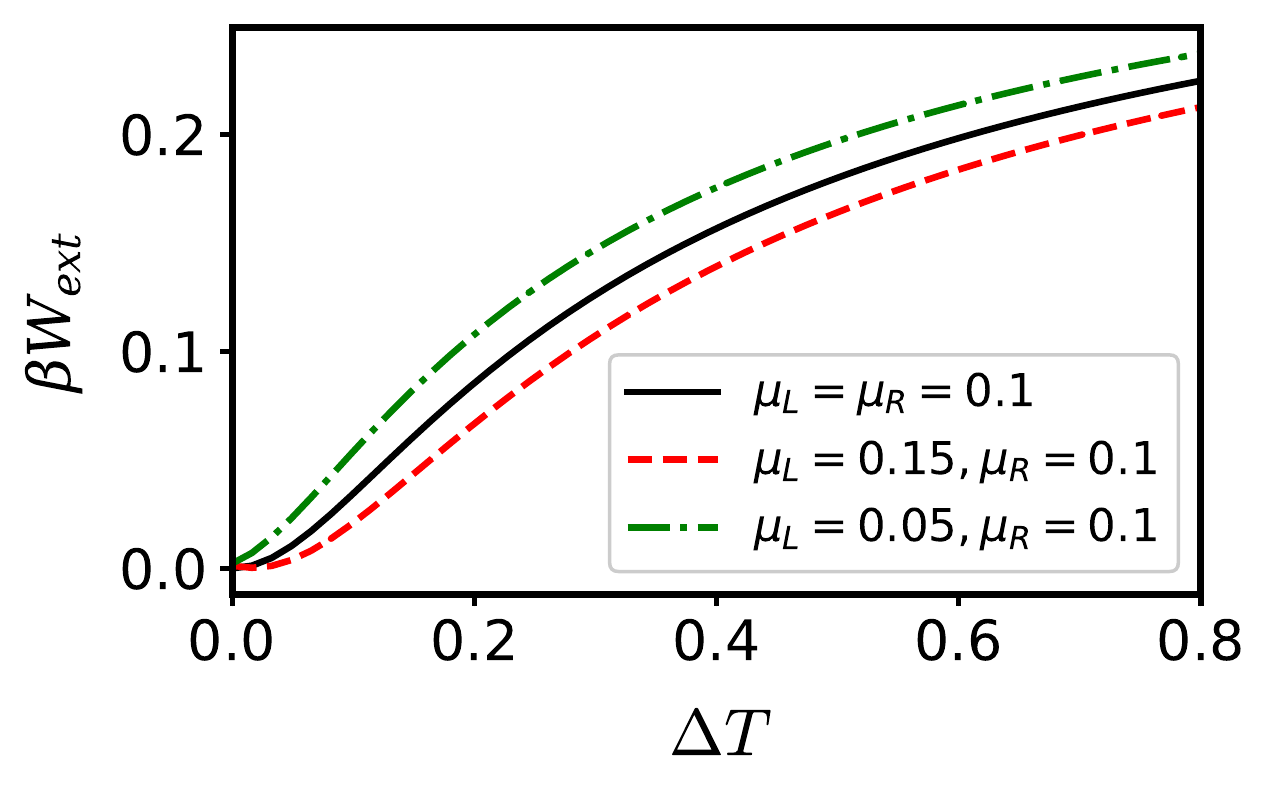}\label{Fig2(a)}}~
	\subfloat[ ]{\includegraphics[scale=0.4]{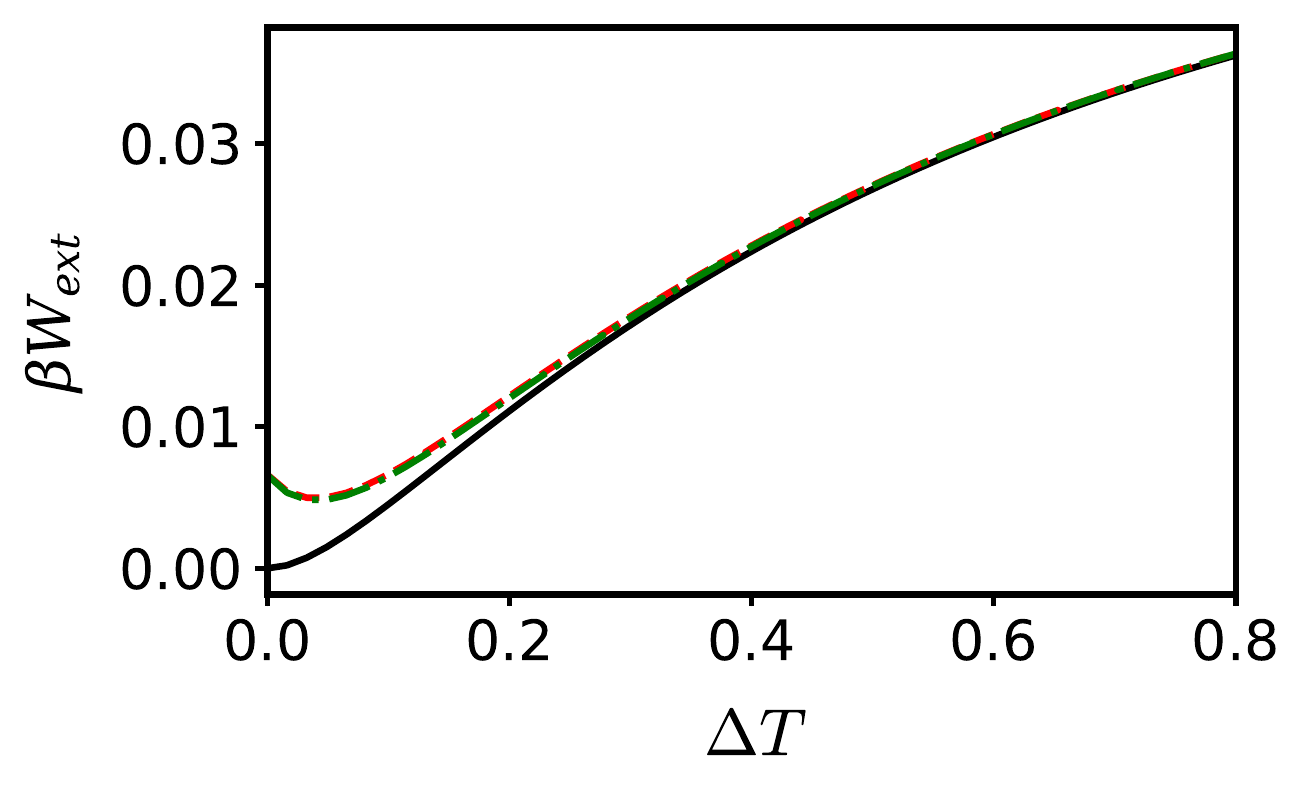}\label{Fig2(b)}}~
	\subfloat[ ]{\includegraphics[scale=0.4]{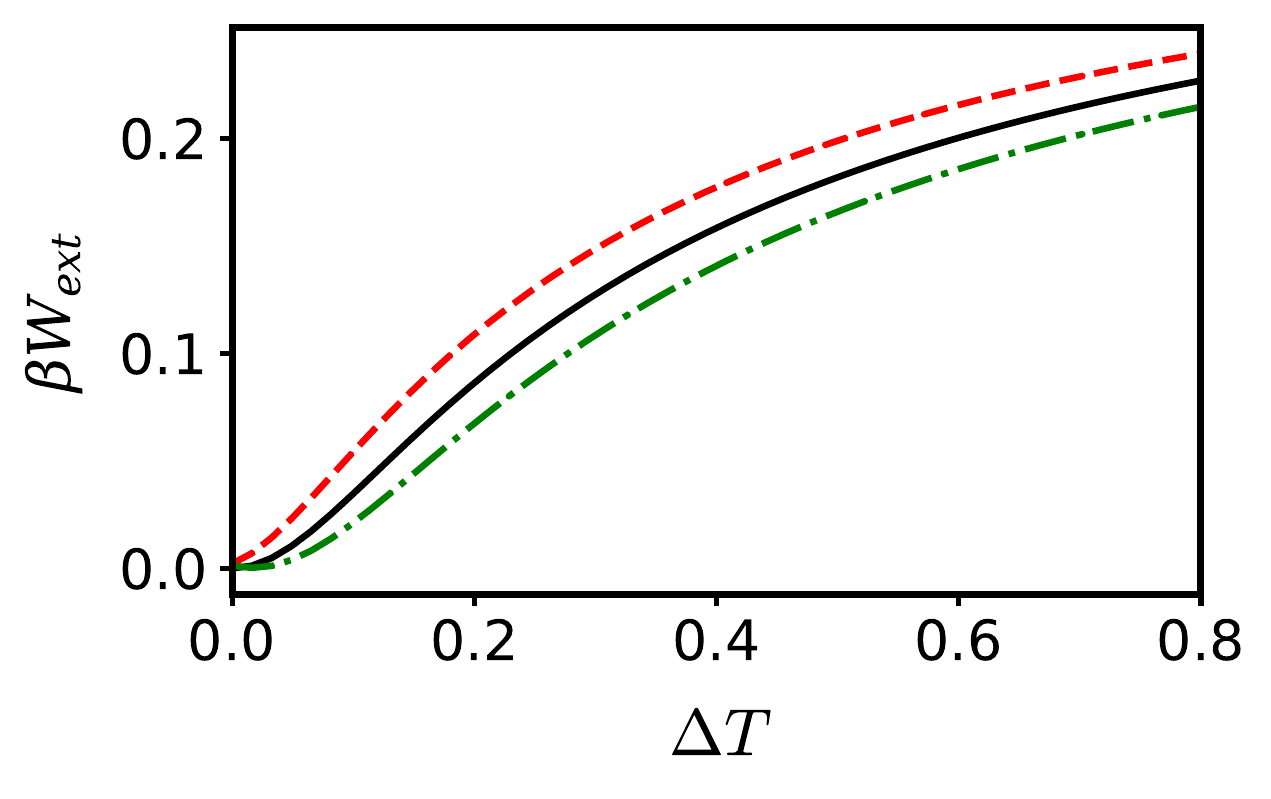}\label{Fig2(c)}}
	\caption{Maximum available work $\beta W_{ext}$ as a function of $\Delta T$ (with additional applied bias), for the different regimes:
		\eqref{Fig2(a)} off-resonant-full $\varepsilon_{QB}=-0.2$, \eqref{Fig2(b)} resonant $\varepsilon_{QB}=0.1$, 
		and, \eqref{Fig2(c)} off-resonant-empty $\varepsilon_{QB}=0.4$. 
		Other parameters are  $\varepsilon_{\alpha}=0,~h_{\alpha}=1.0,~\nu_{\alpha}=0.12$, $T_{eq}=0.1$ and $\mu_{eq}=0.1$.}
	\label{Fig2}
\end{figure*}

\subsubsection{Applied temperature gradient $\Delta T$}
	The exergy results for different temperature gradients are shown in Fig.~\ref{Fig2}.
	As expected, the exergy increases with increasing temperature gradients. However the behavior of $\beta W_{ext} (\beta=\beta_{eq})$ versus $\Delta T$ depends strongly on the position of the energy level of the QB.

For the off-resonant cases, the charge current is dominated by one "type of particle", i.e. by electron for the off-resonant-empty regime and by hole for the off-resonant-full regime. 
	Both thermal and charge currents have opposite sign (compared Fig.~\ref{Fig44(b)} and Fig.~\ref{Fig66(b)}).
	Regardless to the sign of the thermal currents, the currents increase with increasing $\Delta T$
	which therefore leads to an increase of the exergy (see Fig.~\ref{Fig2(a)} and \ref{Fig2(c)}).

For the resonant case, there is a competing contribution from electron and hole processes
	and hence a strong reduction of the thermal charge current (see Figs.~\ref{Fig22(b)}),
	which also corresponds to a reduction of the exergy (see Fig.~\ref{Fig2(b)})
	in comparison to the off-resonant cases.

Adding a small bias $\Delta\mu$ to the temperature gradient $\Delta T$ leads to a modification
	of the charge current. With our convention, a positive (negative) bias $\Delta\mu=\mu_L-\mu_R$ 
	implies electron transfer from left to right (right to the left). This leads to a positive (negative) contribution for the thermal currents.

For the off-resonant-empty regime, an additional positive (negative) bias increases (decreases) the charge current and similarly for the exergy.

For the off-resonant-full regime, an additional positive (negative) bias increases (decreases) the charge current. However, because of the sign of the current, this corresponds to an opposite  behavior for the absolute value of the current. And therefore, an additional positive (negative) bias decreases (increases) the exergy.

\begin{figure*}[t!]
	\centering
	\subfloat[ ]{\includegraphics[scale=0.4]{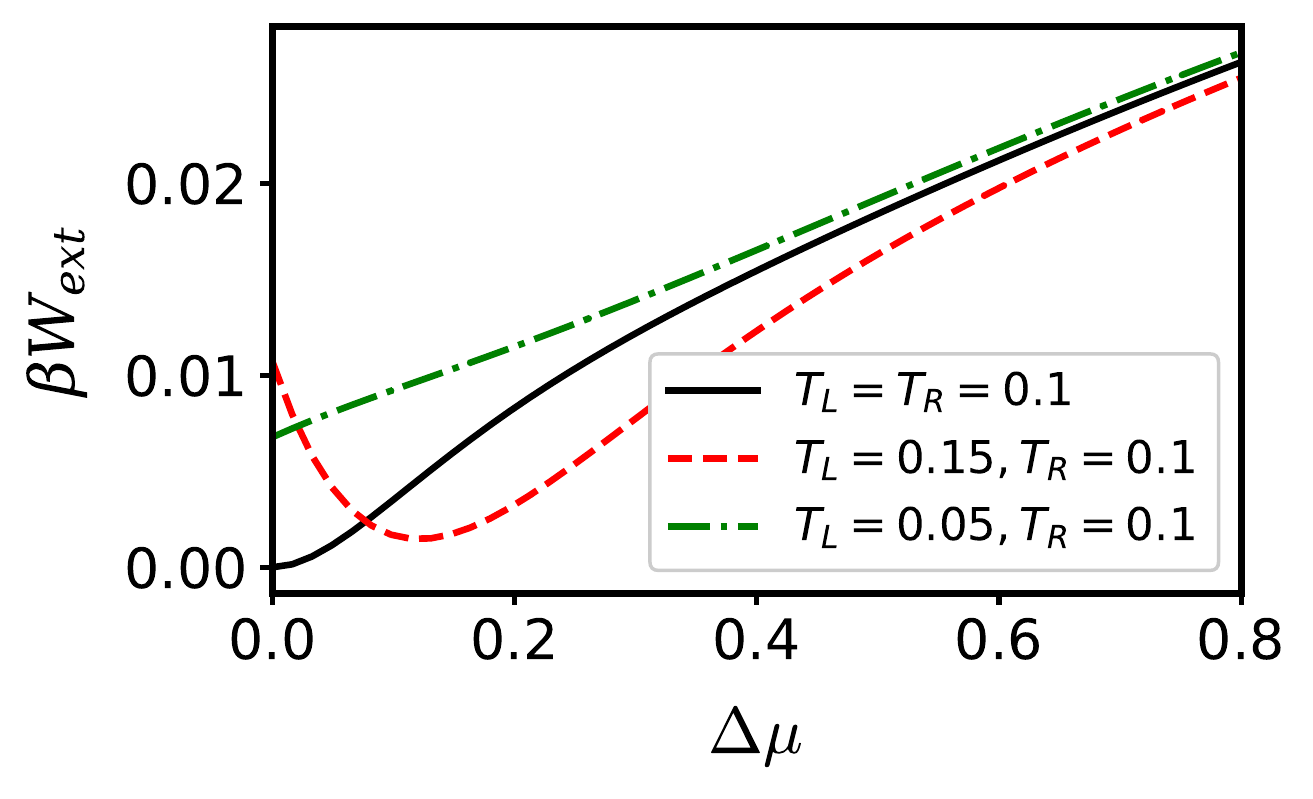}\label{Fig3(a)}}~
	\subfloat[ ]{\includegraphics[scale=0.4]{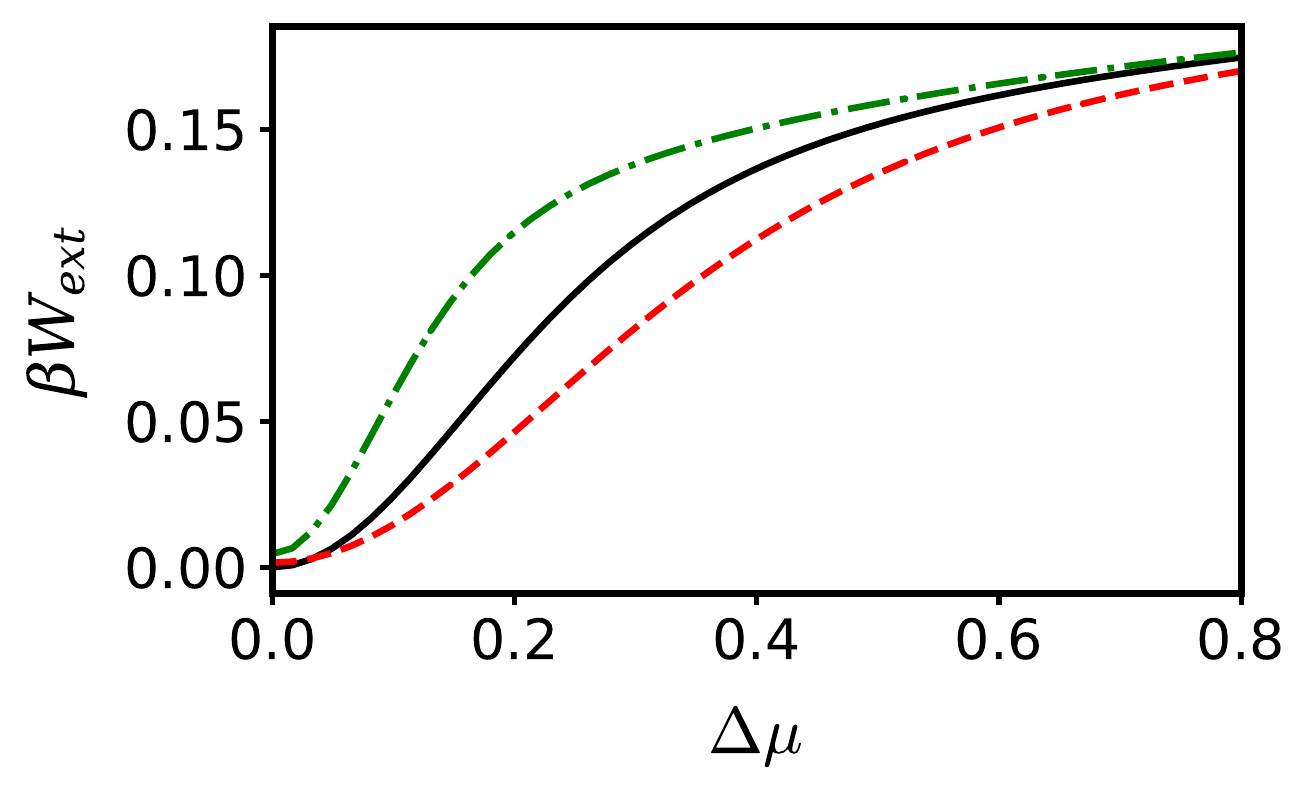}\label{Fig3(b)}}~
	\subfloat[ ]{\includegraphics[scale=0.4]{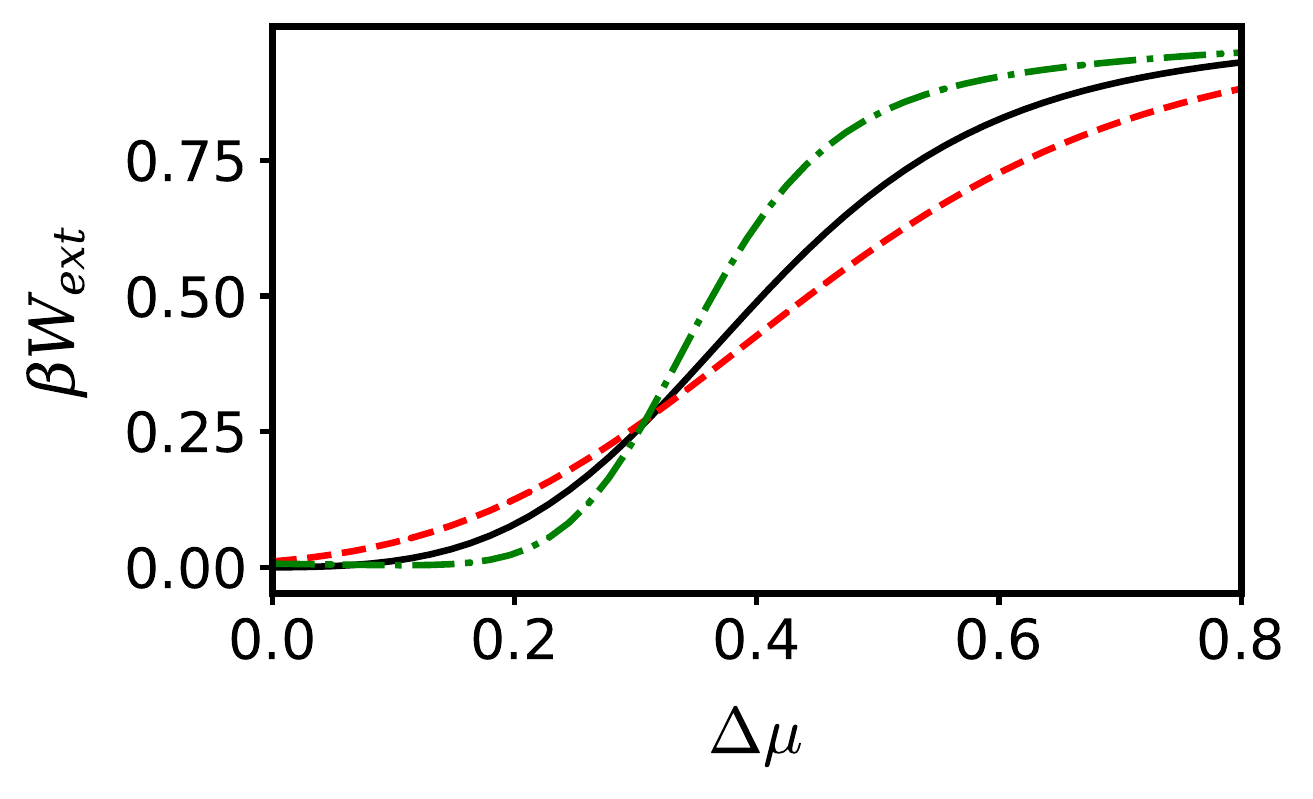}\label{Fig3(c)}}
	\caption{Maximum available work $\beta W_{ext}$ as a function of $\Delta \mu$ (with additional temperature gradient), for the different regimes:
		\eqref{Fig3(a)} off-resonant-full $\varepsilon_{QB}=-0.2$, \eqref{Fig3(b)} resonant $\varepsilon_{QB}=0.1$, 
		and \eqref{Fig3(c)} off-resonant-empty $\varepsilon_{QB}=0.4$.
		The other parameters are the same as in Fig.~\ref{Fig2}.}
	\label{Fig3}
\end{figure*}

\subsubsection{Applied bias $\Delta\mu$}
	We now turn on the effects of small to large applied biases on the exergy (in the presence or not of a small temperature gradient).
	With our set-up, a positive bias favors electron transfer from left to right. 
	Therefore positive biases lead to larger currents in the off-resonant-empty regimes in comparison to the resonant and off-resonant-full regime 
	(compare Fig.~\ref{Fig55(b)} and Fig.~\ref{Fig33(b)} to Fig.~\ref{Fig77(b)}).
	Consequently, the exergy is larger for the off-resonant-empty regime in comparison to the resonant and off-resonant-full regimes (see Fig.~\ref{Fig3}).

Our calculations show that one obtains the largest values for the exergy for positive applied bias (in the off-resonant-empty regime) in comparison to temperature gradients. 
	Based on these considerations, it seems feasible to develop optimal bias-charging protocols for QBs. Moreover, a negative temperature gradient in high bias-charging regimes in Fig.~\ref{Fig3(c)} improves battery performance. As a result, these observations provide a great deal of control over optimal battery charging.

\begin{figure}
	\centering
	{\includegraphics[scale=0.45]{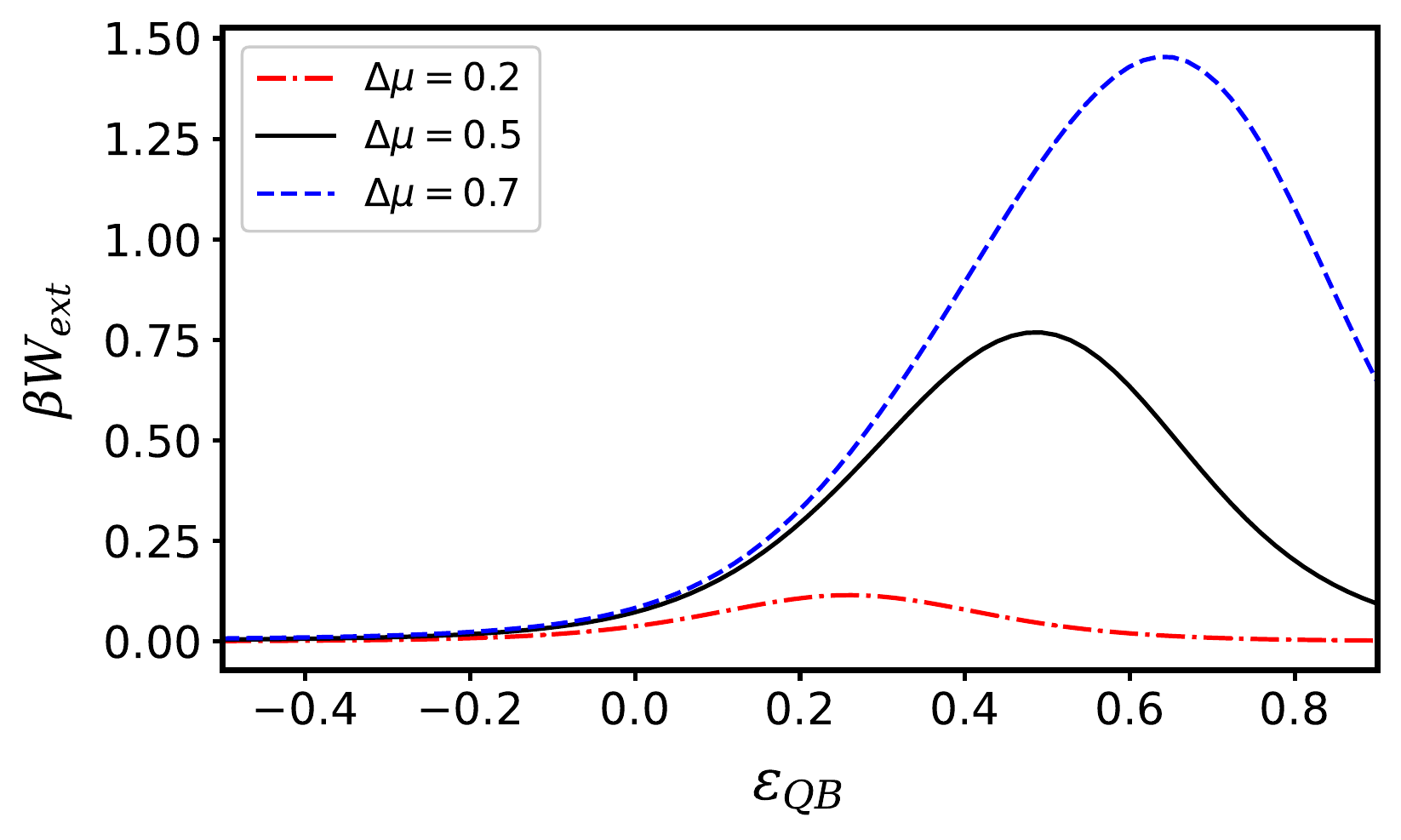}}~
	\caption{Maximum available work $\beta W_{ext}$ versus the energy level $\varepsilon_{QB}$
			for the different applied biases $\Delta\mu$ and $T_L=T_R=T_{eq}$. The exergy has a maximum value for $\varepsilon_{QB}\sim \Delta\mu$. The other parameters are the 
			same as in Fig.~\ref{Fig2}.}
	\label{Wext_e0}
\end{figure}

\begin{figure}
	\centering
	{\includegraphics[scale=0.45]{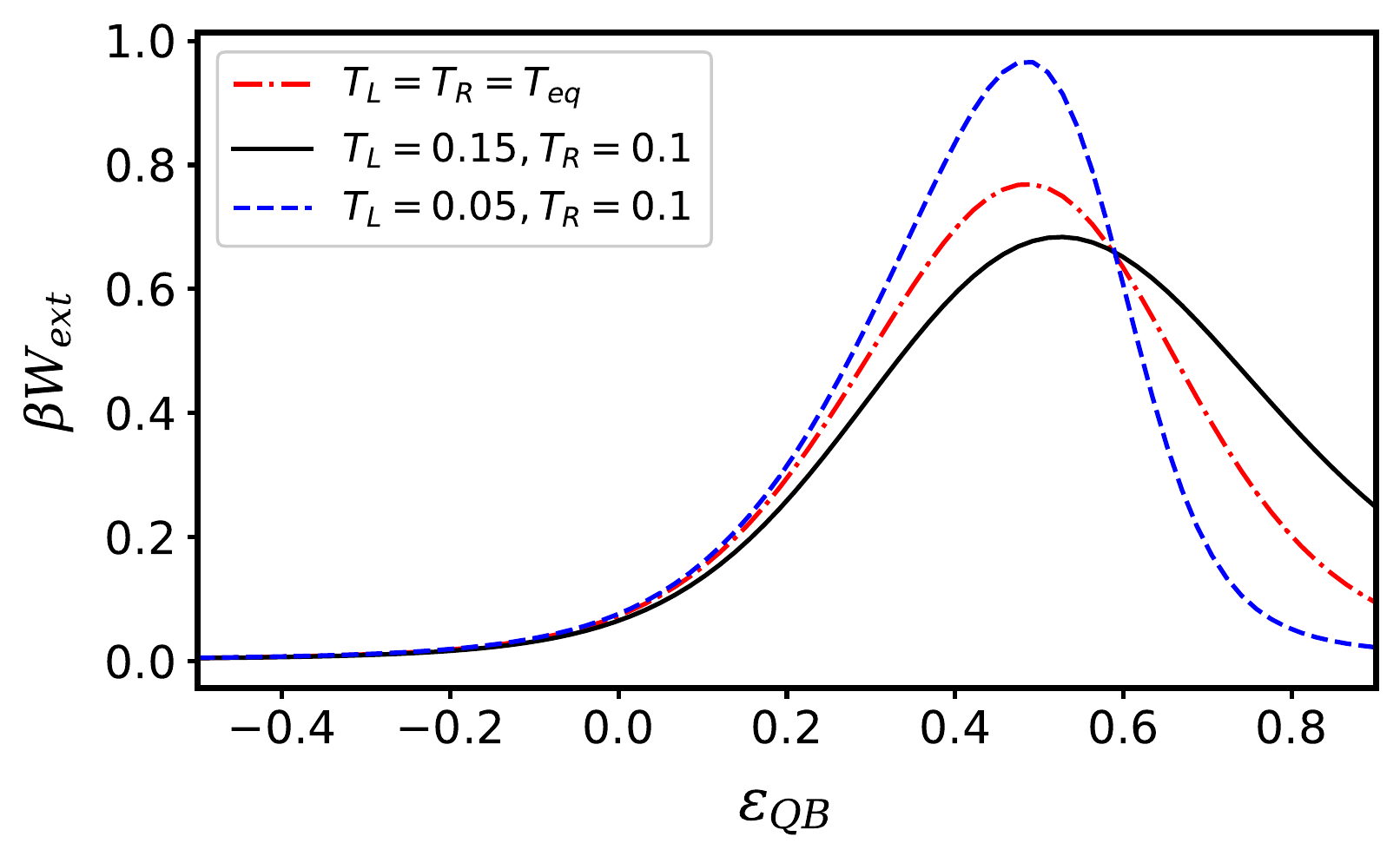}}~
	\caption{Maximum available work $\beta W_{ext}$ versus the energy level $\varepsilon_{QB}$ 
			for an applied bias $\Delta\mu=0.5$ and different small temperature gradients $\Delta T$. The exergy has a maximum value for $\varepsilon_{QB}\sim \Delta\mu$, and 
			the presence of an additional temperature gradient does not affect too much the position of the maximum exergy. The other parameters are the same as in Fig.~\ref{Fig2}.}
	\label{Wext_e0_dT}
\end{figure}

Interestingly, it is possible to optimize further the exergy by adapting the position of QB energy level for the NE conditions (positive applied bias).
	The dependence of the NE exergy versus $\varepsilon_{QB}$ is shown in Fig.~\ref{Wext_e0}.  
	The exergy reaches a maximum value when the QB energy level is located around the value of the
	applied bias $\Delta\mu$.
	Adding a (positive or negative) temperature gradient does not change much the position of the maximum
	of the exergy versus $\varepsilon_{QB}$ as can be seen in Fig.~\ref{Wext_e0_dT}.

\section{Conclusion}\label{V}
We have presented a model for the study of the charging process and thermodynamics of QBs strongly coupled to their environments (heat and particle reservoirs). 
In our model, the battery is charged by applying a temperature gradient and/or bias to create NE steady-state currents of heat and particles. 
Our model uses the standard NE Green's function expressions for the quantum transport. 
We have found optimal regimes for battery charging by studying the battery performance using a NE distribution function approach.

By using this scenario, interesting results were obtained, where the model parameters are delicately balanced. 
It has been shown that the charging of a quantum battery can be significantly improved by increasing the bias (chemical potential difference), 
compared to a temperature gradient, more specifically in the off-resonant-empty transport regime.
Additionally, it is also possible to enhance the battery charging performance by applying 
a small bias to a temperature gradient (or an small additional temperature gradient to an applied bias).

More importantly, one can further increase the exergy of the QB by adapting the position of the QB energy level in the NE conditions corresponding to applied biases. A maximum exergy is obtained for the QB energy level being equal to the applied bias.
A three-terminal device where the QB is gated, in a similar way as in conventional transistors, could be designed to allow optimum exergy, by varying the gate voltage, when a charge current flows between the source and drain electrodes.

\begin{acknowledgments} 
	This work has been supported by the University of Kurdistan. S. Salimi thanks research funded by Iran national science foundation (INSF) under project No.4003162.
\end{acknowledgments}

\appendix

\section{An approximation for the reduced density matrix}\label{appA}

Following the results presented in Ref.~[\onlinecite{Dhar:1}], it is possible to obtain (for small temperature gradients and biases) an explicit expression for the density matrix (in the NE steady-state regime) for the single-level system (QB-region). 
Following Sec.~\ref{III} in Ref.~[\onlinecite{Dhar:1}],
such a density matrix is given by
\begin{align}\label{A1}
	\hat\varrho^{NE}_{QB}=\frac{\exp(-a \hat{d}^\dag \hat{d})}{Z_{QB}}=\frac{\exp(-a~\hat{d}^{\dagger}\hat{d})}{1+\exp(-a)}~.
\end{align}
with $a=\ln(d_{QB}^{-1}-1)$ 
and 
$d_{QB}=\langle \hat{d}^{\dagger}\hat{d}\rangle=\int {{\rm d}\omega}\ A_{QB}(\omega) f_{QB}^{\rm NE}(\omega)$. 

One should note that the definition of $\hat\varrho^{\rm NE}_{QB}$ is completely different from the reduced density matrix $\hat\rho^{\rm NE}_{QB}={\rm Tr}_{L,R}[\hat\rho^{\rm NE}]$ in our approach. 
Some effects of NE conditions for the coupled QB region are taken into account via the quantities $a$ and $d_{QB}$. However $\hat\varrho^{\rm NE}_{QB}$ has clearly the form of a density matrix for an isolated single-level system. 

The elimination of the $L$ and $R$ degrees of freedom, in the calculation of $\hat\rho^{\rm NE}_{QB}$, is a difficult task. 
$\hat\rho^{\rm NE}$ is build from the total Hamiltonian of the coupled system, and the total Hamiltonian does not commute with the 
individual Hamiltonians $H_{L,QB,R}$ because of the coupling operators $H_{int}$ between the QB and $L,R$ regions. 
Therefore the calculations based on $\hat\varrho^{\rm NE}_{QB}$ are only valid in the weak coupling (between the QB and the reservoirs) regime.

From Eq.~\eqref{A1}, we can evaluate the maximum available work in Eq.~\eqref{3} for the QB region
\begin{equation}\label{A2}
	\begin{split}
		\beta W^{\varrho}_{\rm ext}
		&= 
		{\rm Tr}\left[ \hat\varrho^{\rm NE}_{QB} \ln \left( \frac{\hat\varrho^{\rm NE}_{QB}}{\hat\rho^{\rm eq}_{QB}} \right) \right]
	\end{split}
\end{equation}
In the definition of $\hat\varrho^{\rm NE}_{QB}$, the partition function $Z_{QB}$ is the trace $Z_{QB}= {\rm Tr}[\exp(-a \hat{d}^\dag \hat{d})] = 1+\exp(-a)$ taken over only two states corresponding to the QB energy level being empty or occupied by one electron
\begin{equation}
	\begin{split}
		Z_{QB} & = {\rm Tr}[\exp(-a \hat{d}^\dag \hat{d})] \\ 
		& = \sum_{N=0,1} \langle N \vert \exp(-a \hat{d}^\dag \hat{d})\vert N\rangle =  1+\exp(-a) 
	\end{split}
\end{equation}
where $\langle 1 \vert \hat{d}^\dag \hat{d}\vert 1\rangle = 1$ and $\langle 0 \vert \hat{d}^\dag \hat{d}\vert 0\rangle = 0$.


One can also define an entropy from $\hat\varrho^{\rm NE}_{QB}$. By using the same trace as for the partition function, one easily gets
\begin{equation}
\begin{split}
			S^{\rm NE}_{\varrho}
			&= 
			-{\rm Tr}[\hat\varrho^{\rm NE}_{QB} \ln \hat\varrho^{\rm NE}_{QB} ] \\
			& = - \frac{1}{1+e^{-a}}  \ln \left( \frac{1}{1+e^{-a}} \right)
			-     \frac{e^{-a}}{1+e^{-a}}  \ln \left( \frac{e^{-a}}{1+e^{-a}} \right) \\
			& = 
			- (1-d_{QB}) \ln(1-d_{QB}) - d_{QB} \ln(d_{QB}) 
	\end{split}
	\label{SSNE}
\end{equation}


And finally, the exergy in Eq.~\eqref{A2} is given by the following expression
\begin{equation}
	\begin{split}
		\beta W^{\varrho}_{ext}
		&= 
		d_{QB} \ln \left( \frac{d_{QB}}{f^{\rm eq}_{QB}} \right) + (1-d_{QB}) \ln \left( \frac{1-d_{QB}}{1-f^{\rm eq}_{QB}} \right)
	\end{split}
	\label{wSNE}
\end{equation}

Now, a few comments about the differences between our expressions for the NE entropy Eq.~\eqref{10} and maximum available work Eq.~\eqref{9} and Eqs.~\eqref{SSNE}-\eqref{wSNE} are in order.

Our expressions Eq.~\eqref{10} and Eq.~\eqref{9}
are obtained from an energy integration, where each energy corresponding to one ``scattering process''. Such an energy integration does not appear in Eq.~\eqref{SSNE} and Eq.\eqref{wSNE}, or indirectly in the definition of $d_{QB}$.
Moreover, one can interpret the spectral function $A_{QB}$ as a probability distribution since $\int {{\rm d}\omega}\ A_{QB}(\omega) = 1$.
Therefore, one term in our NE entropy expression can be rewritten as follows
	\begin{equation}
		\int {{\rm d}\omega}\ A_{QB}(\omega) 
		\left[ f_{QB}^{\rm NE}(\omega) \ln f_{QB}^{\rm NE}(\omega) \right] =
		\langle  f_{QB}^{\rm NE} \ln f_{QB}^{\rm NE} \rangle_{A_{QB}} \ ,
	\end{equation}
	which is simply an average over the distribution $A_{QB}(\omega)$.

The corresponding term in $S^{\rm NE}_{\varrho}$ Eq.~\eqref{SSNE} is rewritten as 
	\begin{equation}
		d_{QB} \ln d_{QB} = \langle  f_{QB}^{\rm NE} \rangle_{A_{QB}} \ln \langle f_{QB}^{\rm NE} \rangle_{A_{QB}}~.
\end{equation}

From probability theory, the average of a function $\langle f(X)\rangle$ (in the most general case) is never  
equal to the function of the average, i.e. $\langle f(X)\rangle \ne f(\langle X\rangle)$.
Therefore the results from Eq.~\eqref{10} or Eq.~\eqref{9} will always differ from Eq.~\eqref{SSNE} or Eq.~\eqref{wSNE}.

However, there is only one case for which Eq.~\eqref{10} or Eq.~\eqref{9} gives the same results as Eq.~\eqref{SSNE} or Eq.\eqref{wSNE}.
	It is the case of weak coupling to the leads ($\nu_{L,R}\rightarrow 0$) for which
	$A_{QB}(\omega) \rightarrow \delta(\omega - \varepsilon_{QB})$, leading to 
	$\langle  f_{QB}^{\rm NE} \ln f_{QB}^{\rm NE} \rangle_{A_{QB}} =
	f_{QB}^{\rm NE}(\varepsilon_{QB}) \ln f_{QB}^{\rm NE}(\varepsilon_{QB})$ 
	and  
	$\langle  f_{QB}^{\rm NE} \rangle_{A_{QB}}=f_{QB}^{\rm NE}(\varepsilon_{QB})$

\begin{figure}
	\subfloat[ ]{\includegraphics*[scale=0.55]{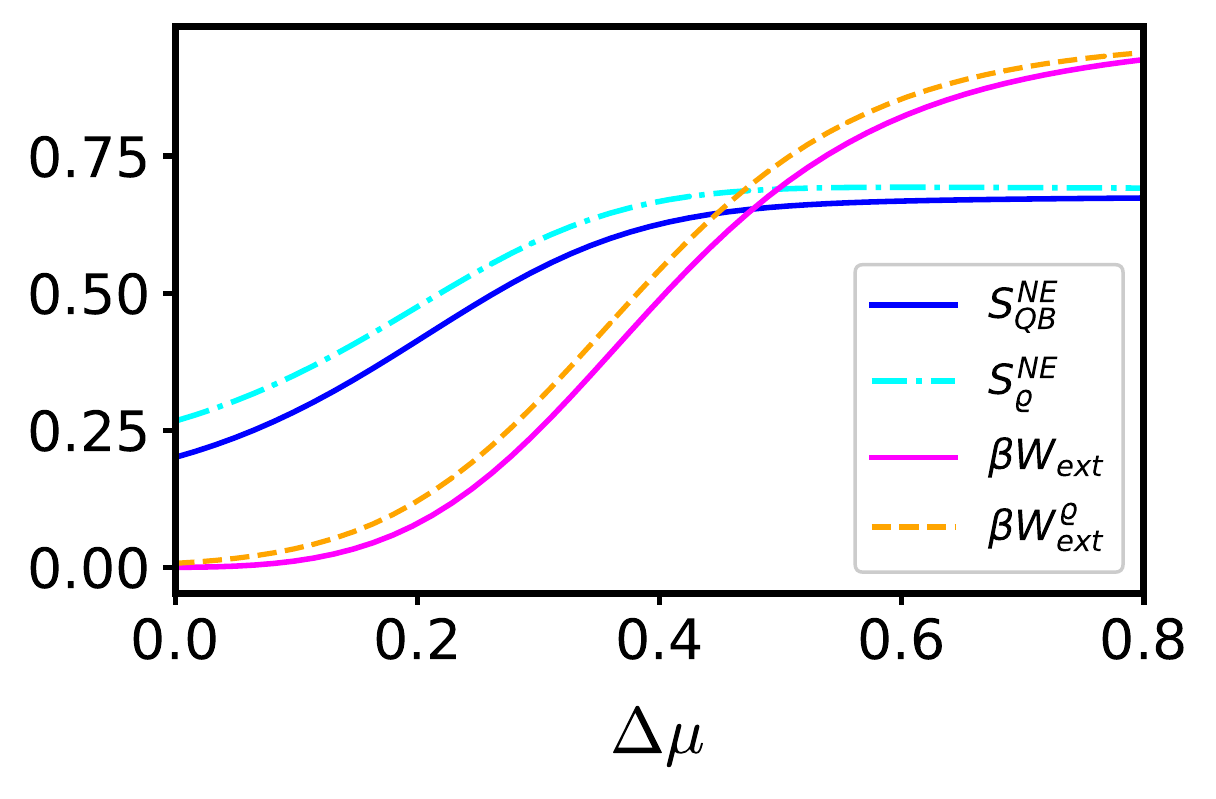}\label{Fig8(a)}}\\\subfloat[ ]{\includegraphics*[scale=0.55]{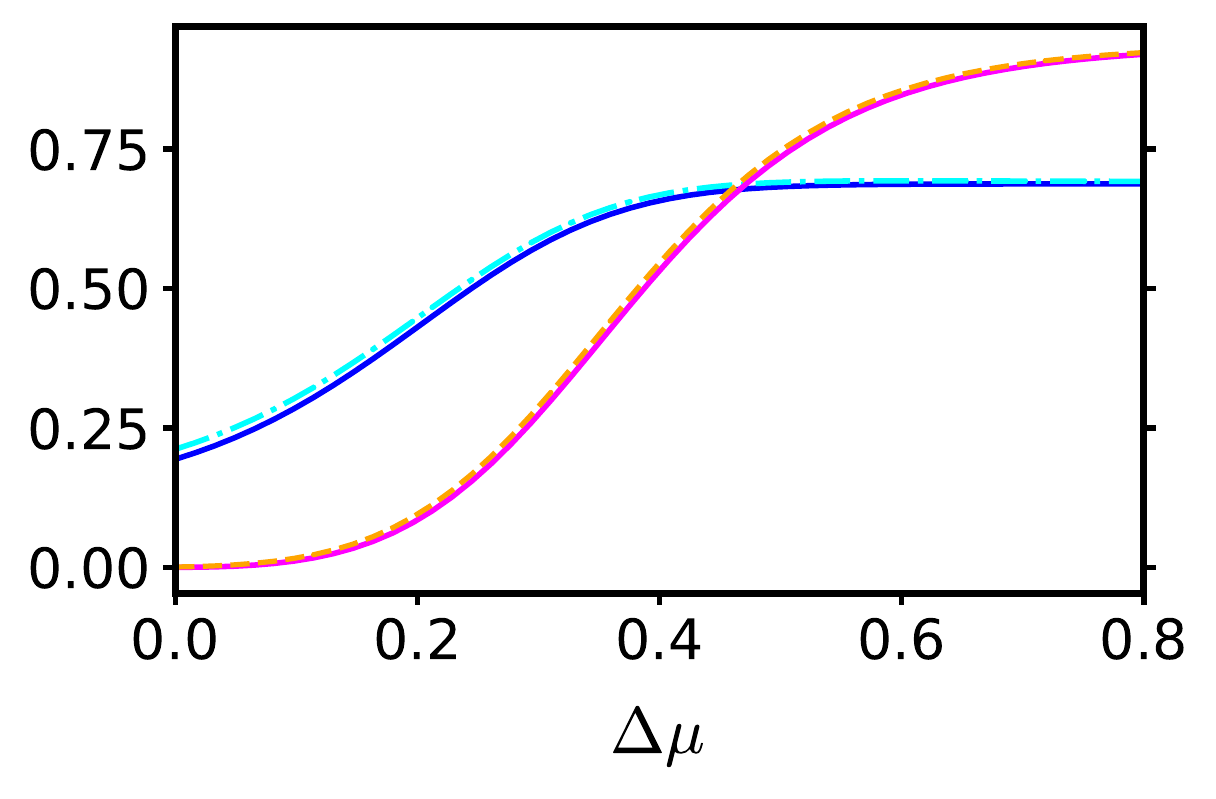}\label{Fig8(b)}}
	\caption{Maximum available work and NE entropy versus the potential difference $\Delta\mu$ in the off-resonant regime $\varepsilon_{QB}=0.3$ 
		calculated
		by using the approximated density matrix and the distribution function. Panel (a): intermediate coupling regime $v_\alpha=0.12$, 
		Panel (b): weak coupling regime $v_\alpha=0.06$. Other parameters are: $\varepsilon_\alpha=0,~ h_\alpha=1.0,~\mu_{eq} =0,~T_{L}=T_{R}=T_{eq}=0.1$.}
	\label{Fig8}
\end{figure}

In Fig.~\ref{Fig8}, we show the maximum available work and NE entropy, as a of function $\Delta\mu$, calculated
from the density matrix $\hat\varrho^{\rm NE}_{QB}$ ($W^{\varrho}_{ext},S^{NE}_\varrho$) and from the correct
distribution function $f^{\rm NE}$ ($W_{ext},S^{NE}_{QB}$). 
Figs.~\ref{Fig8(a)} and \ref{Fig8(b)} correspond to the intermediate coupling regime $\nu_{\alpha}=0.12$ and the weak coupling regime $\nu_{\alpha}=0.06$ respectively. 

Calculations from Eq.~\eqref{9} and Eq.~\eqref{10} differ from the maximum work and NE entropy obtained from Eq.~\eqref{wSNE} and Eq.~\eqref{SSNE}
as expected (in the general cases). 
In the limit of weak coupling to the $L$ and $R$ reservoirs, as expected, both approaches provide the similar results. 
This is to be expected as the effective form of the density matrix $\hat\varrho^{\rm NE}_B$ is that of an isolated (i.e. not coupled to the reservoirs) system.

\section{Results for the charge and energy currents}\label{appB}

The analysis of the $\Delta T$ and $\Delta\mu$ dependence of the currents, entropy and exergy
	leads to the following:

The exergy (divided by the equilibrium temperature) $\beta W_{ext}$ has larger values for an applied bias $\Delta\mu$ (with no temperature gradient $T_L=T_R$) than for an applied temperature gradient $\Delta T$ (with no bias $\mu_L = \mu_R$).

The largest values for $\beta W_{ext}$ are obtained for the off-resonant-empty regime with an applied bias $\Delta\mu$ ($T_L=T_R$), see Fig.~\ref{Fig3}.
	Such values can be increased by applying an additional negative temperature gradient $\Delta T<0$ 
	(where the heat flow goes from right to left $T_L < T_R$) when $\Delta\mu > \varepsilon_{QB} - \mu = 0.3$.
	Below the threshold $\Delta\mu < \varepsilon_{QB} - \mu =0.3$, one can increase the values of $\beta W_{ext}$ by applying an opposite temperature gradient $\Delta T>0$ 
	(where the heat flow goes from left to right $T_L > T_R$).
	Note that the crossover point in the exergy, currents $I_Q$, energy current $J_{E}$ and heat current $J_H^L$, see Figs.~\ref{Fig55(b)},~\ref{Fig55(c)} and \ref{Fig55(d)} respectively, corresponds to the chemical potential $\mu_L$ reaching the maximum of the transmission (and of the spectral function $A_B(\omega)$), i.e. $\mu_L = \varepsilon_{QB}$. 
	Surprisingly,  a reverse temperature gradient in high bias-charging regimes improves battery performance. Consequently, NE heat current can
	boost the battery performance or diminish it. A great deal of control over battery charging is therefore possible.

The values of the charge current $I_Q$ in the off-resonant-full regime for different applied bias $\Delta\mu$, see Fig.~\ref{Fig77(b)}, are rather small.
	This is because the transport occurs via the (descending) tail of the transmission peak
	where the transmission values are small ($\ll 1$). 
	The maximum in transmission $\mathcal{\tau} \approx 1$ occurs for energies around $\omega \sim\varepsilon_{QB}=-0.2$.
	This is also reflected in the small values of the energy and heat current, see Figs.~\ref{Fig77(c)}, \ref{Fig77(d)}
	and of the exergy Fig.~\ref{Fig2(a)}.
	For the off-resonant-empty regime, the applied bias $\Delta\mu$ window swipes the entire transmission peak.
	An increasing bias $\Delta\mu$ leads to an increase in the charge $I_Q$ and energy $J_E$ currents, as well as for the exergy.

As far as the $\Delta\mu$ dependence of the different quantities is concerned, one can see that the NE entropy has smaller amplitudes for the off-resonant-full regime compared to the off-resonant-empty regime, see Fig.~\ref{SNE-mu}. 
	A similar behavior is also obtained for the exergy, compared Fig.~\ref{Fig2(a)} and Fig.~\ref{Fig2(c)}. Given that the exergy or extractable work is a part of entropy production spent to reach NE steady-state, the similar behavior of the NE entropy and of the exergy is reasonable and expected.
\begin{figure*}[t!]
	\centering
	\subfloat[ ]{\includegraphics[scale=0.41]{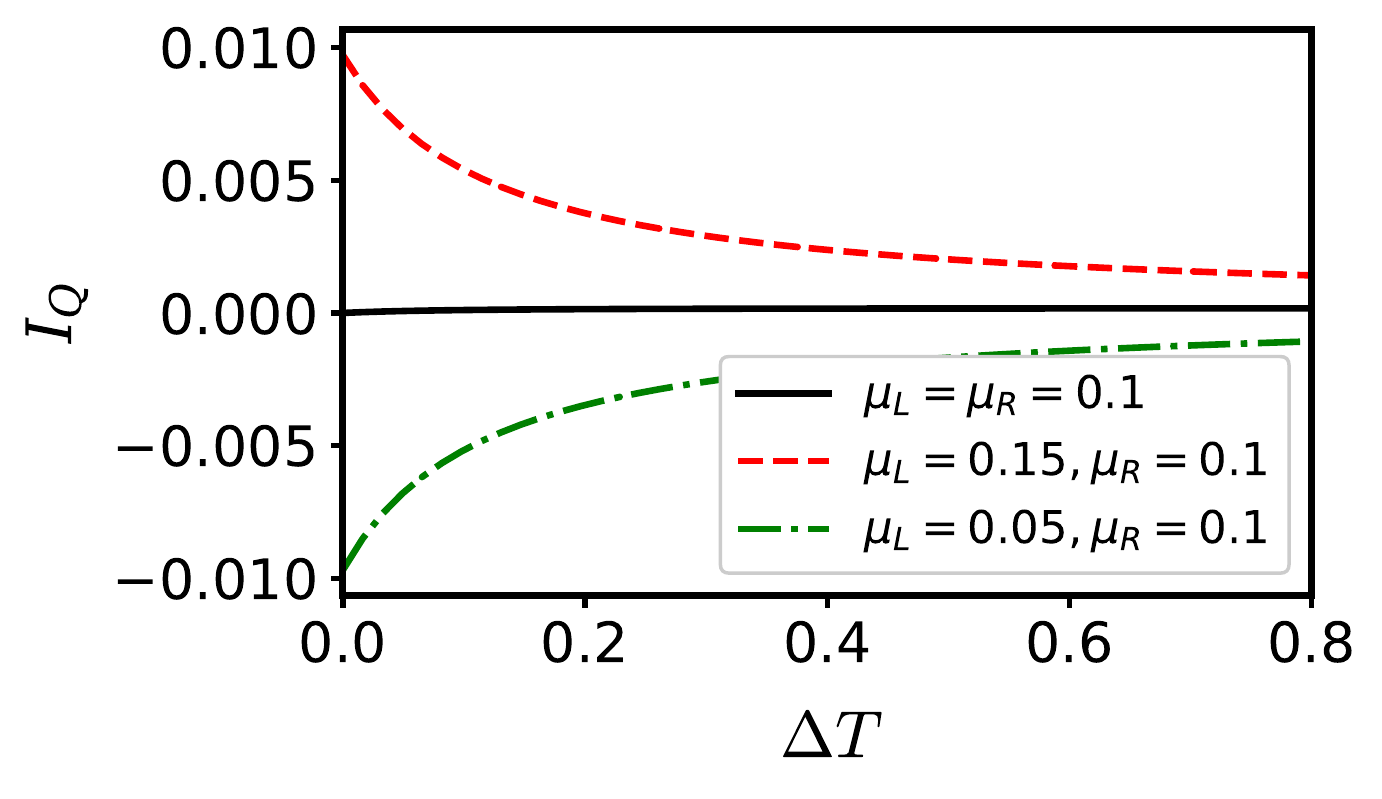}\label{Fig22(b)}}~
	\subfloat[ ]{\includegraphics[scale=0.41]{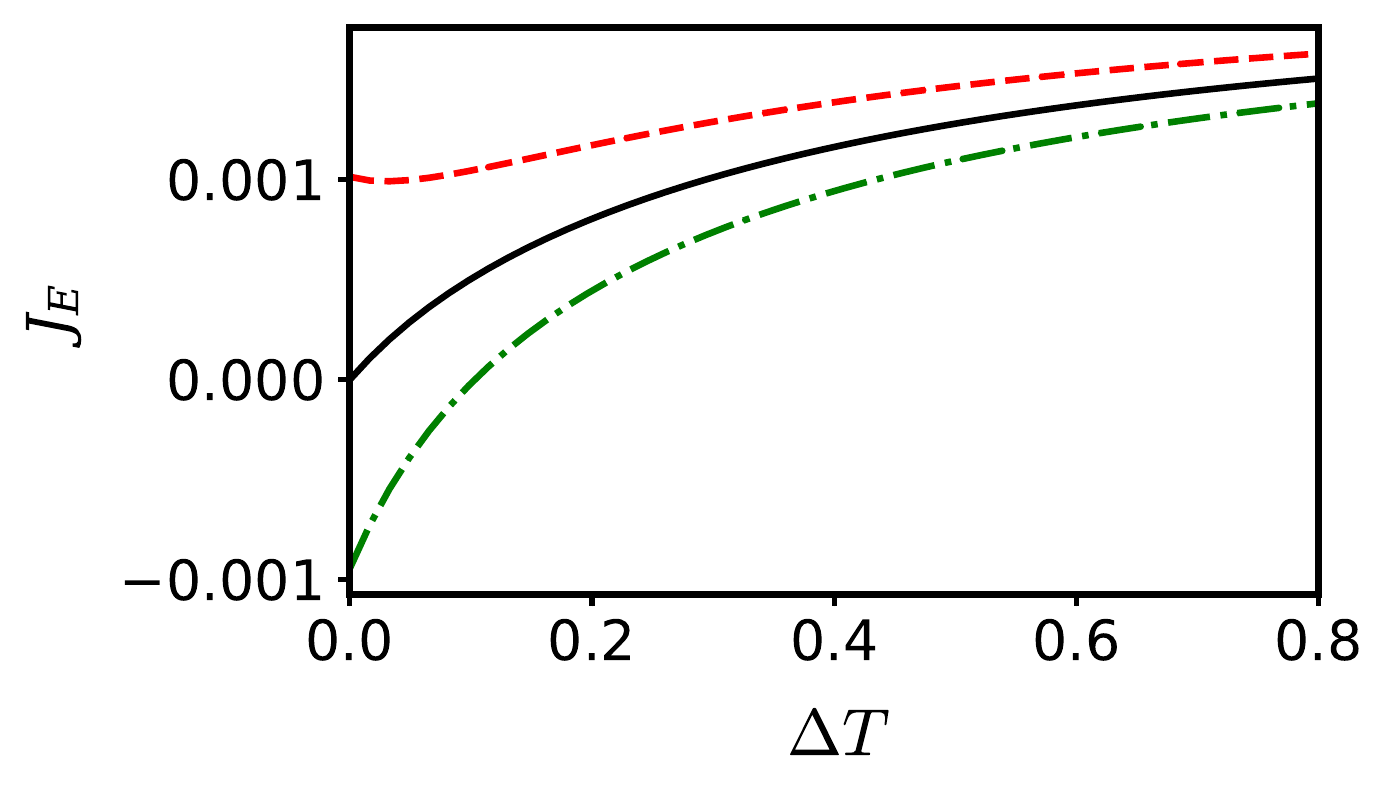}\label{Fig22(c)}}~
	\subfloat[ ]{\includegraphics[scale=0.41]{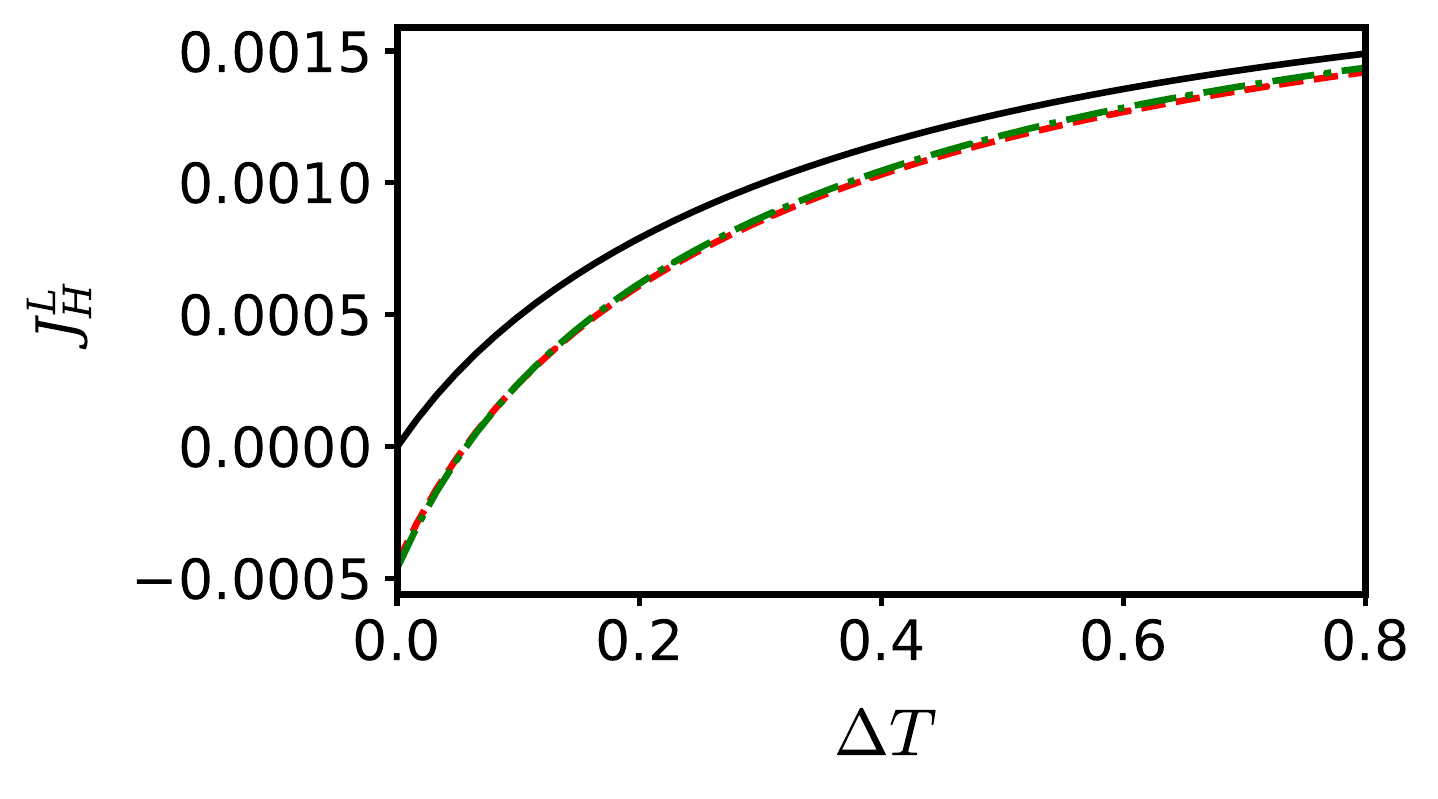}\label{Fig22(d)}}
	\caption{Resonant regime $\varepsilon_{QB}=0.1$. Currents as a function of $\Delta T$. 
		Panel (a): charge current $I_{Q}$. Panel (b): energy current $J_{E}$. Panel (c): left heat current $J^{L}_{H}$. 
		Here $T_{R}=T_{eq}$ and $T_{L}=T_{eq}+\Delta T$. 
		Other parameters are as in Fig.~\ref{Fig2}.
	}
	\label{Fig22}
\end{figure*}

\begin{figure*}[t!]
	\centering
	\subfloat[ ]{\includegraphics[scale=0.42]{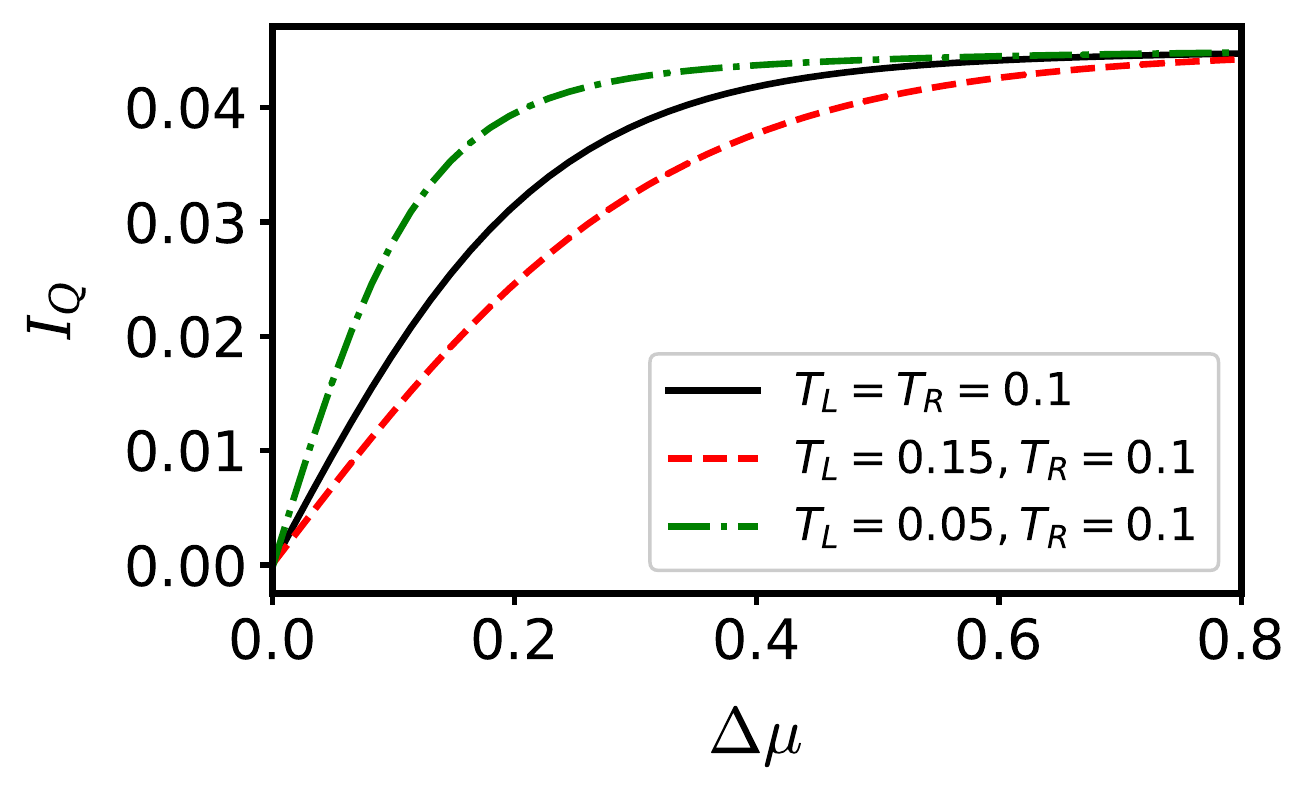}\label{Fig33(b)}}~
	\subfloat[ ]{\includegraphics[scale=0.42]{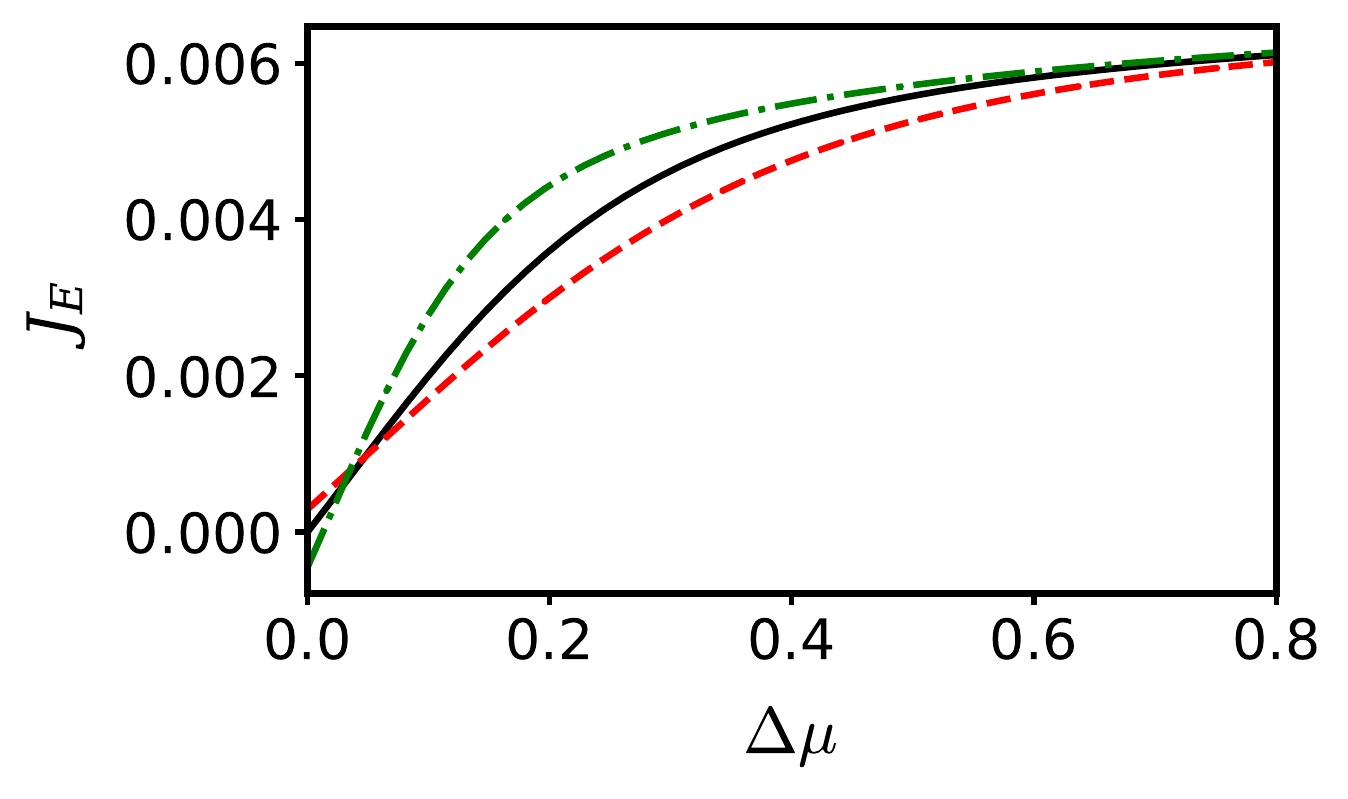}\label{Fig33(c)}}~
	\subfloat[ ]{\includegraphics[scale=0.42]{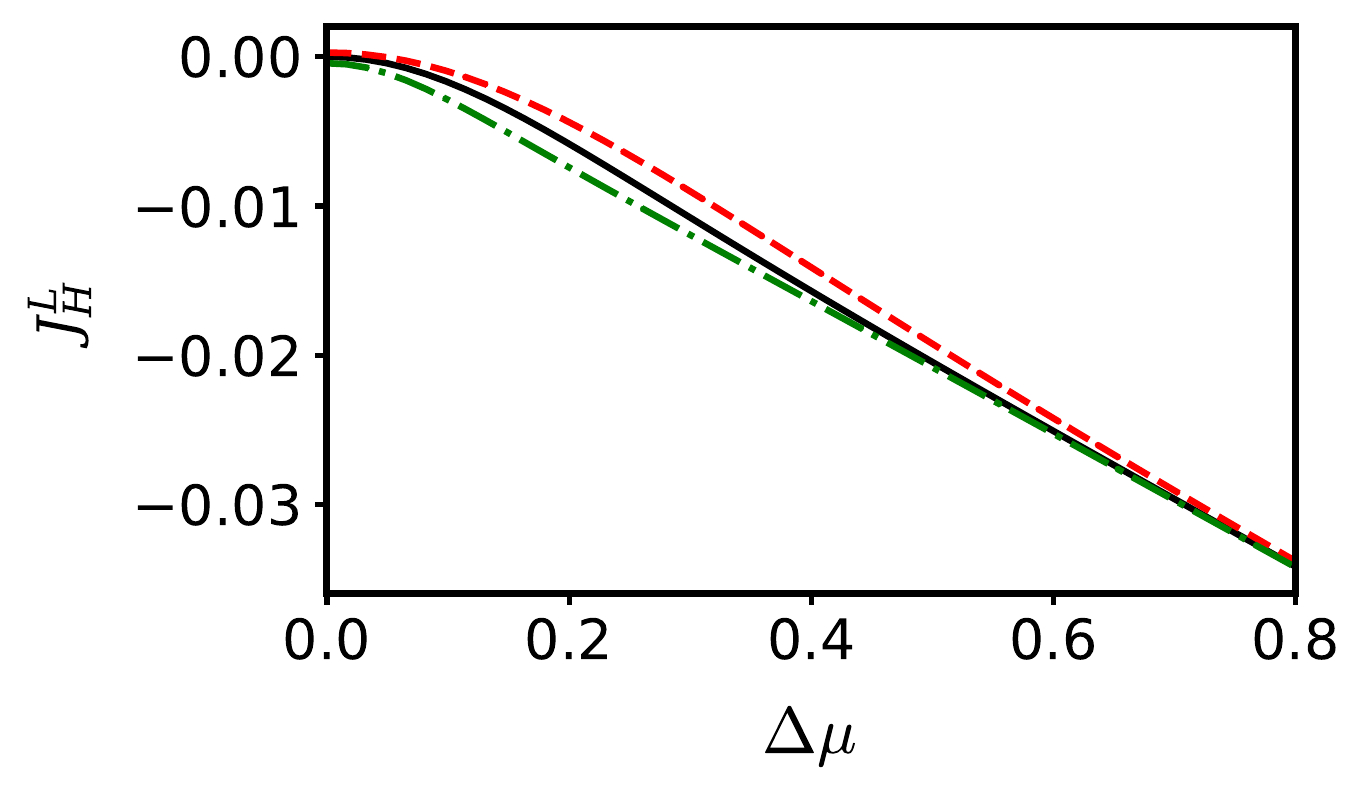}\label{Fig33(d)}}
	\caption{Resonant  regime $\varepsilon_{QB}=0.1$. Currents as a function of $\Delta\mu$.
		Panel (a): charge current $I_{Q}$. Panel (b): energy current $J_{E}$. Panel (c): left heat current $J^{L}_{H}$. 
		Here $\mu_{R}=\mu_{eq}$ and $\mu_{L}=\mu_{eq}+\Delta \mu$. 
		Other parameters are same as Fig.~\ref{Fig2}.}
	\label{Fig33}
\end{figure*}

\begin{figure*}[t!]
	\centering
	\subfloat[ ]{\includegraphics[scale=0.42]{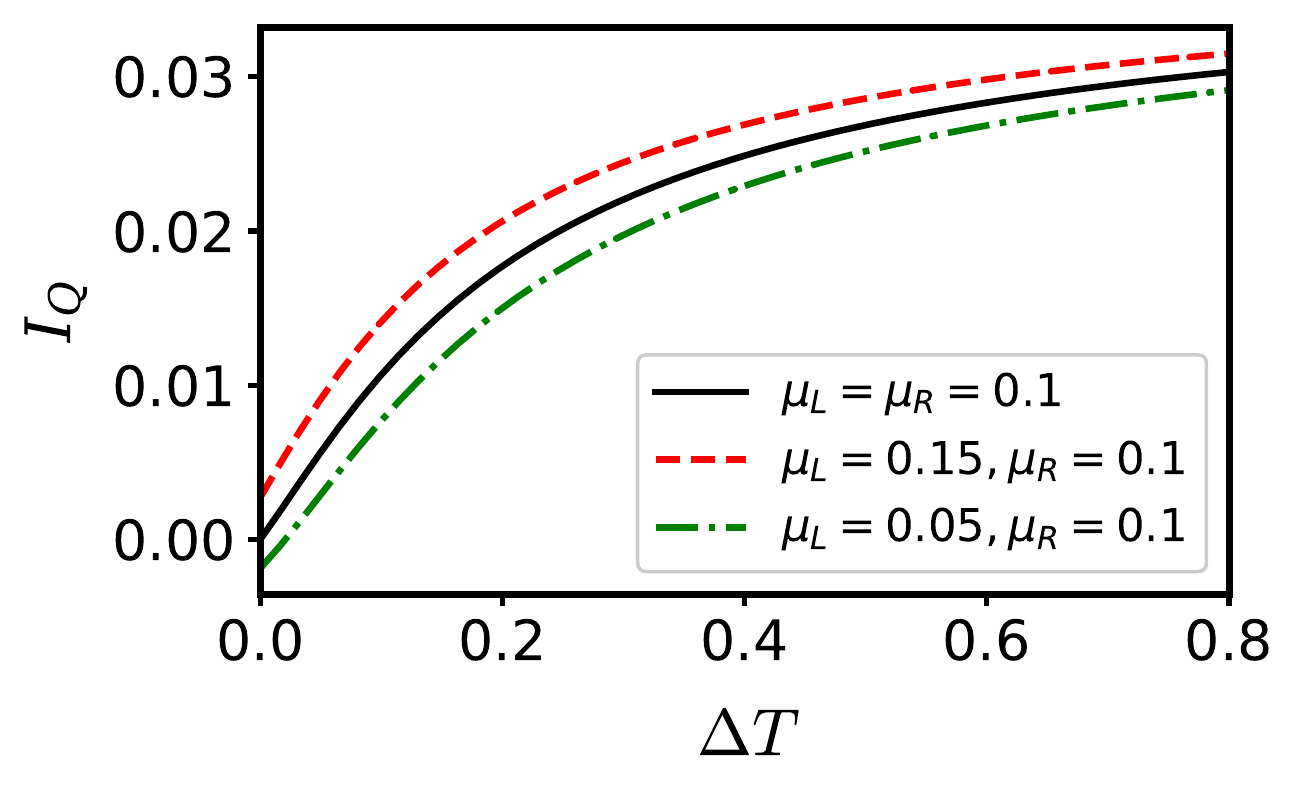}\label{Fig44(b)}}~
	\subfloat[ ]{\includegraphics[scale=0.42]{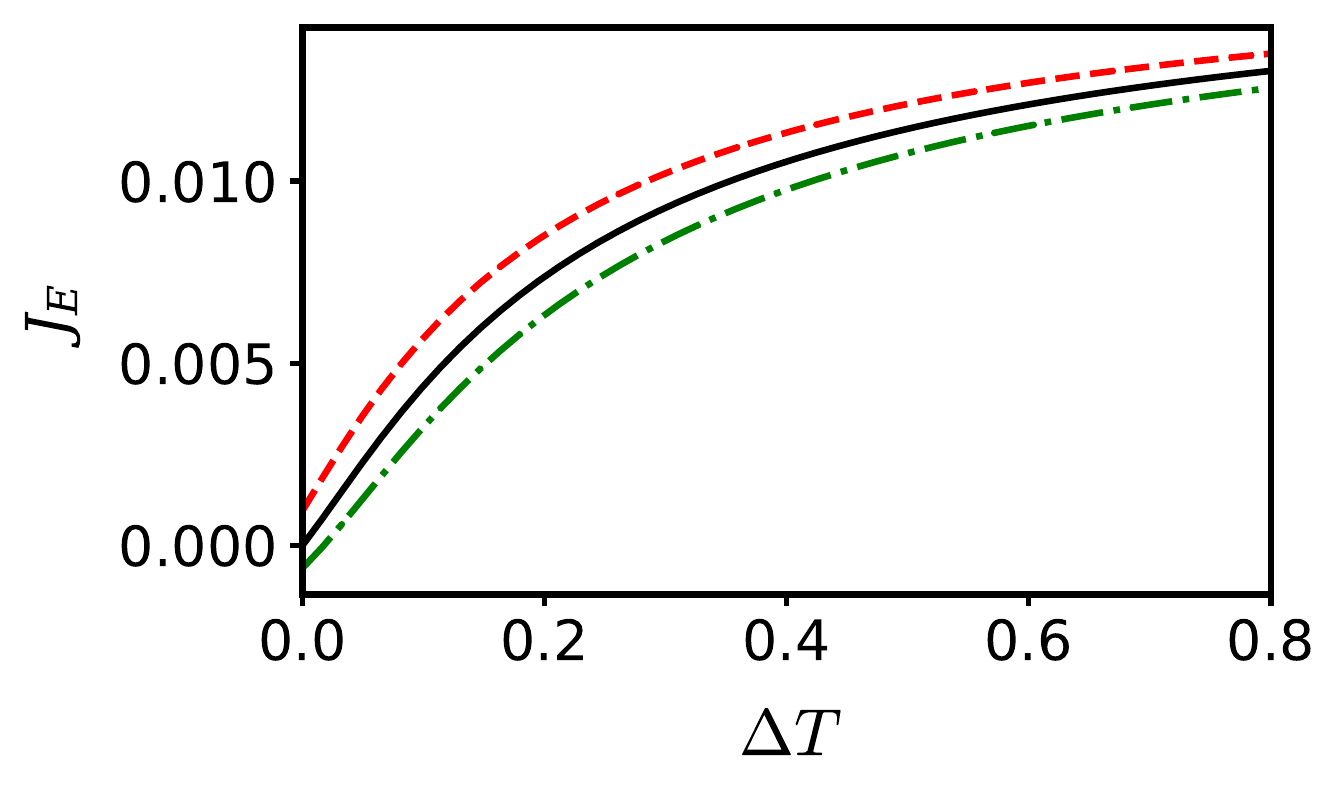}\label{Fig44(c)}}~
	\subfloat[ ]{\includegraphics[scale=0.42]{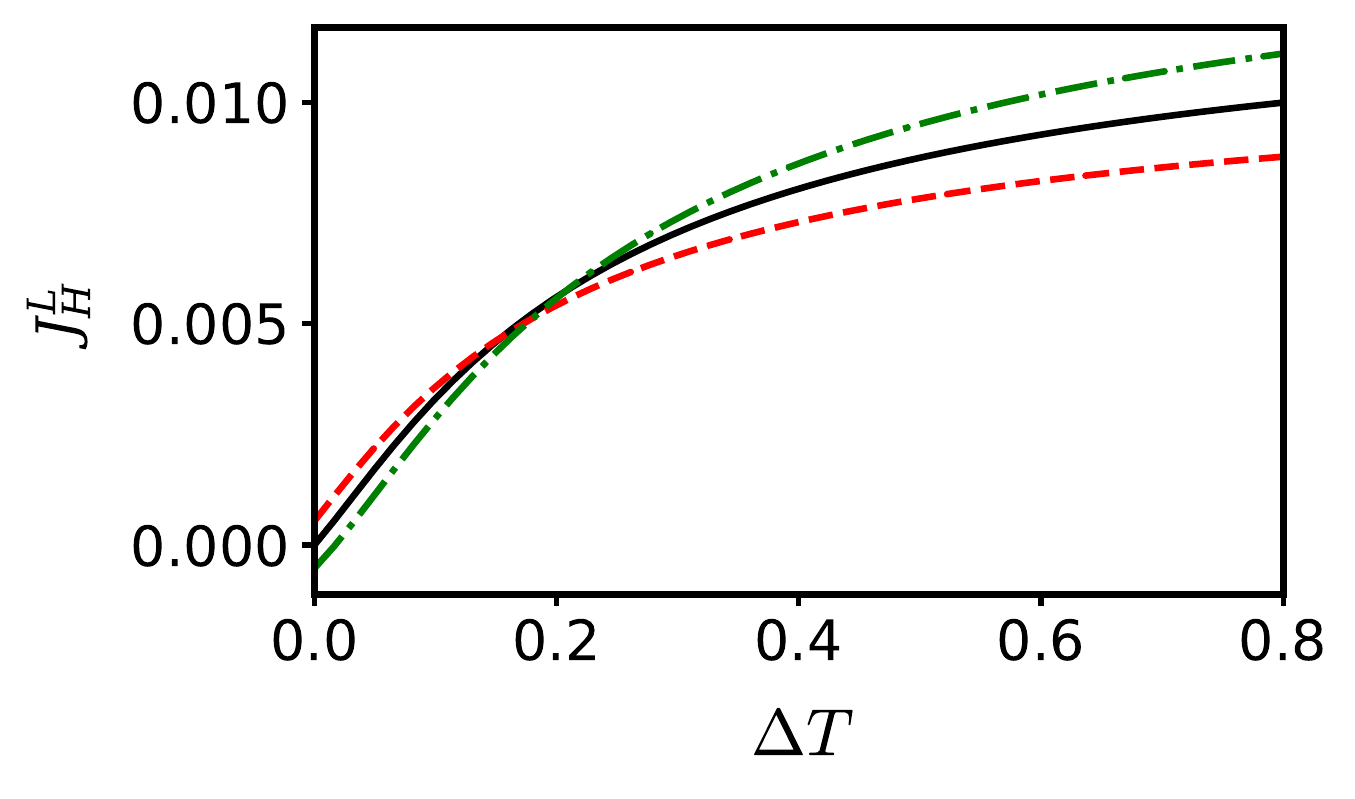}\label{Fig44(d)}}
	\caption{Off-resonant-empty regime $\varepsilon_{QB}=0.4$. Currents as a function of $\Delta T$. 
		Panel (a): charge current $I_{Q}$. Panel (b): energy current $J_{E}$. Panel (c): left heat current $J^{L}_{H}$. 
		With $T_{R}=T_{eq}$ and $T_{L}=T_{eq}+\Delta T$. 
		Other parameters are same as Fig.~\ref{Fig2}.}
	\label{Fig44}
\end{figure*}

\begin{figure*}[t!]
	\centering
	\subfloat[ ]{\includegraphics[scale=0.42]{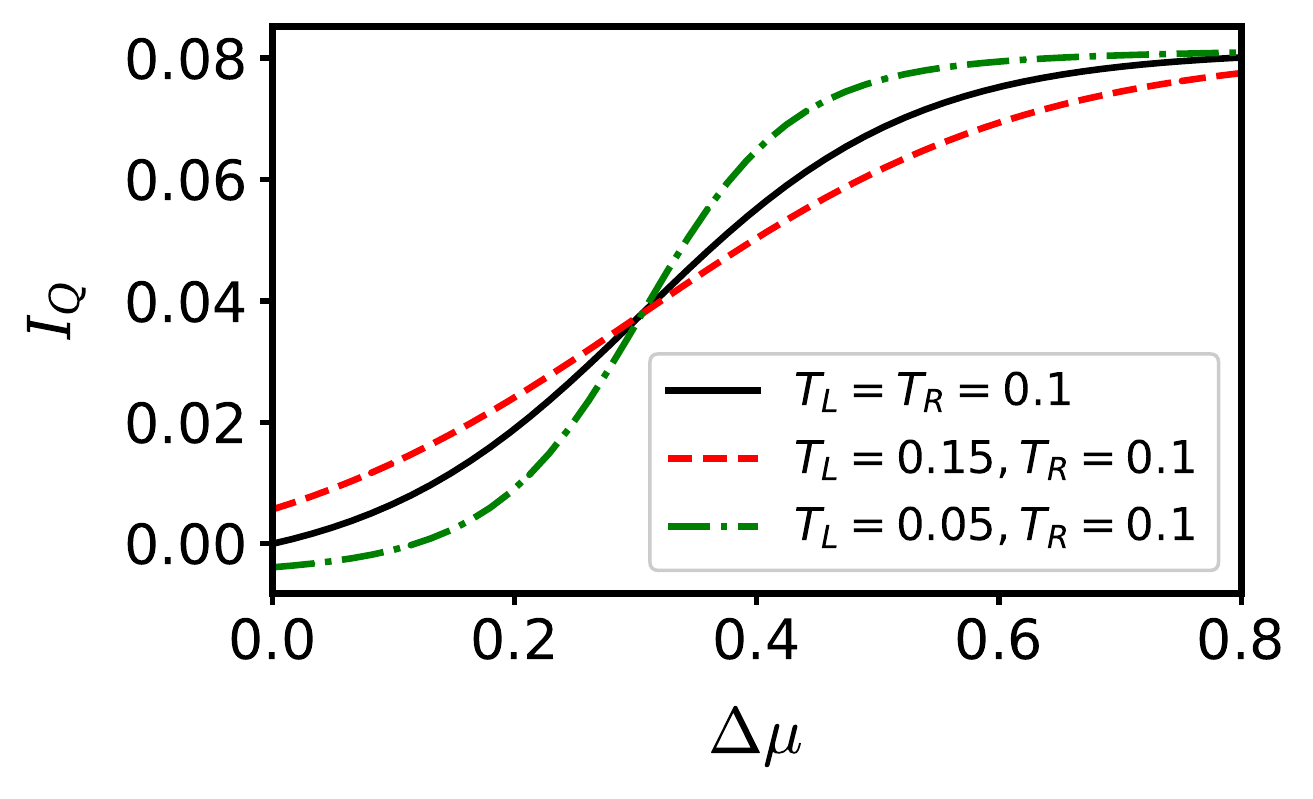}\label{Fig55(b)}}~
	\subfloat[ ]{\includegraphics[scale=0.42]{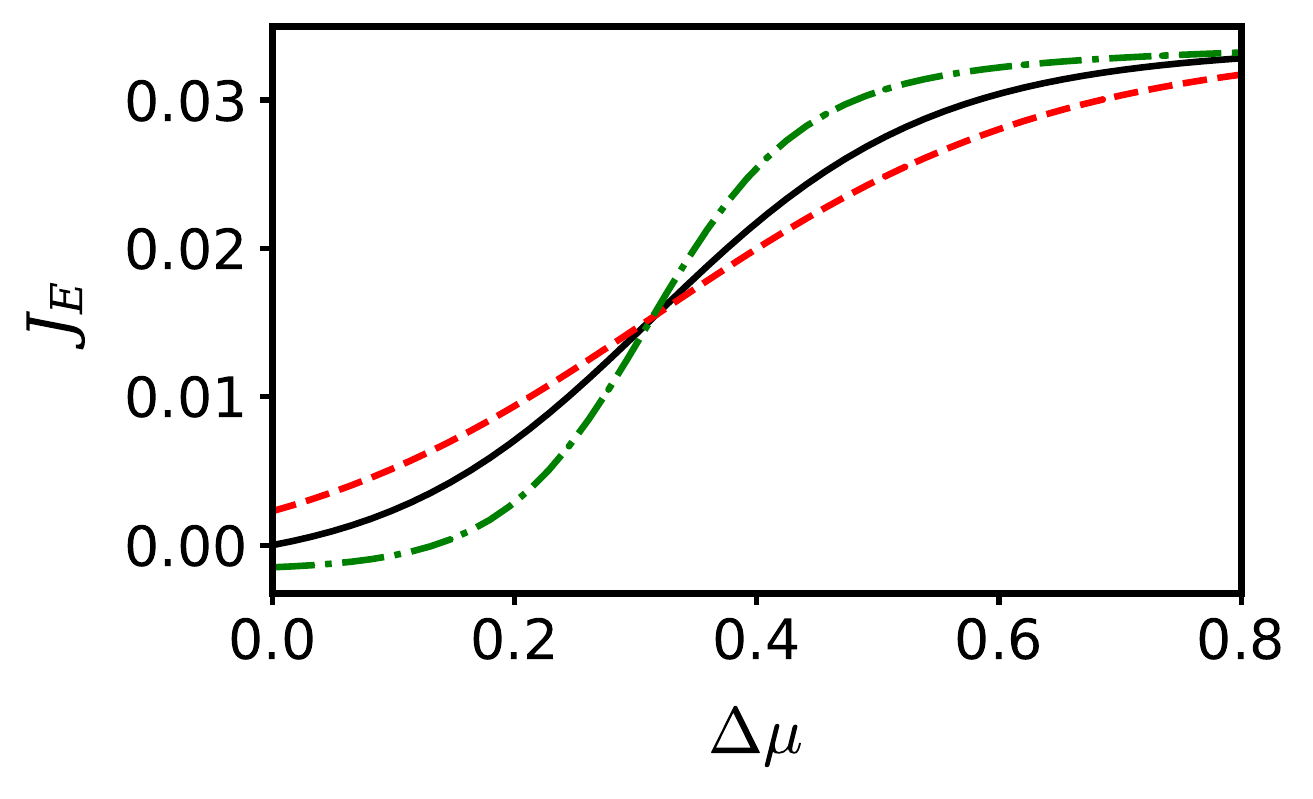}\label{Fig55(c)}}~
	\subfloat[ ]{\includegraphics[scale=0.42]{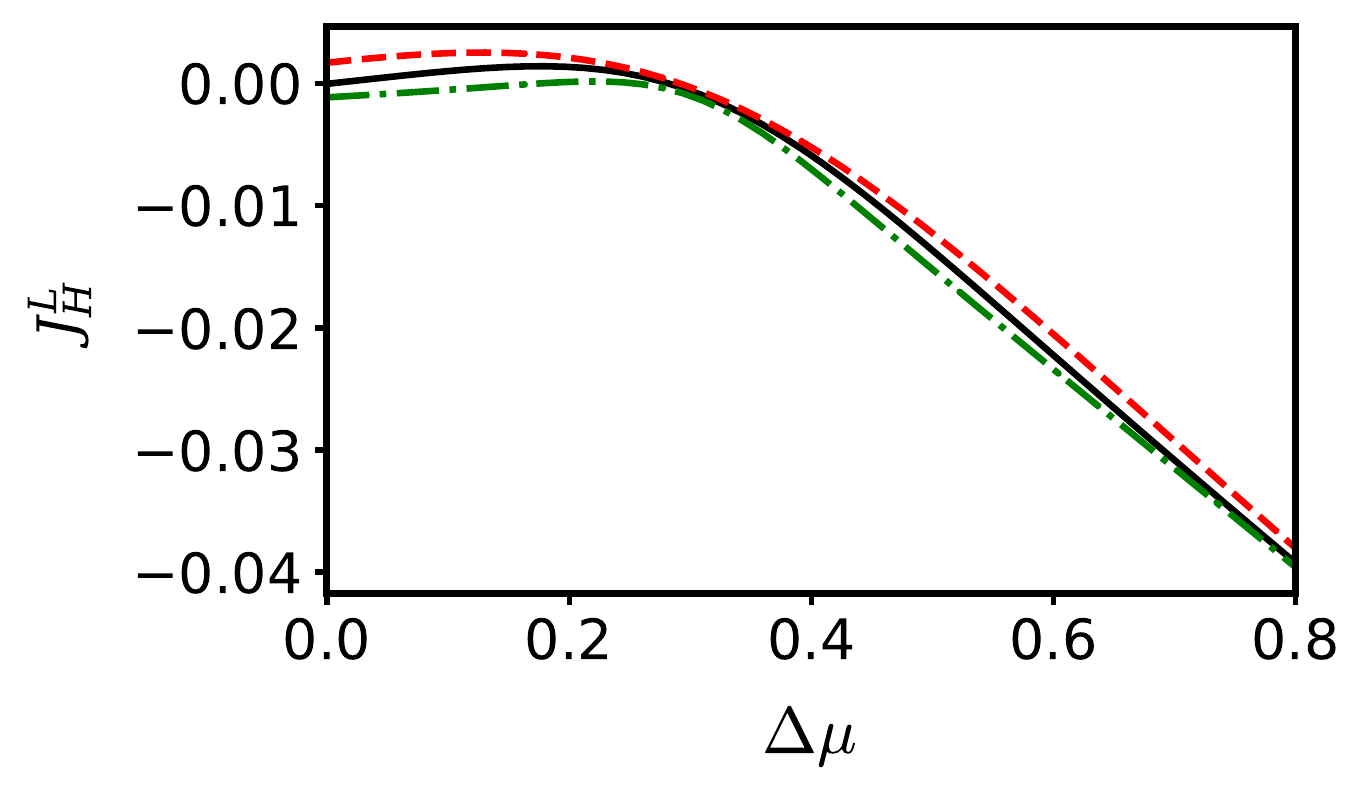}\label{Fig55(d)}}
	\caption{Off-resonant-empty regime $\varepsilon_{QB}=0.4$. Currents as a function of $\Delta\mu$.
		Panel (a): charge current $I_{Q}$. Panel (b): energy current $J_{E}$. Panel (c): left heat current $J^{L}_{H}$. 
		With $\mu_{R}=\mu_{eq}$ and $\mu_{L}=\mu_{eq}+\Delta \mu$. 
		Other parameters are same as Fig.~\ref{Fig2}.}
	\label{Fig55}
\end{figure*}

\begin{figure*}[t!]
	\centering
	\subfloat[ ]{\includegraphics[scale=0.42]{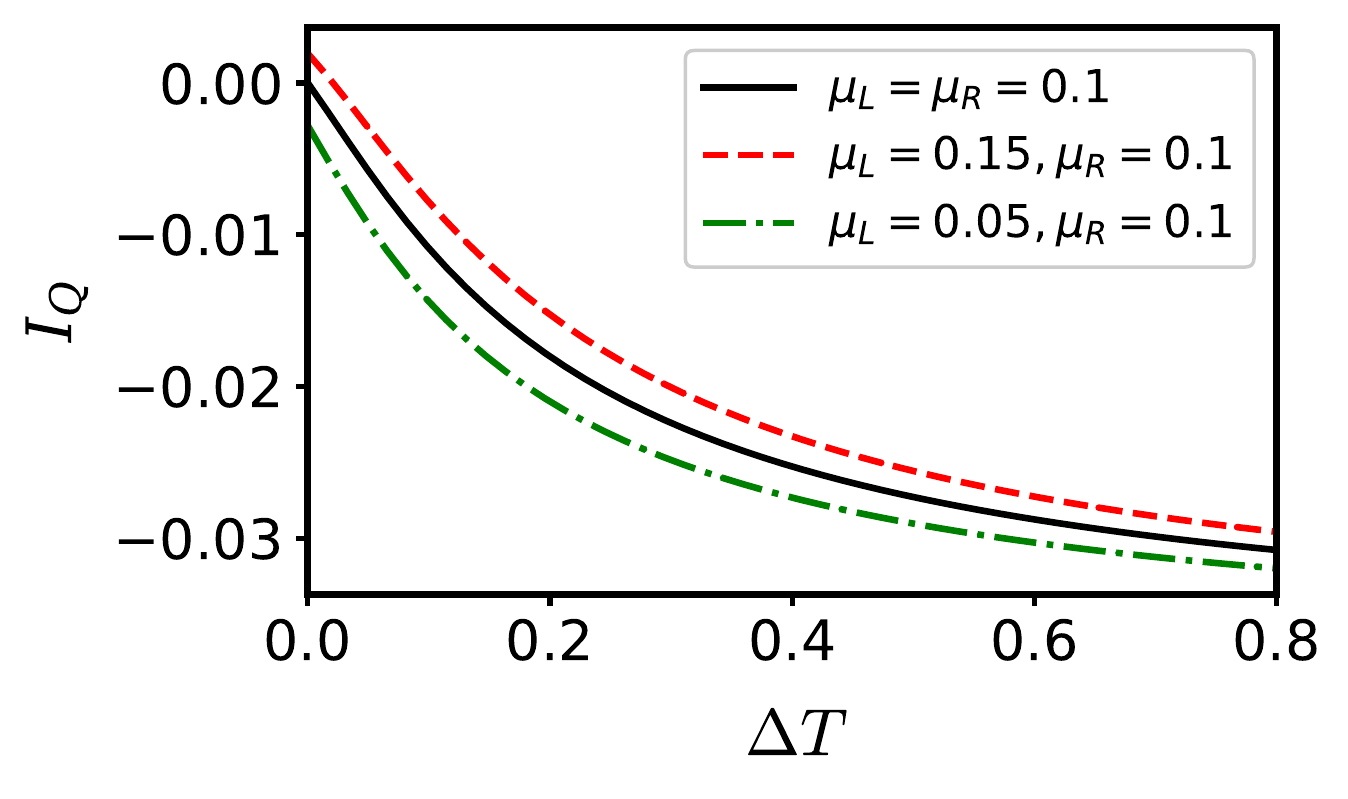}\label{Fig66(b)}}~
	\subfloat[ ]{\includegraphics[scale=0.42]{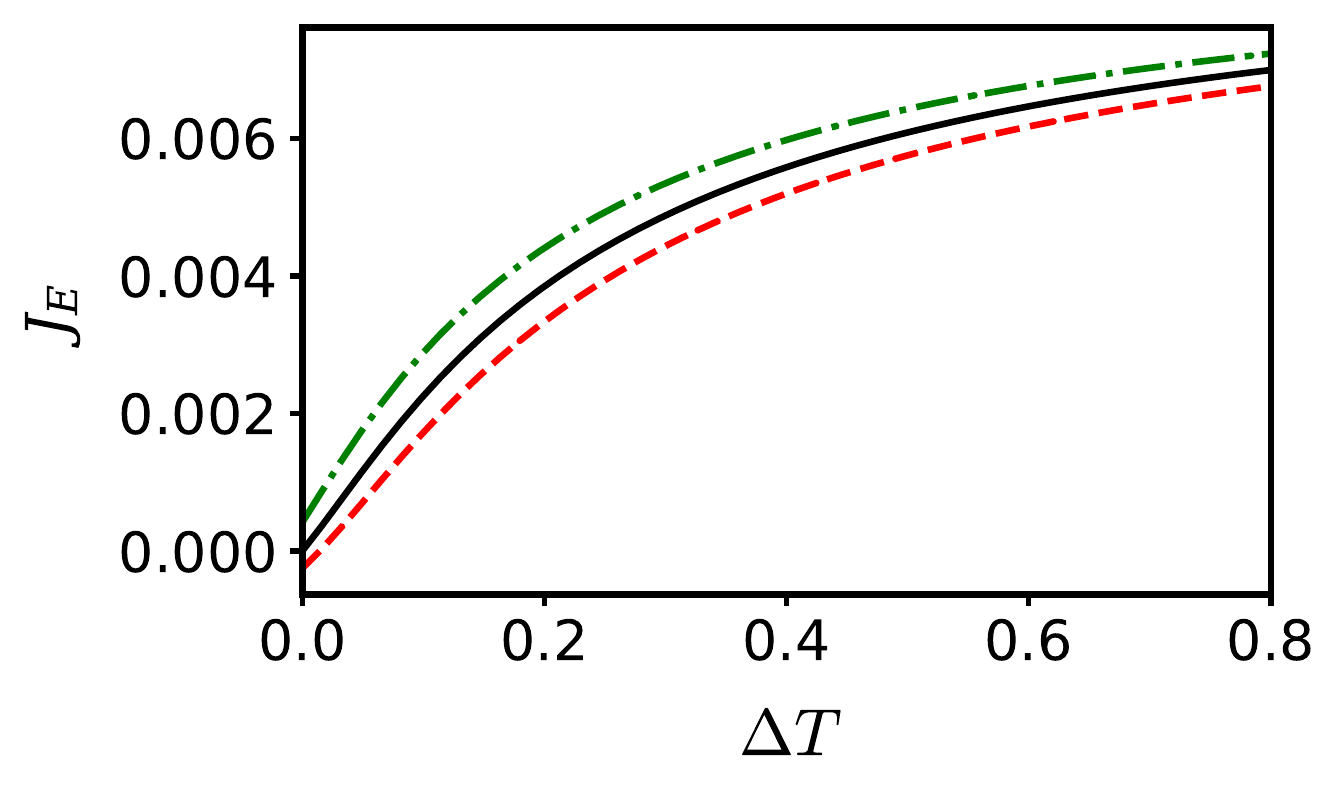}\label{Fig66(c)}}~
	\subfloat[ ]{\includegraphics[scale=0.42]{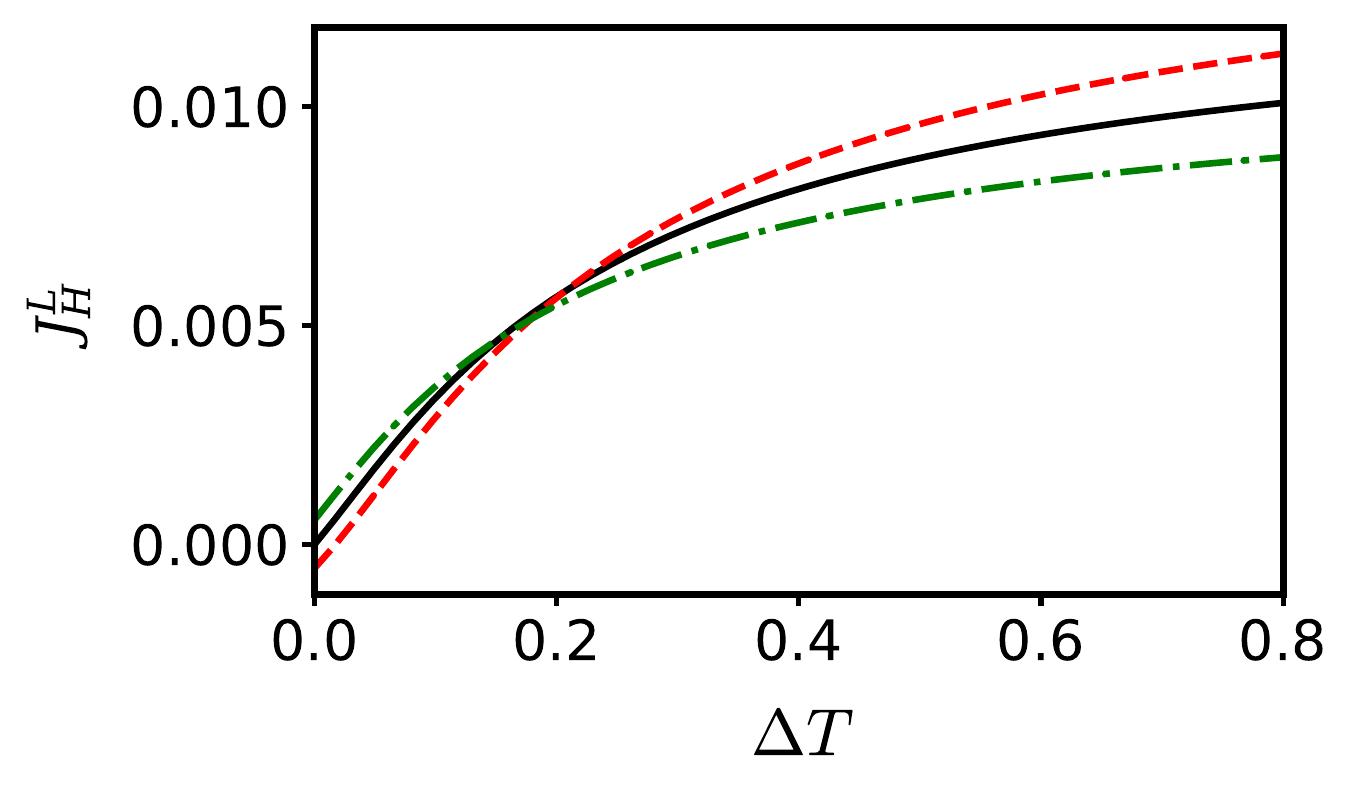}\label{Fig66(d)}}
	\caption{Off-resonant-full regime $\varepsilon_{QB}=-0.2$. Currents as a function of $\Delta T$. 
		Panel (a): charge current $I_{Q}$. Panel (b): energy current $J_{E}$. Panel (c): left heat current $J^{L}_{H}$. 
		Where $T_{R}=T_{eq}$ and $T_{L}=T_{eq}+\Delta T$. 
		Other parameters are same as Fig.~\ref{Fig2}}
	\label{Fig66}
\end{figure*}

\begin{figure*}[t!]
	\centering
	\subfloat[ ]{\includegraphics[scale=0.41]{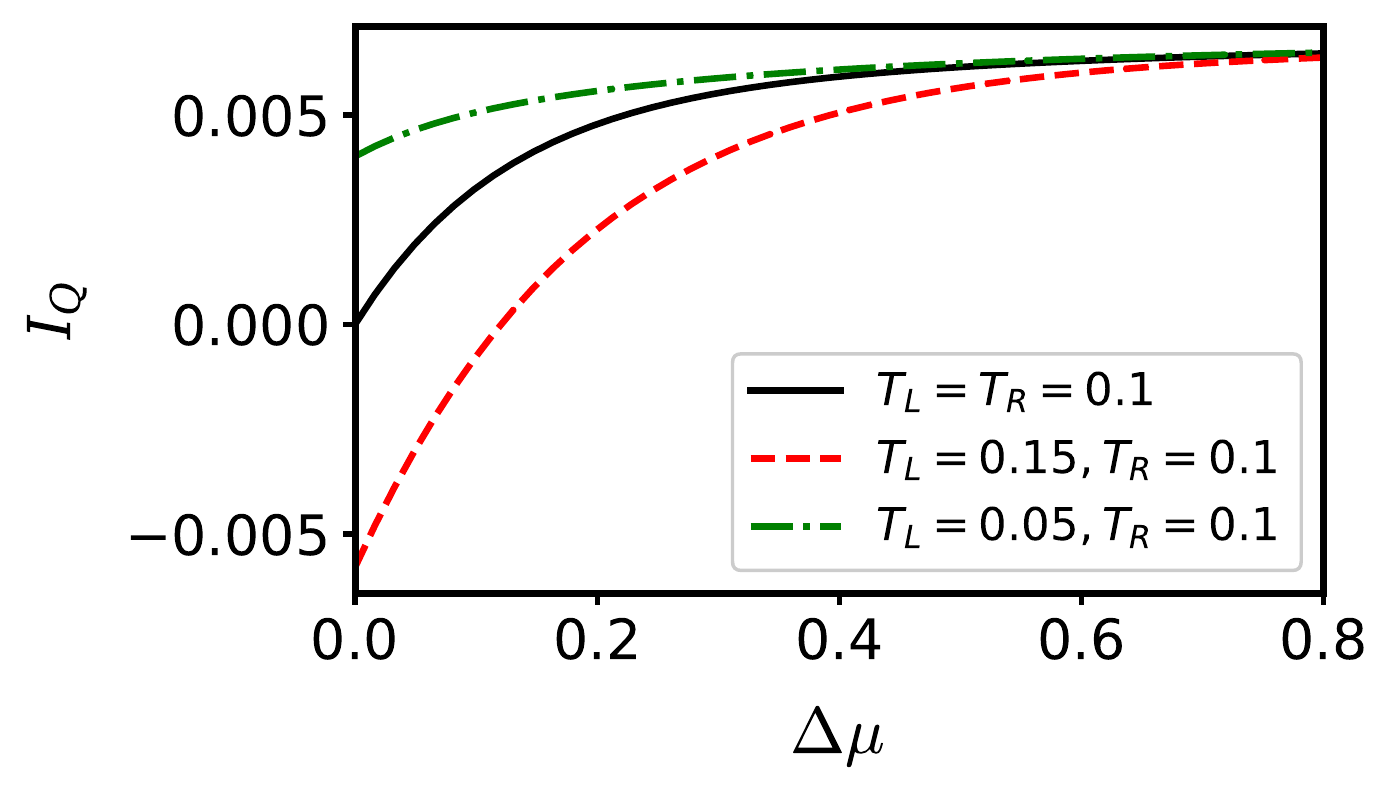}\label{Fig77(b)}}~
	\subfloat[ ]{\includegraphics[scale=0.41]{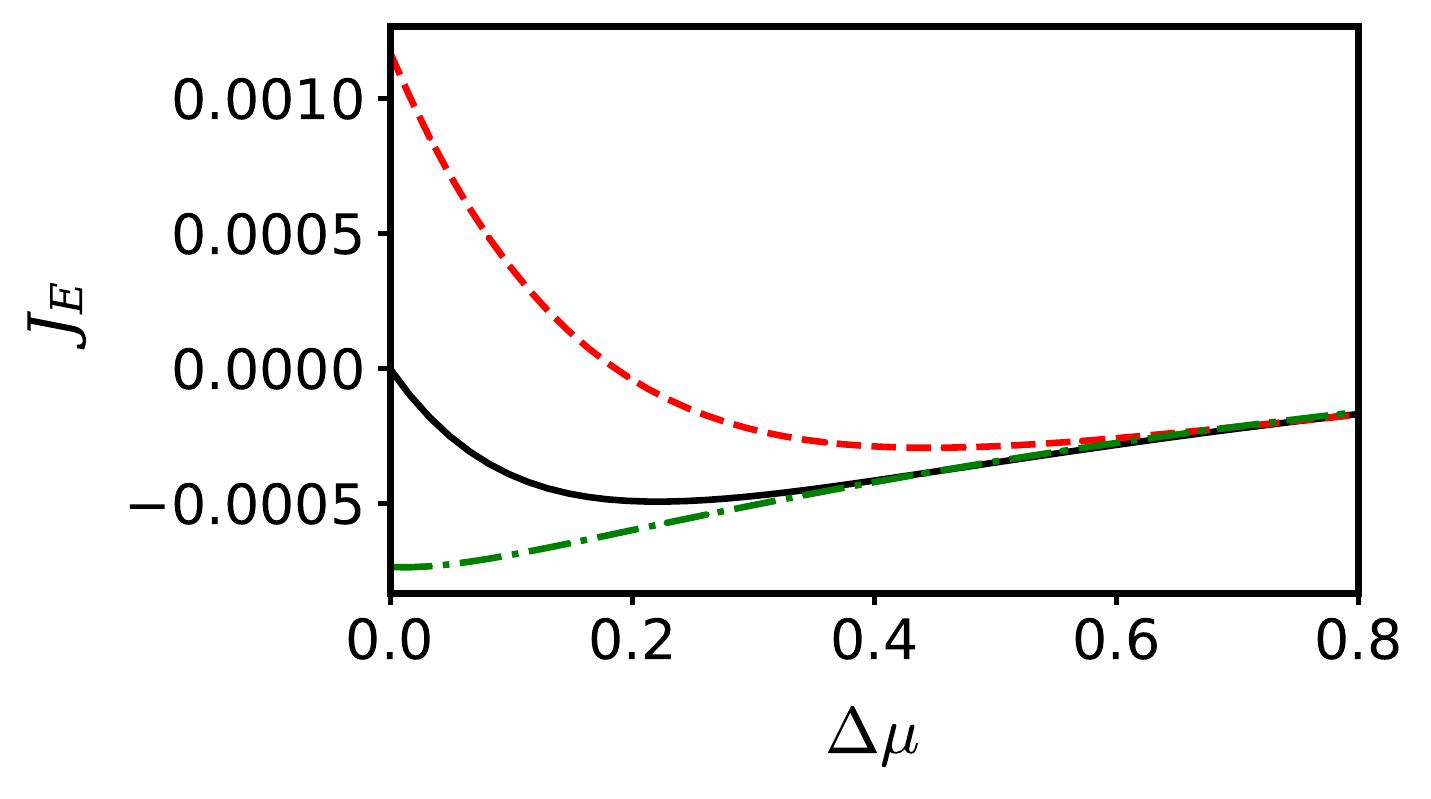}\label{Fig77(c)}}~
	\subfloat[ ]{\includegraphics[scale=0.41]{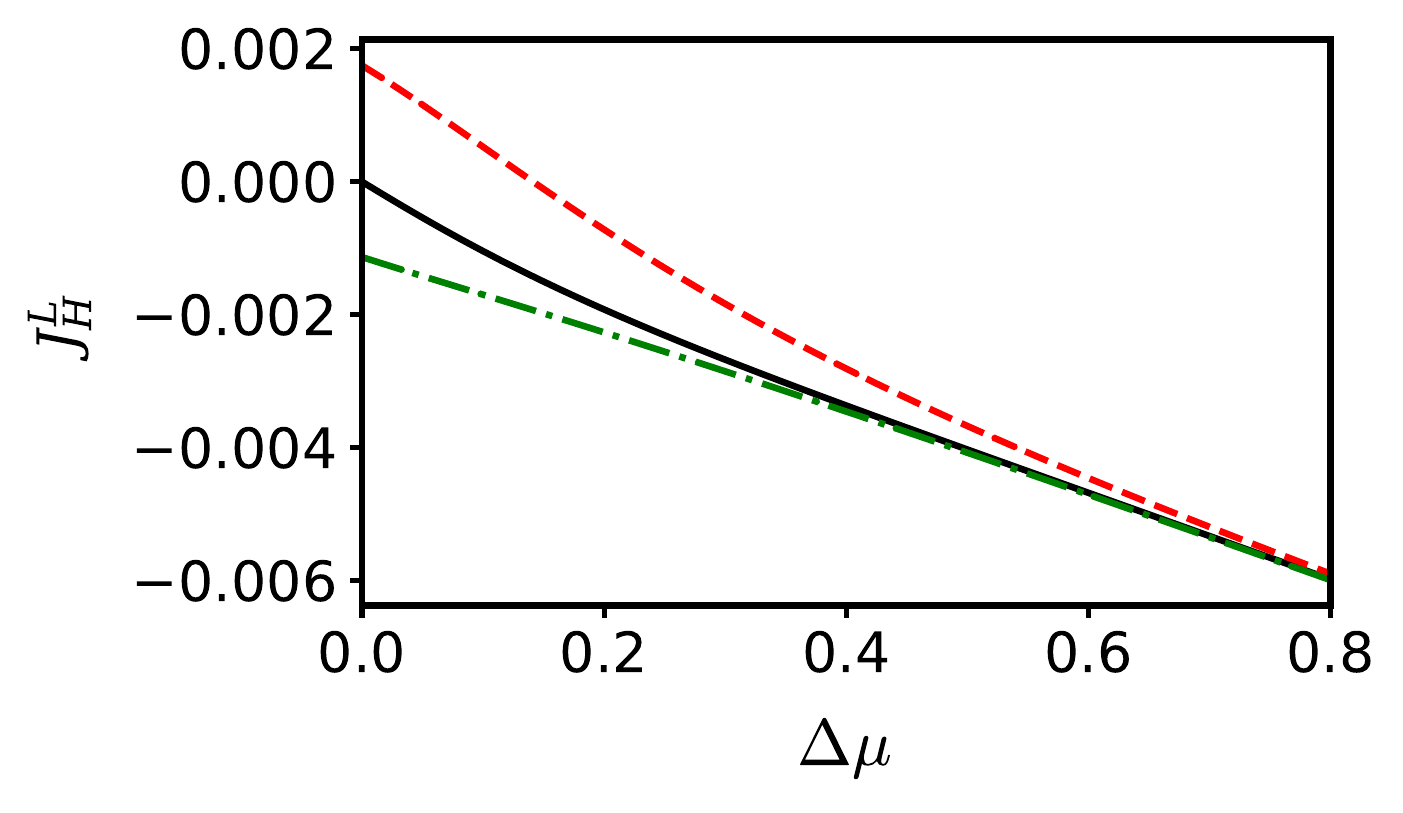}\label{Fig77(d)}}
	\caption{Off-resonant-full regime $\varepsilon_{QB}=-0.2$. Currents as a function of $\Delta\mu$.
		Panel (a): charge current $I_{Q}$. Panel (b): energy current $J_{E}$. Panel (c): left heat current $J^{L}_{H}$.
		Where $\mu_{R}=\mu_{eq}$ and $\mu_{L}=\mu_{eq}+\Delta \mu$. 
		Other parameters are same as Fig.~\ref{Fig2}.}
	\label{Fig77}
\end{figure*}

\begin{figure*}[t!]
	\centering
	\subfloat[ ]{\includegraphics[scale=0.5]{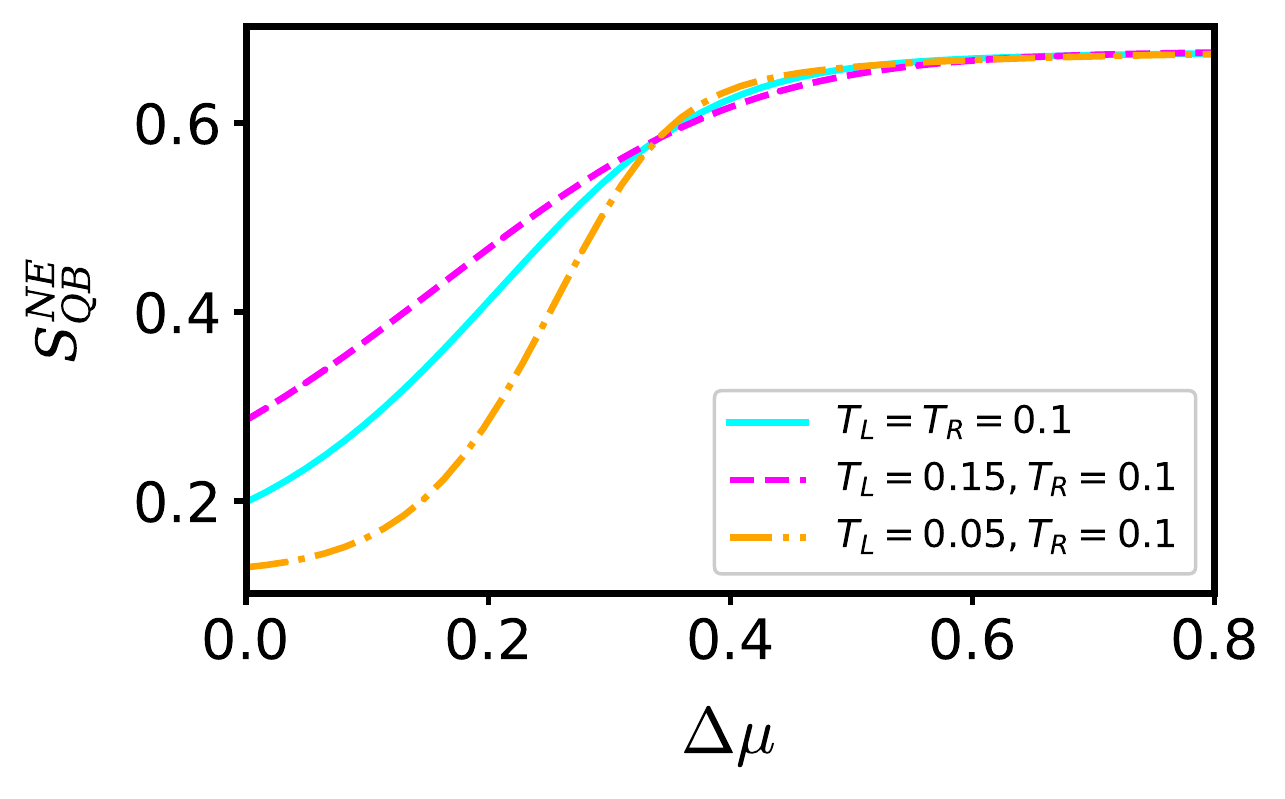}\label{SNE1-empty(a)}}~~~~~~~
	\subfloat[ ]{\includegraphics[scale=0.5]{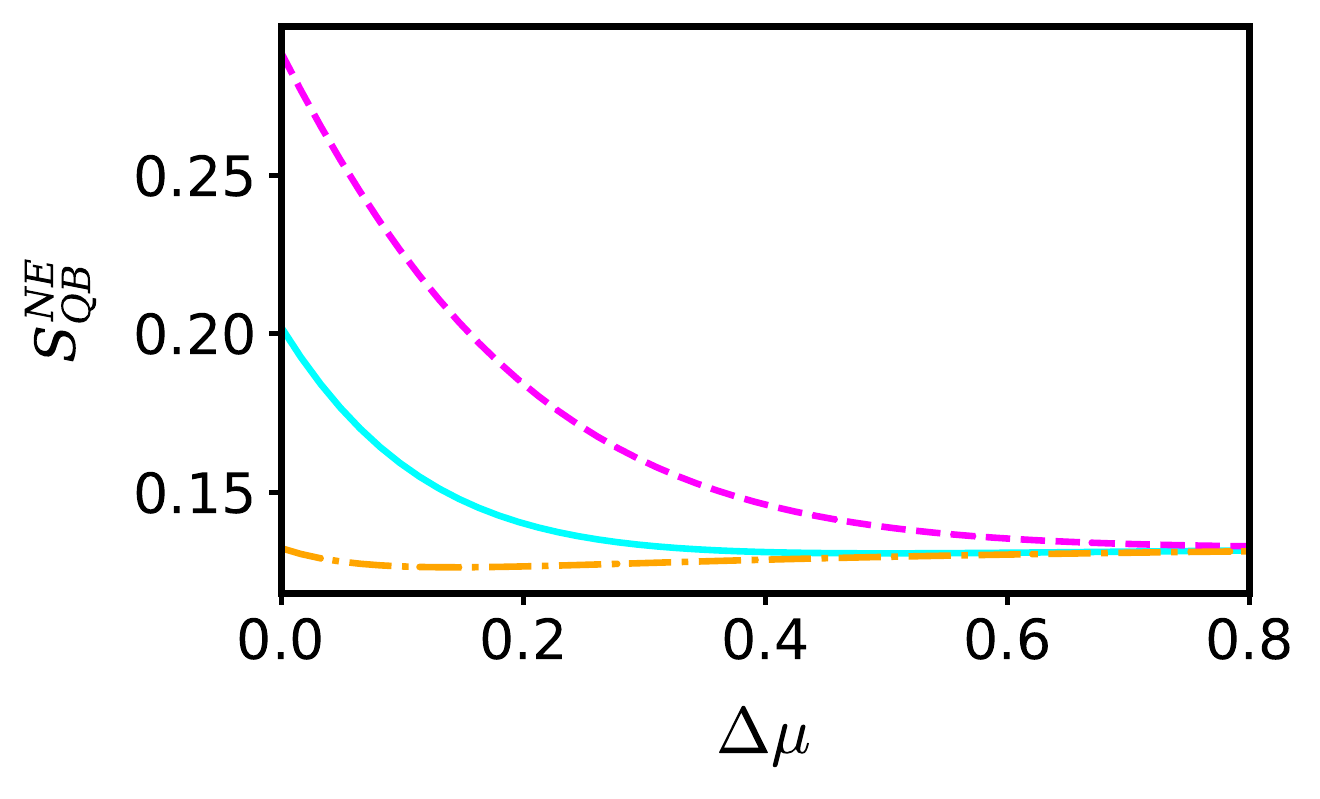}\label{SNE1-full(b)}}
	\caption{Non-equilibrium entropy $S^{NE}_{QB}$ as a function of $\Delta\mu$. Panel (a): Off-resonant-empty regime $\varepsilon_{QB}=0.4$. 
		Panel (b): Off-resonant-full regime $\varepsilon_{QB}=-0.2$.  Other parameters are same as Fig.~\ref{Fig22}.}
	\label{SNE-mu}
\end{figure*}

\newpage


\begin{thebibliography}{65}%
\makeatletter
\providecommand \@ifxundefined [1]{%
 \@ifx{#1\undefined}
}%
\providecommand \@ifnum [1]{%
 \ifnum #1\expandafter \@firstoftwo
 \else \expandafter \@secondoftwo
 \fi
}%
\providecommand \@ifx [1]{%
 \ifx #1\expandafter \@firstoftwo
 \else \expandafter \@secondoftwo
 \fi
}%
\providecommand \natexlab [1]{#1}%
\providecommand \enquote  [1]{``#1''}%
\providecommand \bibnamefont  [1]{#1}%
\providecommand \bibfnamefont [1]{#1}%
\providecommand \citenamefont [1]{#1}%
\providecommand \href@noop [0]{\@secondoftwo}%
\providecommand \href [0]{\begingroup \@sanitize@url \@href}%
\providecommand \@href[1]{\@@startlink{#1}\@@href}%
\providecommand \@@href[1]{\endgroup#1\@@endlink}%
\providecommand \@sanitize@url [0]{\catcode `\\12\catcode `\$12\catcode
  `\&12\catcode `\#12\catcode `\^12\catcode `\_12\catcode `\%12\relax}%
\providecommand \@@startlink[1]{}%
\providecommand \@@endlink[0]{}%
\providecommand \url  [0]{\begingroup\@sanitize@url \@url }%
\providecommand \@url [1]{\endgroup\@href {#1}{\urlprefix }}%
\providecommand \urlprefix  [0]{URL }%
\providecommand \Eprint [0]{\href }%
\providecommand \doibase [0]{https://doi.org/}%
\providecommand \selectlanguage [0]{\@gobble}%
\providecommand \bibinfo  [0]{\@secondoftwo}%
\providecommand \bibfield  [0]{\@secondoftwo}%
\providecommand \translation [1]{[#1]}%
\providecommand \BibitemOpen [0]{}%
\providecommand \bibitemStop [0]{}%
\providecommand \bibitemNoStop [0]{.\EOS\space}%
\providecommand \EOS [0]{\spacefactor3000\relax}%
\providecommand \BibitemShut  [1]{\csname bibitem#1\endcsname}%
\let\auto@bib@innerbib\@empty
\bibitem [{\citenamefont {Alicki}(2018)}]{Alicki2018}%
  \BibitemOpen
  \bibfield  {author} {\bibinfo {author} {\bibfnamefont {R.}~\bibnamefont
  {Alicki}, \bibfnamefont {Robertand~Kosloff}},\ }\bibinfo {title}
  {Introduction to quantum thermodynamics: History and prospects},\ in\ \href
  {https://doi.org/10.1007/978-3-319-99046-0_1} {\emph {\bibinfo {booktitle}
  {Thermodynamics in the Quantum Regime: Fundamental Aspects and New
  Directions}}},\ \bibinfo {editor} {edited by\ \bibinfo {editor}
  {\bibfnamefont {L.~A. G. C. A. J. A.~G.}\ \bibnamefont {Binder},
  \bibfnamefont {Felixand~Correa}}}\ (\bibinfo  {publisher} {Springer
  International Publishing},\ \bibinfo {address} {Cham},\ \bibinfo {year}
  {2018})\ pp.\ \bibinfo {pages} {1--33}\BibitemShut {NoStop}%
\bibitem [{\citenamefont {Nitzan}\ and\ \citenamefont
  {Ratner}(2003)}]{Nitzan:1}%
  \BibitemOpen
  \bibfield  {author} {\bibinfo {author} {\bibfnamefont {A.}~\bibnamefont
  {Nitzan}}\ and\ \bibinfo {author} {\bibfnamefont {M.~A.}\ \bibnamefont
  {Ratner}},\ }\bibfield  {title} {\bibinfo {title} {Electron transport in
  molecular wire junctions},\ }\href {https://doi.org/10.1126/science.1081572}
  {\bibfield  {journal} {\bibinfo  {journal} {Science}\ }\textbf {\bibinfo
  {volume} {300}},\ \bibinfo {pages} {1384} (\bibinfo {year}
  {2003})}\BibitemShut {NoStop}%
\bibitem [{\citenamefont {Pop}(2010)}]{pop:1}%
  \BibitemOpen
  \bibfield  {author} {\bibinfo {author} {\bibfnamefont {E.}~\bibnamefont
  {Pop}},\ }\bibfield  {title} {\bibinfo {title} {Energy dissipation and
  transport in nanoscale devices},\ }\href
  {https://doi.org/https://doi.org/10.1007/s12274-010-1019-z} {\bibfield
  {journal} {\bibinfo  {journal} {Nano Research}\ }\textbf {\bibinfo {volume}
  {3}},\ \bibinfo {pages} {147} (\bibinfo {year} {2010})}\BibitemShut {NoStop}%
\bibitem [{\citenamefont {Bergfield}\ and\ \citenamefont
  {Ratner}(2013)}]{Bergfield:1}%
  \BibitemOpen
  \bibfield  {author} {\bibinfo {author} {\bibfnamefont {J.~P.}\ \bibnamefont
  {Bergfield}}\ and\ \bibinfo {author} {\bibfnamefont {M.~A.}\ \bibnamefont
  {Ratner}},\ }\bibfield  {title} {\bibinfo {title} {Forty years of molecular
  electronics: Non-equilibrium heat and charge transport at the nanoscale},\
  }\href {https://doi.org/https://doi.org/10.1002/pssb.201350048} {\bibfield
  {journal} {\bibinfo  {journal} {physica status solidi (b)}\ }\textbf
  {\bibinfo {volume} {250}},\ \bibinfo {pages} {2249} (\bibinfo {year}
  {2013})}\BibitemShut {NoStop}%
\bibitem [{\citenamefont {Goldhaber-Gordon}\ \emph {et~al.}(1998)\citenamefont
  {Goldhaber-Gordon}, \citenamefont {Shtrikman}, \citenamefont {Mahalu},
  \citenamefont {Abusch-Magder}, \citenamefont {Meirav},\ and\ \citenamefont
  {Kastner}}]{goldhaber:1}%
  \BibitemOpen
  \bibfield  {author} {\bibinfo {author} {\bibfnamefont {D.}~\bibnamefont
  {Goldhaber-Gordon}}, \bibinfo {author} {\bibfnamefont {H.}~\bibnamefont
  {Shtrikman}}, \bibinfo {author} {\bibfnamefont {D.}~\bibnamefont {Mahalu}},
  \bibinfo {author} {\bibfnamefont {D.}~\bibnamefont {Abusch-Magder}}, \bibinfo
  {author} {\bibfnamefont {U.}~\bibnamefont {Meirav}},\ and\ \bibinfo {author}
  {\bibfnamefont {M.}~\bibnamefont {Kastner}},\ }\bibfield  {title} {\bibinfo
  {title} {Kondo effect in a single-electron transistor},\ }\href
  {https://doi.org/https://doi.org/10.1038/34373} {\bibfield  {journal}
  {\bibinfo  {journal} {Nature}\ }\textbf {\bibinfo {volume} {391}},\ \bibinfo
  {pages} {156} (\bibinfo {year} {1998})}\BibitemShut {NoStop}%
\bibitem [{\citenamefont {Cronenwett}\ \emph {et~al.}(1998)\citenamefont
  {Cronenwett}, \citenamefont {Oosterkamp},\ and\ \citenamefont
  {Kouwenhoven}}]{Cronenwett:1}%
  \BibitemOpen
  \bibfield  {author} {\bibinfo {author} {\bibfnamefont {S.~M.}\ \bibnamefont
  {Cronenwett}}, \bibinfo {author} {\bibfnamefont {T.~H.}\ \bibnamefont
  {Oosterkamp}},\ and\ \bibinfo {author} {\bibfnamefont {L.~P.}\ \bibnamefont
  {Kouwenhoven}},\ }\bibfield  {title} {\bibinfo {title} {A tunable kondo
  effect in quantum dots},\ }\href
  {https://www.science.org/doi/abs/10.1126/science.281.5376.540} {\bibfield
  {journal} {\bibinfo  {journal} {Science}\ }\textbf {\bibinfo {volume}
  {281}},\ \bibinfo {pages} {540} (\bibinfo {year} {1998})}\BibitemShut
  {NoStop}%
\bibitem [{\citenamefont {Schmid}\ \emph {et~al.}(1998)\citenamefont {Schmid},
  \citenamefont {Weis}, \citenamefont {Eberl},\ and\ \citenamefont {{v.
  Klitzing}}}]{SCHMID1998182}%
  \BibitemOpen
  \bibfield  {author} {\bibinfo {author} {\bibfnamefont {J.}~\bibnamefont
  {Schmid}}, \bibinfo {author} {\bibfnamefont {J.}~\bibnamefont {Weis}},
  \bibinfo {author} {\bibfnamefont {K.}~\bibnamefont {Eberl}},\ and\ \bibinfo
  {author} {\bibfnamefont {K.}~\bibnamefont {{v. Klitzing}}},\ }\bibfield
  {title} {\bibinfo {title} {A quantum dot in the limit of strong coupling to
  reservoirs},\ }\href
  {https://doi.org/https://doi.org/10.1016/S0921-4526(98)00533-X} {\bibfield
  {journal} {\bibinfo  {journal} {Physica B: Condensed Matter}\ }\textbf
  {\bibinfo {volume} {256-258}},\ \bibinfo {pages} {182} (\bibinfo {year}
  {1998})}\BibitemShut {NoStop}%
\bibitem [{\citenamefont {Lerner}\ \emph {et~al.}(2004)\citenamefont {Lerner},
  \citenamefont {Altshuler},\ and\ \citenamefont {Gefen}}]{lerner:1}%
  \BibitemOpen
  \bibfield  {author} {\bibinfo {author} {\bibfnamefont {I.}~\bibnamefont
  {Lerner}}, \bibinfo {author} {\bibfnamefont {B.}~\bibnamefont {Altshuler}},\
  and\ \bibinfo {author} {\bibfnamefont {Y.}~\bibnamefont {Gefen}},\ }\href
  {https://books.google.com/books?id=DXXUW8qIU-0C} {\emph {\bibinfo {title}
  {Fundamental Problems of Mesoscopic Physics: Interactions and
  Decoherence}}},\ NATO Science Series II: Mathematics, Physics and Chemistry\
  (\bibinfo  {publisher} {Springer Netherlands},\ \bibinfo {year}
  {2004})\BibitemShut {NoStop}%
\bibitem [{\citenamefont {Bouchiat}\ \emph {et~al.}(2004)\citenamefont
  {Bouchiat}, \citenamefont {Dalibard}, \citenamefont {Gefen}, \citenamefont
  {Gu{\'e}ron},\ and\ \citenamefont {Montambaux}}]{Glazman:1}%
  \BibitemOpen
  \bibfield  {author} {\bibinfo {author} {\bibfnamefont {H.}~\bibnamefont
  {Bouchiat}}, \bibinfo {author} {\bibfnamefont {J.}~\bibnamefont {Dalibard}},
  \bibinfo {author} {\bibfnamefont {Y.}~\bibnamefont {Gefen}}, \bibinfo
  {author} {\bibfnamefont {S.}~\bibnamefont {Gu{\'e}ron}},\ and\ \bibinfo
  {author} {\bibfnamefont {G.}~\bibnamefont {Montambaux}},\ }\href
  {https://cds.cern.ch/record/992286} {\emph {\bibinfo {title} {Nanophysics:
  Coherence and Transport}}}\ (\bibinfo  {publisher} {Elsevier},\ \bibinfo
  {address} {San Diego, CA},\ \bibinfo {year} {2004})\BibitemShut {NoStop}%
\bibitem [{\citenamefont {Campaioli}\ \emph {et~al.}(2017)\citenamefont
  {Campaioli}, \citenamefont {Pollock}, \citenamefont {Binder}, \citenamefont
  {C\'eleri}, \citenamefont {Goold}, \citenamefont {Vinjanampathy},\ and\
  \citenamefont {Modi}}]{Campaioli:1}%
  \BibitemOpen
  \bibfield  {author} {\bibinfo {author} {\bibfnamefont {F.}~\bibnamefont
  {Campaioli}}, \bibinfo {author} {\bibfnamefont {F.~A.}\ \bibnamefont
  {Pollock}}, \bibinfo {author} {\bibfnamefont {F.~C.}\ \bibnamefont {Binder}},
  \bibinfo {author} {\bibfnamefont {L.}~\bibnamefont {C\'eleri}}, \bibinfo
  {author} {\bibfnamefont {J.}~\bibnamefont {Goold}}, \bibinfo {author}
  {\bibfnamefont {S.}~\bibnamefont {Vinjanampathy}},\ and\ \bibinfo {author}
  {\bibfnamefont {K.}~\bibnamefont {Modi}},\ }\bibfield  {title} {\bibinfo
  {title} {Enhancing the charging power of quantum batteries},\ }\href
  {https://doi.org/10.1103/PhysRevLett.118.150601} {\bibfield  {journal}
  {\bibinfo  {journal} {Phys. Rev. Lett.}\ }\textbf {\bibinfo {volume} {118}},\
  \bibinfo {pages} {150601} (\bibinfo {year} {2017})}\BibitemShut {NoStop}%
\bibitem [{\citenamefont {Andolina}\ \emph {et~al.}(2019)\citenamefont
  {Andolina}, \citenamefont {Keck}, \citenamefont {Mari}, \citenamefont
  {Campisi}, \citenamefont {Giovannetti},\ and\ \citenamefont
  {Polini}}]{Andolina:1}%
  \BibitemOpen
  \bibfield  {author} {\bibinfo {author} {\bibfnamefont {G.~M.}\ \bibnamefont
  {Andolina}}, \bibinfo {author} {\bibfnamefont {M.}~\bibnamefont {Keck}},
  \bibinfo {author} {\bibfnamefont {A.}~\bibnamefont {Mari}}, \bibinfo {author}
  {\bibfnamefont {M.}~\bibnamefont {Campisi}}, \bibinfo {author} {\bibfnamefont
  {V.}~\bibnamefont {Giovannetti}},\ and\ \bibinfo {author} {\bibfnamefont
  {M.}~\bibnamefont {Polini}},\ }\bibfield  {title} {\bibinfo {title}
  {Extractable work, the role of correlations, and asymptotic freedom in
  quantum batteries},\ }\href {https://doi.org/10.1103/PhysRevLett.122.047702}
  {\bibfield  {journal} {\bibinfo  {journal} {Phys. Rev. Lett.}\ }\textbf
  {\bibinfo {volume} {122}},\ \bibinfo {pages} {047702} (\bibinfo {year}
  {2019})}\BibitemShut {NoStop}%
\bibitem [{\citenamefont {Kamin}\ \emph
  {et~al.}(2020{\natexlab{a}})\citenamefont {Kamin}, \citenamefont {Tabesh},
  \citenamefont {Salimi},\ and\ \citenamefont {Santos}}]{Kamin:3}%
  \BibitemOpen
  \bibfield  {author} {\bibinfo {author} {\bibfnamefont {F.~H.}\ \bibnamefont
  {Kamin}}, \bibinfo {author} {\bibfnamefont {F.~T.}\ \bibnamefont {Tabesh}},
  \bibinfo {author} {\bibfnamefont {S.}~\bibnamefont {Salimi}},\ and\ \bibinfo
  {author} {\bibfnamefont {A.~C.}\ \bibnamefont {Santos}},\ }\bibfield  {title}
  {\bibinfo {title} {Entanglement, coherence, and charging process of quantum
  batteries},\ }\href {https://doi.org/10.1103/PhysRevE.102.052109} {\bibfield
  {journal} {\bibinfo  {journal} {Phys. Rev. E}\ }\textbf {\bibinfo {volume}
  {102}},\ \bibinfo {pages} {052109} (\bibinfo {year}
  {2020}{\natexlab{a}})}\BibitemShut {NoStop}%
\bibitem [{\citenamefont {Cruz}\ \emph {et~al.}(2022)\citenamefont {Cruz},
  \citenamefont {Anka}, \citenamefont {Reis}, \citenamefont {Bachelard},\ and\
  \citenamefont {Santos}}]{Cruz:1}%
  \BibitemOpen
  \bibfield  {author} {\bibinfo {author} {\bibfnamefont {C.}~\bibnamefont
  {Cruz}}, \bibinfo {author} {\bibfnamefont {M.~F.}\ \bibnamefont {Anka}},
  \bibinfo {author} {\bibfnamefont {M.~S.}\ \bibnamefont {Reis}}, \bibinfo
  {author} {\bibfnamefont {R.}~\bibnamefont {Bachelard}},\ and\ \bibinfo
  {author} {\bibfnamefont {A.~C.}\ \bibnamefont {Santos}},\ }\bibfield  {title}
  {\bibinfo {title} {Quantum battery based on quantum discord at room
  temperature},\ }\href {https://doi.org/10.1088/2058-9565/ac57f3} {\bibfield
  {journal} {\bibinfo  {journal} {Quantum Science and Technology}\ }\textbf
  {\bibinfo {volume} {7}},\ \bibinfo {pages} {025020} (\bibinfo {year}
  {2022})}\BibitemShut {NoStop}%
\bibitem [{\citenamefont {Rossini}\ \emph {et~al.}(2020)\citenamefont
  {Rossini}, \citenamefont {Andolina}, \citenamefont {Rosa}, \citenamefont
  {Carrega},\ and\ \citenamefont {Polini}}]{Rossini:2020}%
  \BibitemOpen
  \bibfield  {author} {\bibinfo {author} {\bibfnamefont {D.}~\bibnamefont
  {Rossini}}, \bibinfo {author} {\bibfnamefont {G.~M.}\ \bibnamefont
  {Andolina}}, \bibinfo {author} {\bibfnamefont {D.}~\bibnamefont {Rosa}},
  \bibinfo {author} {\bibfnamefont {M.}~\bibnamefont {Carrega}},\ and\ \bibinfo
  {author} {\bibfnamefont {M.}~\bibnamefont {Polini}},\ }\bibfield  {title}
  {\bibinfo {title} {Quantum advantage in the charging process of
  sachdev-ye-kitaev batteries},\ }\href
  {https://doi.org/10.1103/PhysRevLett.125.236402} {\bibfield  {journal}
  {\bibinfo  {journal} {Phys. Rev. Lett.}\ }\textbf {\bibinfo {volume} {125}},\
  \bibinfo {pages} {236402} (\bibinfo {year} {2020})}\BibitemShut {NoStop}%
\bibitem [{\citenamefont {Gyhm}\ \emph {et~al.}(2022)\citenamefont {Gyhm},
  \citenamefont {\ifmmode~\check{S}\else \v{S}\fi{}afr\'anek},\ and\
  \citenamefont {Rosa}}]{Gyhm:2022}%
  \BibitemOpen
  \bibfield  {author} {\bibinfo {author} {\bibfnamefont {J.-Y.}\ \bibnamefont
  {Gyhm}}, \bibinfo {author} {\bibfnamefont {D.}~\bibnamefont
  {\ifmmode~\check{S}\else \v{S}\fi{}afr\'anek}},\ and\ \bibinfo {author}
  {\bibfnamefont {D.}~\bibnamefont {Rosa}},\ }\bibfield  {title} {\bibinfo
  {title} {Quantum charging advantage cannot be extensive without global
  operations},\ }\href {https://doi.org/10.1103/PhysRevLett.128.140501}
  {\bibfield  {journal} {\bibinfo  {journal} {Phys. Rev. Lett.}\ }\textbf
  {\bibinfo {volume} {128}},\ \bibinfo {pages} {140501} (\bibinfo {year}
  {2022})}\BibitemShut {NoStop}%
\bibitem [{\citenamefont {Bozkurt}\ \emph {et~al.}(2018)\citenamefont
  {Bozkurt}, \citenamefont {Pekerten},\ and\ \citenamefont
  {Adagideli}}]{Bozkurt:2018}%
  \BibitemOpen
  \bibfield  {author} {\bibinfo {author} {\bibfnamefont {A.~M.}\ \bibnamefont
  {Bozkurt}}, \bibinfo {author} {\bibfnamefont {B.}~\bibnamefont {Pekerten}},\
  and\ \bibinfo {author} {\bibfnamefont {I.}~\bibnamefont {Adagideli}},\
  }\bibfield  {title} {\bibinfo {title} {Work extraction and landauer's
  principle in a quantum spin hall device},\ }\href
  {https://doi.org/10.1103/PhysRevB.97.245414} {\bibfield  {journal} {\bibinfo
  {journal} {Phys. Rev. B}\ }\textbf {\bibinfo {volume} {97}},\ \bibinfo
  {pages} {245414} (\bibinfo {year} {2018})}\BibitemShut {NoStop}%
\bibitem [{\citenamefont {Ferraro}\ \emph {et~al.}(2018)\citenamefont
  {Ferraro}, \citenamefont {Campisi}, \citenamefont {Andolina}, \citenamefont
  {Pellegrini},\ and\ \citenamefont {Polini}}]{Ferraro:1}%
  \BibitemOpen
  \bibfield  {author} {\bibinfo {author} {\bibfnamefont {D.}~\bibnamefont
  {Ferraro}}, \bibinfo {author} {\bibfnamefont {M.}~\bibnamefont {Campisi}},
  \bibinfo {author} {\bibfnamefont {G.~M.}\ \bibnamefont {Andolina}}, \bibinfo
  {author} {\bibfnamefont {V.}~\bibnamefont {Pellegrini}},\ and\ \bibinfo
  {author} {\bibfnamefont {M.}~\bibnamefont {Polini}},\ }\bibfield  {title}
  {\bibinfo {title} {High-power collective charging of a solid-state quantum
  battery},\ }\href {https://doi.org/10.1103/PhysRevLett.120.117702} {\bibfield
   {journal} {\bibinfo  {journal} {Phys. Rev. Lett.}\ }\textbf {\bibinfo
  {volume} {120}},\ \bibinfo {pages} {117702} (\bibinfo {year}
  {2018})}\BibitemShut {NoStop}%
\bibitem [{\citenamefont {Rossini}\ \emph {et~al.}(2019)\citenamefont
  {Rossini}, \citenamefont {Andolina},\ and\ \citenamefont
  {Polini}}]{Rossini:1}%
  \BibitemOpen
  \bibfield  {author} {\bibinfo {author} {\bibfnamefont {D.}~\bibnamefont
  {Rossini}}, \bibinfo {author} {\bibfnamefont {G.~M.}\ \bibnamefont
  {Andolina}},\ and\ \bibinfo {author} {\bibfnamefont {M.}~\bibnamefont
  {Polini}},\ }\bibfield  {title} {\bibinfo {title} {Many-body localized
  quantum batteries},\ }\href {https://doi.org/10.1103/PhysRevB.100.115142}
  {\bibfield  {journal} {\bibinfo  {journal} {Phys. Rev. B}\ }\textbf {\bibinfo
  {volume} {100}},\ \bibinfo {pages} {115142} (\bibinfo {year}
  {2019})}\BibitemShut {NoStop}%
\bibitem [{\citenamefont {Crescente}\ \emph {et~al.}(2020)\citenamefont
  {Crescente}, \citenamefont {Carrega}, \citenamefont {Sassetti},\ and\
  \citenamefont {Ferraro}}]{Crescente:1}%
  \BibitemOpen
  \bibfield  {author} {\bibinfo {author} {\bibfnamefont {A.}~\bibnamefont
  {Crescente}}, \bibinfo {author} {\bibfnamefont {M.}~\bibnamefont {Carrega}},
  \bibinfo {author} {\bibfnamefont {M.}~\bibnamefont {Sassetti}},\ and\
  \bibinfo {author} {\bibfnamefont {D.}~\bibnamefont {Ferraro}},\ }\bibfield
  {title} {\bibinfo {title} {Ultrafast charging in a two-photon dicke quantum
  battery},\ }\href {https://doi.org/10.1103/PhysRevB.102.245407} {\bibfield
  {journal} {\bibinfo  {journal} {Phys. Rev. B}\ }\textbf {\bibinfo {volume}
  {102}},\ \bibinfo {pages} {245407} (\bibinfo {year} {2020})}\BibitemShut
  {NoStop}%
\bibitem [{\citenamefont {Le}\ \emph {et~al.}(2018)\citenamefont {Le},
  \citenamefont {Levinsen}, \citenamefont {Modi}, \citenamefont {Parish},\ and\
  \citenamefont {Pollock}}]{Le:1}%
  \BibitemOpen
  \bibfield  {author} {\bibinfo {author} {\bibfnamefont {T.~P.}\ \bibnamefont
  {Le}}, \bibinfo {author} {\bibfnamefont {J.}~\bibnamefont {Levinsen}},
  \bibinfo {author} {\bibfnamefont {K.}~\bibnamefont {Modi}}, \bibinfo {author}
  {\bibfnamefont {M.~M.}\ \bibnamefont {Parish}},\ and\ \bibinfo {author}
  {\bibfnamefont {F.~A.}\ \bibnamefont {Pollock}},\ }\bibfield  {title}
  {\bibinfo {title} {Spin-chain model of a many-body quantum battery},\ }\href
  {https://doi.org/10.1103/PhysRevA.97.022106} {\bibfield  {journal} {\bibinfo
  {journal} {Phys. Rev. A}\ }\textbf {\bibinfo {volume} {97}},\ \bibinfo
  {pages} {022106} (\bibinfo {year} {2018})}\BibitemShut {NoStop}%
\bibitem [{\citenamefont {Chen}\ \emph {et~al.}(2020)\citenamefont {Chen},
  \citenamefont {Zhan}, \citenamefont {Shao}, \citenamefont {Zhang},
  \citenamefont {Zhang},\ and\ \citenamefont {Wang}}]{Chen:1}%
  \BibitemOpen
  \bibfield  {author} {\bibinfo {author} {\bibfnamefont {J.}~\bibnamefont
  {Chen}}, \bibinfo {author} {\bibfnamefont {L.}~\bibnamefont {Zhan}}, \bibinfo
  {author} {\bibfnamefont {L.}~\bibnamefont {Shao}}, \bibinfo {author}
  {\bibfnamefont {X.}~\bibnamefont {Zhang}}, \bibinfo {author} {\bibfnamefont
  {Y.}~\bibnamefont {Zhang}},\ and\ \bibinfo {author} {\bibfnamefont
  {X.}~\bibnamefont {Wang}},\ }\bibfield  {title} {\bibinfo {title} {Charging
  quantum batteries with a general harmonic driving field},\ }\href
  {https://doi.org/https://doi.org/10.1002/andp.201900487} {\bibfield
  {journal} {\bibinfo  {journal} {Annalen der Physik}\ }\textbf {\bibinfo
  {volume} {532}},\ \bibinfo {pages} {1900487} (\bibinfo {year}
  {2020})}\BibitemShut {NoStop}%
\bibitem [{\citenamefont {Bai}\ and\ \citenamefont {An}(2020)}]{Bai:1}%
  \BibitemOpen
  \bibfield  {author} {\bibinfo {author} {\bibfnamefont {S.-Y.}\ \bibnamefont
  {Bai}}\ and\ \bibinfo {author} {\bibfnamefont {J.-H.}\ \bibnamefont {An}},\
  }\bibfield  {title} {\bibinfo {title} {Floquet engineering to reactivate a
  dissipative quantum battery},\ }\href
  {https://doi.org/10.1103/PhysRevA.102.060201} {\bibfield  {journal} {\bibinfo
   {journal} {Phys. Rev. A}\ }\textbf {\bibinfo {volume} {102}},\ \bibinfo
  {pages} {060201} (\bibinfo {year} {2020})}\BibitemShut {NoStop}%
\bibitem [{\citenamefont {Pirmoradian}\ and\ \citenamefont
  {M\o{}lmer}(2019)}]{pirmoradian:1}%
  \BibitemOpen
  \bibfield  {author} {\bibinfo {author} {\bibfnamefont {F.}~\bibnamefont
  {Pirmoradian}}\ and\ \bibinfo {author} {\bibfnamefont {K.}~\bibnamefont
  {M\o{}lmer}},\ }\bibfield  {title} {\bibinfo {title} {Aging of a quantum
  battery},\ }\href {https://doi.org/10.1103/PhysRevA.100.043833} {\bibfield
  {journal} {\bibinfo  {journal} {Phys. Rev. A}\ }\textbf {\bibinfo {volume}
  {100}},\ \bibinfo {pages} {043833} (\bibinfo {year} {2019})}\BibitemShut
  {NoStop}%
\bibitem [{\citenamefont {Barra}(2019)}]{Barra:1}%
  \BibitemOpen
  \bibfield  {author} {\bibinfo {author} {\bibfnamefont {F.}~\bibnamefont
  {Barra}},\ }\bibfield  {title} {\bibinfo {title} {Dissipative charging of a
  quantum battery},\ }\href {https://doi.org/10.1103/PhysRevLett.122.210601}
  {\bibfield  {journal} {\bibinfo  {journal} {Phys. Rev. Lett.}\ }\textbf
  {\bibinfo {volume} {122}},\ \bibinfo {pages} {210601} (\bibinfo {year}
  {2019})}\BibitemShut {NoStop}%
\bibitem [{\citenamefont {Farina}\ \emph {et~al.}(2019)\citenamefont {Farina},
  \citenamefont {Andolina}, \citenamefont {Mari}, \citenamefont {Polini},\ and\
  \citenamefont {Giovannetti}}]{Farina:1}%
  \BibitemOpen
  \bibfield  {author} {\bibinfo {author} {\bibfnamefont {D.}~\bibnamefont
  {Farina}}, \bibinfo {author} {\bibfnamefont {G.~M.}\ \bibnamefont
  {Andolina}}, \bibinfo {author} {\bibfnamefont {A.}~\bibnamefont {Mari}},
  \bibinfo {author} {\bibfnamefont {M.}~\bibnamefont {Polini}},\ and\ \bibinfo
  {author} {\bibfnamefont {V.}~\bibnamefont {Giovannetti}},\ }\bibfield
  {title} {\bibinfo {title} {Charger-mediated energy transfer for quantum
  batteries: An open-system approach},\ }\href
  {https://doi.org/10.1103/PhysRevB.99.035421} {\bibfield  {journal} {\bibinfo
  {journal} {Phys. Rev. B}\ }\textbf {\bibinfo {volume} {99}},\ \bibinfo
  {pages} {035421} (\bibinfo {year} {2019})}\BibitemShut {NoStop}%
\bibitem [{\citenamefont {Liu}\ \emph {et~al.}(2019)\citenamefont {Liu},
  \citenamefont {Segal},\ and\ \citenamefont {Hanna}}]{Liu:1}%
  \BibitemOpen
  \bibfield  {author} {\bibinfo {author} {\bibfnamefont {J.}~\bibnamefont
  {Liu}}, \bibinfo {author} {\bibfnamefont {D.}~\bibnamefont {Segal}},\ and\
  \bibinfo {author} {\bibfnamefont {G.}~\bibnamefont {Hanna}},\ }\bibfield
  {title} {\bibinfo {title} {Loss-free excitonic quantum battery},\ }\href
  {https://doi.org/10.1021/acs.jpcc.9b06373} {\bibfield  {journal} {\bibinfo
  {journal} {The Journal of Physical Chemistry C}\ }\textbf {\bibinfo {volume}
  {123}},\ \bibinfo {pages} {18303} (\bibinfo {year} {2019})}\BibitemShut
  {NoStop}%
\bibitem [{\citenamefont {Quach}\ and\ \citenamefont {Munro}(2020)}]{Quach:1}%
  \BibitemOpen
  \bibfield  {author} {\bibinfo {author} {\bibfnamefont {J.~Q.}\ \bibnamefont
  {Quach}}\ and\ \bibinfo {author} {\bibfnamefont {W.~J.}\ \bibnamefont
  {Munro}},\ }\bibfield  {title} {\bibinfo {title} {Using dark states to charge
  and stabilize open quantum batteries},\ }\href
  {https://doi.org/10.1103/PhysRevApplied.14.024092} {\bibfield  {journal}
  {\bibinfo  {journal} {Phys. Rev. Applied}\ }\textbf {\bibinfo {volume}
  {14}},\ \bibinfo {pages} {024092} (\bibinfo {year} {2020})}\BibitemShut
  {NoStop}%
\bibitem [{\citenamefont {Kamin}\ \emph
  {et~al.}(2020{\natexlab{b}})\citenamefont {Kamin}, \citenamefont {Tabesh},
  \citenamefont {Salimi}, \citenamefont {Kheirandish},\ and\ \citenamefont
  {Santos}}]{Kamin:1}%
  \BibitemOpen
  \bibfield  {author} {\bibinfo {author} {\bibfnamefont {F.~H.}\ \bibnamefont
  {Kamin}}, \bibinfo {author} {\bibfnamefont {F.~T.}\ \bibnamefont {Tabesh}},
  \bibinfo {author} {\bibfnamefont {S.}~\bibnamefont {Salimi}}, \bibinfo
  {author} {\bibfnamefont {F.}~\bibnamefont {Kheirandish}},\ and\ \bibinfo
  {author} {\bibfnamefont {A.~C.}\ \bibnamefont {Santos}},\ }\bibfield  {title}
  {\bibinfo {title} {Non-markovian effects on charging and self-discharging
  process of quantum batteries},\ }\href
  {https://doi.org/10.1088/1367-2630/ab9ee2} {\bibfield  {journal} {\bibinfo
  {journal} {New Journal of Physics}\ }\textbf {\bibinfo {volume} {22}},\
  \bibinfo {pages} {083007} (\bibinfo {year} {2020}{\natexlab{b}})}\BibitemShut
  {NoStop}%
\bibitem [{\citenamefont {Tabesh}\ \emph {et~al.}(2020)\citenamefont {Tabesh},
  \citenamefont {Kamin},\ and\ \citenamefont {Salimi}}]{Tabesh:1}%
  \BibitemOpen
  \bibfield  {author} {\bibinfo {author} {\bibfnamefont {F.~T.}\ \bibnamefont
  {Tabesh}}, \bibinfo {author} {\bibfnamefont {F.~H.}\ \bibnamefont {Kamin}},\
  and\ \bibinfo {author} {\bibfnamefont {S.}~\bibnamefont {Salimi}},\
  }\bibfield  {title} {\bibinfo {title} {Environment-mediated charging process
  of quantum batteries},\ }\href {https://doi.org/10.1103/PhysRevA.102.052223}
  {\bibfield  {journal} {\bibinfo  {journal} {Phys. Rev. A}\ }\textbf {\bibinfo
  {volume} {102}},\ \bibinfo {pages} {052223} (\bibinfo {year}
  {2020})}\BibitemShut {NoStop}%
\bibitem [{\citenamefont {Carrega}\ \emph {et~al.}(2020)\citenamefont
  {Carrega}, \citenamefont {Crescente}, \citenamefont {Ferraro},\ and\
  \citenamefont {Sassetti}}]{Carrega:1}%
  \BibitemOpen
  \bibfield  {author} {\bibinfo {author} {\bibfnamefont {M.}~\bibnamefont
  {Carrega}}, \bibinfo {author} {\bibfnamefont {A.}~\bibnamefont {Crescente}},
  \bibinfo {author} {\bibfnamefont {D.}~\bibnamefont {Ferraro}},\ and\ \bibinfo
  {author} {\bibfnamefont {M.}~\bibnamefont {Sassetti}},\ }\bibfield  {title}
  {\bibinfo {title} {Dissipative dynamics of an open quantum battery},\ }\href
  {https://doi.org/10.1088/1367-2630/abaa01} {\bibfield  {journal} {\bibinfo
  {journal} {New Journal of Physics}\ }\textbf {\bibinfo {volume} {22}},\
  \bibinfo {pages} {083085} (\bibinfo {year} {2020})}\BibitemShut {NoStop}%
\bibitem [{\citenamefont {Quach}\ \emph {et~al.}(2022)\citenamefont {Quach},
  \citenamefont {McGhee}, \citenamefont {Ganzer}, \citenamefont {Rouse},
  \citenamefont {Lovett}, \citenamefont {Gauger}, \citenamefont {Keeling},
  \citenamefont {Cerullo}, \citenamefont {Lidzey},\ and\ \citenamefont
  {Virgili}}]{James:1}%
  \BibitemOpen
  \bibfield  {author} {\bibinfo {author} {\bibfnamefont {J.~Q.}\ \bibnamefont
  {Quach}}, \bibinfo {author} {\bibfnamefont {K.~E.}\ \bibnamefont {McGhee}},
  \bibinfo {author} {\bibfnamefont {L.}~\bibnamefont {Ganzer}}, \bibinfo
  {author} {\bibfnamefont {D.~M.}\ \bibnamefont {Rouse}}, \bibinfo {author}
  {\bibfnamefont {B.~W.}\ \bibnamefont {Lovett}}, \bibinfo {author}
  {\bibfnamefont {E.~M.}\ \bibnamefont {Gauger}}, \bibinfo {author}
  {\bibfnamefont {J.}~\bibnamefont {Keeling}}, \bibinfo {author} {\bibfnamefont
  {G.}~\bibnamefont {Cerullo}}, \bibinfo {author} {\bibfnamefont {D.~G.}\
  \bibnamefont {Lidzey}},\ and\ \bibinfo {author} {\bibfnamefont
  {T.}~\bibnamefont {Virgili}},\ }\bibfield  {title} {\bibinfo {title}
  {Superabsorption in an organic microcavity: Toward a quantum battery},\
  }\href {https://doi.org/10.1126/sciadv.abk3160} {\bibfield  {journal}
  {\bibinfo  {journal} {Science Advances}\ }\textbf {\bibinfo {volume} {8}},\
  \bibinfo {pages} {eabk3160} (\bibinfo {year} {2022})}\BibitemShut {NoStop}%
\bibitem [{\citenamefont {Wenniger}\ \emph {et~al.}(2022)\citenamefont
  {Wenniger}, \citenamefont {Thomas}, \citenamefont {Maffei}, \citenamefont
  {Wein}, \citenamefont {Pont}, \citenamefont {Harouri}, \citenamefont
  {Lema{\^\i}tre}, \citenamefont {Sagnes}, \citenamefont {Somaschi},
  \citenamefont {Auff{\`e}ves} \emph {et~al.}}]{wenniger:1}%
  \BibitemOpen
  \bibfield  {author} {\bibinfo {author} {\bibfnamefont {I.}~\bibnamefont
  {Wenniger}}, \bibinfo {author} {\bibfnamefont {S.}~\bibnamefont {Thomas}},
  \bibinfo {author} {\bibfnamefont {M.}~\bibnamefont {Maffei}}, \bibinfo
  {author} {\bibfnamefont {S.}~\bibnamefont {Wein}}, \bibinfo {author}
  {\bibfnamefont {M.}~\bibnamefont {Pont}}, \bibinfo {author} {\bibfnamefont
  {A.}~\bibnamefont {Harouri}}, \bibinfo {author} {\bibfnamefont
  {A.}~\bibnamefont {Lema{\^\i}tre}}, \bibinfo {author} {\bibfnamefont
  {I.}~\bibnamefont {Sagnes}}, \bibinfo {author} {\bibfnamefont
  {N.}~\bibnamefont {Somaschi}}, \bibinfo {author} {\bibfnamefont
  {A.}~\bibnamefont {Auff{\`e}ves}}, \emph {et~al.},\ }\bibfield  {title}
  {\bibinfo {title} {Coherence-powered charge and discharge of a quantum
  battery},\ }\href {https://arxiv.org/abs/2202.01109} {\bibfield  {journal}
  {\bibinfo  {journal} {arXiv:2202.01109}\ } (\bibinfo {year}
  {2022})}\BibitemShut {NoStop}%
\bibitem [{\citenamefont {Hu}\ \emph {et~al.}(2022)\citenamefont {Hu},
  \citenamefont {Qiu}, \citenamefont {Souza}, \citenamefont {Yuan},
  \citenamefont {Zhou}, \citenamefont {Zhang}, \citenamefont {Chu},
  \citenamefont {Pan}, \citenamefont {Hu}, \citenamefont {Li}, \citenamefont
  {Xu}, \citenamefont {Zhong}, \citenamefont {Liu}, \citenamefont {Yan},
  \citenamefont {Tan}, \citenamefont {Bachelard}, \citenamefont {Villas-Boas},
  \citenamefont {Santos},\ and\ \citenamefont {Yu}}]{Hu:1}%
  \BibitemOpen
  \bibfield  {author} {\bibinfo {author} {\bibfnamefont {C.-K.}\ \bibnamefont
  {Hu}}, \bibinfo {author} {\bibfnamefont {J.}~\bibnamefont {Qiu}}, \bibinfo
  {author} {\bibfnamefont {P.~J.~P.}\ \bibnamefont {Souza}}, \bibinfo {author}
  {\bibfnamefont {J.}~\bibnamefont {Yuan}}, \bibinfo {author} {\bibfnamefont
  {Y.}~\bibnamefont {Zhou}}, \bibinfo {author} {\bibfnamefont {L.}~\bibnamefont
  {Zhang}}, \bibinfo {author} {\bibfnamefont {J.}~\bibnamefont {Chu}}, \bibinfo
  {author} {\bibfnamefont {X.}~\bibnamefont {Pan}}, \bibinfo {author}
  {\bibfnamefont {L.}~\bibnamefont {Hu}}, \bibinfo {author} {\bibfnamefont
  {J.}~\bibnamefont {Li}}, \bibinfo {author} {\bibfnamefont {Y.}~\bibnamefont
  {Xu}}, \bibinfo {author} {\bibfnamefont {Y.}~\bibnamefont {Zhong}}, \bibinfo
  {author} {\bibfnamefont {S.}~\bibnamefont {Liu}}, \bibinfo {author}
  {\bibfnamefont {F.}~\bibnamefont {Yan}}, \bibinfo {author} {\bibfnamefont
  {D.}~\bibnamefont {Tan}}, \bibinfo {author} {\bibfnamefont {R.}~\bibnamefont
  {Bachelard}}, \bibinfo {author} {\bibfnamefont {C.~J.}\ \bibnamefont
  {Villas-Boas}}, \bibinfo {author} {\bibfnamefont {A.~C.}\ \bibnamefont
  {Santos}},\ and\ \bibinfo {author} {\bibfnamefont {D.}~\bibnamefont {Yu}},\
  }\bibfield  {title} {\bibinfo {title} {Optimal charging of a superconducting
  quantum battery},\ }\href {https://doi.org/10.1088/2058-9565/ac8444}
  {\bibfield  {journal} {\bibinfo  {journal} {Quantum Science and Technology}\
  }\textbf {\bibinfo {volume} {7}},\ \bibinfo {pages} {045018} (\bibinfo {year}
  {2022})}\BibitemShut {NoStop}%
\bibitem [{\citenamefont {Zheng}\ \emph {et~al.}(2022)\citenamefont {Zheng},
  \citenamefont {Ning}, \citenamefont {Yang}, \citenamefont {Xia},\ and\
  \citenamefont {Zheng}}]{Zheng:1}%
  \BibitemOpen
  \bibfield  {author} {\bibinfo {author} {\bibfnamefont {R.-H.}\ \bibnamefont
  {Zheng}}, \bibinfo {author} {\bibfnamefont {W.}~\bibnamefont {Ning}},
  \bibinfo {author} {\bibfnamefont {Z.-B.}\ \bibnamefont {Yang}}, \bibinfo
  {author} {\bibfnamefont {Y.}~\bibnamefont {Xia}},\ and\ \bibinfo {author}
  {\bibfnamefont {S.-B.}\ \bibnamefont {Zheng}},\ }\bibfield  {title} {\bibinfo
  {title} {Demonstration of dynamical control of three-level open systems with
  a superconducting qutrit},\ }\href {https://doi.org/10.1088/1367-2630/ac788f}
  {\bibfield  {journal} {\bibinfo  {journal} {New Journal of Physics}\ }\textbf
  {\bibinfo {volume} {24}},\ \bibinfo {pages} {063031} (\bibinfo {year}
  {2022})}\BibitemShut {NoStop}%
\bibitem [{\citenamefont {Gemme}\ \emph {et~al.}(2022)\citenamefont {Gemme},
  \citenamefont {Grossi}, \citenamefont {Ferraro}, \citenamefont {Vallecorsa},\
  and\ \citenamefont {Sassetti}}]{Gemme:1}%
  \BibitemOpen
  \bibfield  {author} {\bibinfo {author} {\bibfnamefont {G.}~\bibnamefont
  {Gemme}}, \bibinfo {author} {\bibfnamefont {M.}~\bibnamefont {Grossi}},
  \bibinfo {author} {\bibfnamefont {D.}~\bibnamefont {Ferraro}}, \bibinfo
  {author} {\bibfnamefont {S.}~\bibnamefont {Vallecorsa}},\ and\ \bibinfo
  {author} {\bibfnamefont {M.}~\bibnamefont {Sassetti}},\ }\bibfield  {title}
  {\bibinfo {title} {Ibm quantum platforms: A quantum battery perspective},\
  }\bibfield  {journal} {\bibinfo  {journal} {Batteries}\ }\textbf {\bibinfo
  {volume} {8}},\ \href {https://doi.org/10.3390/batteries8050043}
  {10.3390/batteries8050043} (\bibinfo {year} {2022})\BibitemShut {NoStop}%
\bibitem [{\citenamefont {Joshi}\ and\ \citenamefont {Mahesh}(2022)}]{Joshi:1}%
  \BibitemOpen
  \bibfield  {author} {\bibinfo {author} {\bibfnamefont {J.}~\bibnamefont
  {Joshi}}\ and\ \bibinfo {author} {\bibfnamefont {T.~S.}\ \bibnamefont
  {Mahesh}},\ }\bibfield  {title} {\bibinfo {title} {Experimental investigation
  of a quantum battery using star-topology nmr spin systems},\ }\href
  {https://doi.org/10.1103/PhysRevA.106.042601} {\bibfield  {journal} {\bibinfo
   {journal} {Phys. Rev. A}\ }\textbf {\bibinfo {volume} {106}},\ \bibinfo
  {pages} {042601} (\bibinfo {year} {2022})}\BibitemShut {NoStop}%
\bibitem [{\citenamefont {Santos}\ \emph {et~al.}(2019)\citenamefont {Santos},
  \citenamefont {\ifmmode~\mbox{\c{C}}\else \c{C}\fi{}akmak}, \citenamefont
  {Campbell},\ and\ \citenamefont {Zinner}}]{Santos:1}%
  \BibitemOpen
  \bibfield  {author} {\bibinfo {author} {\bibfnamefont {A.~C.}\ \bibnamefont
  {Santos}}, \bibinfo {author} {\bibfnamefont {B.~i. e. i. f. m.~c.}\
  \bibnamefont {\ifmmode~\mbox{\c{C}}\else \c{C}\fi{}akmak}}, \bibinfo {author}
  {\bibfnamefont {S.}~\bibnamefont {Campbell}},\ and\ \bibinfo {author}
  {\bibfnamefont {N.~T.}\ \bibnamefont {Zinner}},\ }\bibfield  {title}
  {\bibinfo {title} {Stable adiabatic quantum batteries},\ }\href
  {https://doi.org/10.1103/PhysRevE.100.032107} {\bibfield  {journal} {\bibinfo
   {journal} {Phys. Rev. E}\ }\textbf {\bibinfo {volume} {100}},\ \bibinfo
  {pages} {032107} (\bibinfo {year} {2019})}\BibitemShut {NoStop}%
\bibitem [{\citenamefont {Kamin}\ \emph {et~al.}(2021)\citenamefont {Kamin},
  \citenamefont {Salimi},\ and\ \citenamefont {Santos}}]{Kamin:2}%
  \BibitemOpen
  \bibfield  {author} {\bibinfo {author} {\bibfnamefont {F.~H.}\ \bibnamefont
  {Kamin}}, \bibinfo {author} {\bibfnamefont {S.}~\bibnamefont {Salimi}},\ and\
  \bibinfo {author} {\bibfnamefont {A.~C.}\ \bibnamefont {Santos}},\ }\bibfield
   {title} {\bibinfo {title} {Exergy of passive states: Waste energy after
  ergotropy extraction},\ }\href {https://doi.org/10.1103/PhysRevE.104.034134}
  {\bibfield  {journal} {\bibinfo  {journal} {Phys. Rev. E}\ }\textbf {\bibinfo
  {volume} {104}},\ \bibinfo {pages} {034134} (\bibinfo {year}
  {2021})}\BibitemShut {NoStop}%
\bibitem [{\citenamefont {Datta}(1997)}]{Datta:1}%
  \BibitemOpen
  \bibfield  {author} {\bibinfo {author} {\bibfnamefont {S.}~\bibnamefont
  {Datta}},\ }\href {https://books.google.com/books?id=28BC-ofEhvUC} {\emph
  {\bibinfo {title} {Electronic Transport in Mesoscopic Systems}}},\ Cambridge
  Studies in Semiconductor Physi\ (\bibinfo  {publisher} {Cambridge University
  Press},\ \bibinfo {year} {1997})\BibitemShut {NoStop}%
\bibitem [{\citenamefont {{Di Ventra}}(2008)}]{Ventra:1}%
  \BibitemOpen
  \bibfield  {author} {\bibinfo {author} {\bibfnamefont {M.}~\bibnamefont {{Di
  Ventra}}},\ }\href {https://ui.adsabs.harvard.edu/abs/2008etns.book.....D}
  {\emph {\bibinfo {title} {Electrical Transport in Nanoscale Systems}}}\
  (\bibinfo {year} {2008})\BibitemShut {NoStop}%
\bibitem [{\citenamefont {Goldsmid}\ \emph {et~al.}(2010)\citenamefont
  {Goldsmid} \emph {et~al.}}]{Goldsmid:1}%
  \BibitemOpen
  \bibfield  {author} {\bibinfo {author} {\bibfnamefont {H.~J.}\ \bibnamefont
  {Goldsmid}} \emph {et~al.},\ }\href
  {https://doi.org/https://doi.org/10.1007/978-3-662-49256-7} {\emph {\bibinfo
  {title} {Introduction to thermoelectricity}}},\ Vol.\ \bibinfo {volume}
  {121}\ (\bibinfo {year} {2010})\BibitemShut {NoStop}%
\bibitem [{\citenamefont {Rowe}(2018)}]{Rowe:1}%
  \BibitemOpen
  \bibfield  {author} {\bibinfo {author} {\bibfnamefont {D.~M.}\ \bibnamefont
  {Rowe}},\ }\href {https://doi.org/https://doi.org/10.1201/9781420038903}
  {\emph {\bibinfo {title} {Thermoelectrics handbook: macro to nano}}}\
  (\bibinfo  {publisher} {CRC press},\ \bibinfo {year} {2018})\BibitemShut
  {NoStop}%
\bibitem [{\citenamefont {K{\=o}moto}\ and\ \citenamefont
  {Mori}(2013)}]{komoto:1}%
  \BibitemOpen
  \bibfield  {author} {\bibinfo {author} {\bibfnamefont {K.}~\bibnamefont
  {K{\=o}moto}}\ and\ \bibinfo {author} {\bibfnamefont {T.}~\bibnamefont
  {Mori}},\ }\href {https://doi.org/https://doi.org/10.1007/978-3-642-37537-8}
  {\emph {\bibinfo {title} {Thermoelectric Nanomaterials: Materials Design and
  Applications}}}\ (\bibinfo  {publisher} {Springer},\ \bibinfo {year}
  {2013})\BibitemShut {NoStop}%
\bibitem [{\citenamefont {Dresselhaus}\ \emph {et~al.}(2007)\citenamefont
  {Dresselhaus}, \citenamefont {Chen}, \citenamefont {Tang}, \citenamefont
  {Yang}, \citenamefont {Lee}, \citenamefont {Wang}, \citenamefont {Ren},
  \citenamefont {Fleurial},\ and\ \citenamefont {Gogna}}]{Dresselhaus:1}%
  \BibitemOpen
  \bibfield  {author} {\bibinfo {author} {\bibfnamefont {M.}~\bibnamefont
  {Dresselhaus}}, \bibinfo {author} {\bibfnamefont {G.}~\bibnamefont {Chen}},
  \bibinfo {author} {\bibfnamefont {M.}~\bibnamefont {Tang}}, \bibinfo {author}
  {\bibfnamefont {R.}~\bibnamefont {Yang}}, \bibinfo {author} {\bibfnamefont
  {H.}~\bibnamefont {Lee}}, \bibinfo {author} {\bibfnamefont {D.}~\bibnamefont
  {Wang}}, \bibinfo {author} {\bibfnamefont {Z.}~\bibnamefont {Ren}}, \bibinfo
  {author} {\bibfnamefont {J.-P.}\ \bibnamefont {Fleurial}},\ and\ \bibinfo
  {author} {\bibfnamefont {P.}~\bibnamefont {Gogna}},\ }\bibfield  {title}
  {\bibinfo {title} {New directions for low-dimensional thermoelectric
  materials},\ }\href {https://doi.org/https://doi.org/10.1002/adma.200600527}
  {\bibfield  {journal} {\bibinfo  {journal} {Advanced Materials}\ }\textbf
  {\bibinfo {volume} {19}},\ \bibinfo {pages} {1043} (\bibinfo {year}
  {2007})}\BibitemShut {NoStop}%
\bibitem [{\citenamefont {Maciejko}\ \emph {et~al.}(2006)\citenamefont
  {Maciejko}, \citenamefont {Wang},\ and\ \citenamefont {Guo}}]{Maciejko:1}%
  \BibitemOpen
  \bibfield  {author} {\bibinfo {author} {\bibfnamefont {J.}~\bibnamefont
  {Maciejko}}, \bibinfo {author} {\bibfnamefont {J.}~\bibnamefont {Wang}},\
  and\ \bibinfo {author} {\bibfnamefont {H.}~\bibnamefont {Guo}},\ }\bibfield
  {title} {\bibinfo {title} {Time-dependent quantum transport far from
  equilibrium: An exact nonlinear response theory},\ }\href
  {https://doi.org/10.1103/PhysRevB.74.085324} {\bibfield  {journal} {\bibinfo
  {journal} {Phys. Rev. B}\ }\textbf {\bibinfo {volume} {74}},\ \bibinfo
  {pages} {085324} (\bibinfo {year} {2006})}\BibitemShut {NoStop}%
\bibitem [{\citenamefont {Stefanucci}\ and\ \citenamefont {van
  Leeuwen}(2013)}]{Stefanucci:1}%
  \BibitemOpen
  \bibfield  {author} {\bibinfo {author} {\bibfnamefont {G.}~\bibnamefont
  {Stefanucci}}\ and\ \bibinfo {author} {\bibfnamefont {R.}~\bibnamefont {van
  Leeuwen}},\ }\href {https://books.google.com/books?id=6GsrjPFXLDYC} {\emph
  {\bibinfo {title} {Nonequilibrium Many-Body Theory of Quantum Systems: A
  Modern Introduction}}}\ (\bibinfo  {publisher} {Cambridge University Press},\
  \bibinfo {year} {2013})\BibitemShut {NoStop}%
\bibitem [{\citenamefont {McLennan}(1959)}]{McLennan:1}%
  \BibitemOpen
  \bibfield  {author} {\bibinfo {author} {\bibfnamefont {J.~A.}\ \bibnamefont
  {McLennan}},\ }\bibfield  {title} {\bibinfo {title} {Statistical mechanics of
  the steady state},\ }\href {https://doi.org/10.1103/PhysRev.115.1405}
  {\bibfield  {journal} {\bibinfo  {journal} {Phys. Rev.}\ }\textbf {\bibinfo
  {volume} {115}},\ \bibinfo {pages} {1405} (\bibinfo {year}
  {1959})}\BibitemShut {NoStop}%
\bibitem [{\citenamefont {McLennan~Jr.}(1963)}]{McLennan:2}%
  \BibitemOpen
  \bibfield  {author} {\bibinfo {author} {\bibfnamefont {J.~A.}\ \bibnamefont
  {McLennan~Jr.}},\ }\bibinfo {title} {The formal statistical theory of
  transport processes},\ in\ \href
  {https://doi.org/https://doi.org/10.1002/9780470143513.ch6} {\emph {\bibinfo
  {booktitle} {Advances in Chemical Physics}}}\ (\bibinfo  {publisher} {John
  Wiley {\&} Sons, Ltd},\ \bibinfo {year} {1963})\ pp.\ \bibinfo {pages}
  {261--317}\BibitemShut {NoStop}%
\bibitem [{\citenamefont {Zubarev}(1994)}]{Zubarev:1}%
  \BibitemOpen
  \bibfield  {author} {\bibinfo {author} {\bibfnamefont {D.}~\bibnamefont
  {Zubarev}},\ }\bibfield  {title} {\bibinfo {title} {Nonequilibrium
  statistical operator as a generalization of gibbs distribution for
  nonequilibrium case},\ }\bibfield  {journal} {\bibinfo  {journal} {Cond.
  Matt. Phys}\ }\textbf {\bibinfo {volume} {4}},\ \href
  {https://doi.org/10.5488/CMP.4.7} {10.5488/CMP.4.7} (\bibinfo {year}
  {1994})\BibitemShut {NoStop}%
\bibitem [{\citenamefont {Zubarev}(1970)}]{Zubarev:2}%
  \BibitemOpen
  \bibfield  {author} {\bibinfo {author} {\bibfnamefont {D.~N.}\ \bibnamefont
  {Zubarev}},\ }\bibfield  {title} {\bibinfo {title} {The method of the
  non-equilibrium statistical operator and its applications. i},\ }\href
  {https://doi.org/https://doi.org/10.1002/prop.19700180302} {\bibfield
  {journal} {\bibinfo  {journal} {Fortschritte der Physik}\ }\textbf {\bibinfo
  {volume} {18}},\ \bibinfo {pages} {125} (\bibinfo {year} {1970})}\BibitemShut
  {NoStop}%
\bibitem [{\citenamefont {Hershfield}(1993)}]{Hershfield:1}%
  \BibitemOpen
  \bibfield  {author} {\bibinfo {author} {\bibfnamefont {S.}~\bibnamefont
  {Hershfield}},\ }\bibfield  {title} {\bibinfo {title} {Reformulation of
  steady state nonequilibrium quantum statistical mechanics},\ }\href
  {https://doi.org/10.1103/PhysRevLett.70.2134} {\bibfield  {journal} {\bibinfo
   {journal} {Phys. Rev. Lett.}\ }\textbf {\bibinfo {volume} {70}},\ \bibinfo
  {pages} {2134} (\bibinfo {year} {1993})}\BibitemShut {NoStop}%
\bibitem [{\citenamefont {Ness}(2013)}]{Ness:1}%
  \BibitemOpen
  \bibfield  {author} {\bibinfo {author} {\bibfnamefont {H.}~\bibnamefont
  {Ness}},\ }\bibfield  {title} {\bibinfo {title} {Nonequilibrium density
  matrix for quantum transport: Hershfield approach as a mclennan-zubarev form
  of the statistical operator},\ }\href
  {https://doi.org/10.1103/PhysRevE.88.022121} {\bibfield  {journal} {\bibinfo
  {journal} {Phys. Rev. E}\ }\textbf {\bibinfo {volume} {88}},\ \bibinfo
  {pages} {022121} (\bibinfo {year} {2013})}\BibitemShut {NoStop}%
\bibitem [{\citenamefont {Ness}(2014{\natexlab{a}})}]{Ness:1a}%
  \BibitemOpen
  \bibfield  {author} {\bibinfo {author} {\bibfnamefont {H.}~\bibnamefont
  {Ness}},\ }\bibfield  {title} {\bibinfo {title} {Nonequilibrium density
  matrix in quantum open systems: Generalization for simultaneous heat and
  charge steady-state transport},\ }\href
  {https://doi.org/10.1103/PhysRevE.90.062119} {\bibfield  {journal} {\bibinfo
  {journal} {Phys. Rev. E}\ }\textbf {\bibinfo {volume} {90}},\ \bibinfo
  {pages} {062119} (\bibinfo {year} {2014}{\natexlab{a}})}\BibitemShut
  {NoStop}%
\bibitem [{\citenamefont {Gell-Mann}\ and\ \citenamefont
  {Goldberger}(1953)}]{Gell-Mann:1}%
  \BibitemOpen
  \bibfield  {author} {\bibinfo {author} {\bibfnamefont {M.}~\bibnamefont
  {Gell-Mann}}\ and\ \bibinfo {author} {\bibfnamefont {M.~L.}\ \bibnamefont
  {Goldberger}},\ }\bibfield  {title} {\bibinfo {title} {The formal theory of
  scattering},\ }\href {https://doi.org/10.1103/PhysRev.91.398} {\bibfield
  {journal} {\bibinfo  {journal} {Phys. Rev.}\ }\textbf {\bibinfo {volume}
  {91}},\ \bibinfo {pages} {398} (\bibinfo {year} {1953})}\BibitemShut
  {NoStop}%
\bibitem [{\citenamefont {B{\"o}hm}\ and\ \citenamefont
  {Loewe}(2013)}]{Bohm:1}%
  \BibitemOpen
  \bibfield  {author} {\bibinfo {author} {\bibfnamefont {A.}~\bibnamefont
  {B{\"o}hm}}\ and\ \bibinfo {author} {\bibfnamefont {M.}~\bibnamefont
  {Loewe}},\ }\href {https://books.google.com/books?id=njT6CAAAQBAJ} {\emph
  {\bibinfo {title} {Quantum Mechanics: Foundations and Applications}}},\
  Theoretical and Mathematical Physics\ (\bibinfo  {publisher} {Springer Berlin
  Heidelberg},\ \bibinfo {year} {2013})\BibitemShut {NoStop}%
\bibitem [{\citenamefont {Baute}\ \emph {et~al.}(2001)\citenamefont {Baute},
  \citenamefont {Egusquiza},\ and\ \citenamefont {Muga}}]{Baute:1}%
  \BibitemOpen
  \bibfield  {author} {\bibinfo {author} {\bibfnamefont {A.~D.}\ \bibnamefont
  {Baute}}, \bibinfo {author} {\bibfnamefont {I.~L.}\ \bibnamefont
  {Egusquiza}},\ and\ \bibinfo {author} {\bibfnamefont {J.~G.}\ \bibnamefont
  {Muga}},\ }\bibfield  {title} {\bibinfo {title} {Moller operators and
  lippmann-schwinger equations for steplike potentials},\ }\href
  {https://doi.org/10.1088/0305-4470/34/26/305} {\bibfield  {journal} {\bibinfo
   {journal} {Journal of Physics A: Mathematical and General}\ }\textbf
  {\bibinfo {volume} {34}},\ \bibinfo {pages} {5341} (\bibinfo {year}
  {2001})}\BibitemShut {NoStop}%
\bibitem [{\citenamefont {De~Groot}\ and\ \citenamefont
  {Mazur}(2013)}]{meixner:1}%
  \BibitemOpen
  \bibfield  {author} {\bibinfo {author} {\bibfnamefont {S.~R.}\ \bibnamefont
  {De~Groot}}\ and\ \bibinfo {author} {\bibfnamefont {P.}~\bibnamefont
  {Mazur}},\ }\href {https://books.google.com/books?id=mfFyG9jfaMYC} {\emph
  {\bibinfo {title} {Non-equilibrium thermodynamics}}}\ (\bibinfo  {publisher}
  {Courier Corporation},\ \bibinfo {year} {2013})\BibitemShut {NoStop}%
\bibitem [{\citenamefont {Bruch}\ \emph {et~al.}(2018)\citenamefont {Bruch},
  \citenamefont {Lewenkopf},\ and\ \citenamefont {von Oppen}}]{Bruch:2}%
  \BibitemOpen
  \bibfield  {author} {\bibinfo {author} {\bibfnamefont {A.}~\bibnamefont
  {Bruch}}, \bibinfo {author} {\bibfnamefont {C.}~\bibnamefont {Lewenkopf}},\
  and\ \bibinfo {author} {\bibfnamefont {F.}~\bibnamefont {von Oppen}},\
  }\bibfield  {title} {\bibinfo {title} {Landauer-b{\"u}ttiker approach to
  strongly coupled quantum thermodynamics: Inside-outside duality of entropy
  evolution},\ }\href
  {https://doi.org/https://doi.org/10.1103/PhysRevLett.120.107701} {\bibfield
  {journal} {\bibinfo  {journal} {Physical review letters}\ }\textbf {\bibinfo
  {volume} {120}},\ \bibinfo {pages} {107701} (\bibinfo {year}
  {2018})}\BibitemShut {NoStop}%
\bibitem [{\citenamefont {Ness}(2017)}]{Ness:2}%
  \BibitemOpen
  \bibfield  {author} {\bibinfo {author} {\bibfnamefont {H.}~\bibnamefont
  {Ness}},\ }\bibfield  {title} {\bibinfo {title} {Nonequilibrium
  thermodynamics and steady state density matrix for quantum open systems},\
  }\href {https://www.mdpi.com/1099-4300/19/4/158} {\bibfield  {journal}
  {\bibinfo  {journal} {Entropy}\ }\textbf {\bibinfo {volume} {19}} (\bibinfo
  {year} {2017})}\BibitemShut {NoStop}%
\bibitem [{\citenamefont {Guarnieri}\ \emph {et~al.}(2019)\citenamefont
  {Guarnieri}, \citenamefont {Landi}, \citenamefont {Clark},\ and\
  \citenamefont {Goold}}]{Guarnieri:1}%
  \BibitemOpen
  \bibfield  {author} {\bibinfo {author} {\bibfnamefont {G.}~\bibnamefont
  {Guarnieri}}, \bibinfo {author} {\bibfnamefont {G.~T.}\ \bibnamefont
  {Landi}}, \bibinfo {author} {\bibfnamefont {S.~R.}\ \bibnamefont {Clark}},\
  and\ \bibinfo {author} {\bibfnamefont {J.}~\bibnamefont {Goold}},\ }\bibfield
   {title} {\bibinfo {title} {Thermodynamics of precision in quantum
  nonequilibrium steady states},\ }\href
  {https://doi.org/10.1103/PhysRevResearch.1.033021} {\bibfield  {journal}
  {\bibinfo  {journal} {Phys. Rev. Research}\ }\textbf {\bibinfo {volume}
  {1}},\ \bibinfo {pages} {033021} (\bibinfo {year} {2019})}\BibitemShut
  {NoStop}%
\bibitem [{\citenamefont {Schl{\"o}gl}(2013)}]{Friedrich:1}%
  \BibitemOpen
  \bibfield  {author} {\bibinfo {author} {\bibfnamefont {F.}~\bibnamefont
  {Schl{\"o}gl}},\ }\href {https://books.google.com/books?id=e2\_SBgAAQBAJ}
  {\emph {\bibinfo {title} {Probability and Heat: Fundamentals of
  Thermostatistics}}}\ (\bibinfo  {publisher} {Vieweg+Teubner Verlag},\
  \bibinfo {year} {2013})\BibitemShut {NoStop}%
\bibitem [{\citenamefont {Esposito}\ \emph {et~al.}(2015)\citenamefont
  {Esposito}, \citenamefont {Ochoa},\ and\ \citenamefont
  {Galperin}}]{Esposito:1}%
  \BibitemOpen
  \bibfield  {author} {\bibinfo {author} {\bibfnamefont {M.}~\bibnamefont
  {Esposito}}, \bibinfo {author} {\bibfnamefont {M.~A.}\ \bibnamefont
  {Ochoa}},\ and\ \bibinfo {author} {\bibfnamefont {M.}~\bibnamefont
  {Galperin}},\ }\bibfield  {title} {\bibinfo {title} {Quantum thermodynamics:
  A nonequilibrium green's function approach},\ }\href
  {https://doi.org/10.1103/PhysRevLett.114.080602} {\bibfield  {journal}
  {\bibinfo  {journal} {Phys. Rev. Lett.}\ }\textbf {\bibinfo {volume} {114}},\
  \bibinfo {pages} {080602} (\bibinfo {year} {2015})}\BibitemShut {NoStop}%
\bibitem [{\citenamefont {Meir}\ and\ \citenamefont
  {Wingreen}(1992)}]{Meir:1992}%
  \BibitemOpen
  \bibfield  {author} {\bibinfo {author} {\bibfnamefont {Y.}~\bibnamefont
  {Meir}}\ and\ \bibinfo {author} {\bibfnamefont {N.~S.}\ \bibnamefont
  {Wingreen}},\ }\bibfield  {title} {\bibinfo {title} {Landauer formula for the
  current through an interacting electron region},\ }\href
  {https://doi.org/10.1103/PhysRevLett.68.2512} {\bibfield  {journal} {\bibinfo
   {journal} {Phys. Rev. Lett.}\ }\textbf {\bibinfo {volume} {68}},\ \bibinfo
  {pages} {2512} (\bibinfo {year} {1992})}\BibitemShut {NoStop}%
\bibitem [{\citenamefont {Ness}(2014{\natexlab{b}})}]{Ness:2014}%
  \BibitemOpen
  \bibfield  {author} {\bibinfo {author} {\bibfnamefont {H.}~\bibnamefont
  {Ness}},\ }\bibfield  {title} {\bibinfo {title} {Nonequilibrium distribution
  functions for quantum transport: Universality and approximation for the
  steady state regime},\ }\href {https://doi.org/10.1103/PhysRevB.89.045409}
  {\bibfield  {journal} {\bibinfo  {journal} {Phys. Rev. B}\ }\textbf {\bibinfo
  {volume} {89}},\ \bibinfo {pages} {045409} (\bibinfo {year}
  {2014}{\natexlab{b}})}\BibitemShut {NoStop}%
\bibitem [{\citenamefont {Dhar}\ \emph {et~al.}(2012)\citenamefont {Dhar},
  \citenamefont {Saito},\ and\ \citenamefont {H\"anggi}}]{Dhar:1}%
  \BibitemOpen
  \bibfield  {author} {\bibinfo {author} {\bibfnamefont {A.}~\bibnamefont
  {Dhar}}, \bibinfo {author} {\bibfnamefont {K.}~\bibnamefont {Saito}},\ and\
  \bibinfo {author} {\bibfnamefont {P.}~\bibnamefont {H\"anggi}},\ }\bibfield
  {title} {\bibinfo {title} {Nonequilibrium density-matrix description of
  steady-state quantum transport},\ }\href
  {https://doi.org/10.1103/PhysRevE.85.011126} {\bibfield  {journal} {\bibinfo
  {journal} {Phys. Rev. E}\ }\textbf {\bibinfo {volume} {85}},\ \bibinfo
  {pages} {011126} (\bibinfo {year} {2012})}\BibitemShut {NoStop}%
\end{thebibliography}
\end{document}